\documentclass[twocolumn,nofootinbib,preprintnumbers,floatfix]{revtex4}

\usepackage{color}
\usepackage{graphicx}
\usepackage{amsmath}
\usepackage{amssymb}
\usepackage{xspace}
\usepackage[small]{subfigure}
\usepackage[hyperfootnotes=false]{hyperref}

\newcommand{\eq}[1]{Eq.~\eqref{eq:#1}}
\newcommand{\eqs}[2]{Eqs.~\eqref{eq:#1} and \eqref{eq:#2}}
\renewcommand{\sec}[1]{Sec.~\ref{sec:#1}}

\newcommand{\subsec}[1]{Sec.~\ref{subsec:#1}}
\newcommand{\fig}[1]{Fig.~\ref{fig:#1}}
\newcommand{\figs}[2]{Figs.~\ref{fig:#1} and \ref{fig:#2}}

\newcommand{\abs}[1]{\lvert#1\rvert}
\newcommand{\Abs}[1]{\bigl\lvert#1\bigr\rvert}
\newcommand{\ord}[1]{\mathcal{O}(#1)}
\newcommand{\ORd}[1]{\mathcal{O}\Bigl(#1\Bigr)}
\newcommand{\mae}[3]{\langle#1\rvert#2\rvert#3\rangle}
\newcommand{\Mae}[3]{\bigl\langle#1\bigr\rvert#2\bigr\rvert#3\bigr\rangle}
\newcommand{\MAe}[3]{\Bigl\langle#1\Bigr\rvert#2\Bigr\rvert#3\Bigr\rangle}
\newcommand{\ket}[1]{\lvert#1\rangle}

\newcommand{\intlim}[3]{\int_{#1}^{#2}\! \df #3 \,}

\newcommand{\df}{\mathrm{d}}
\newcommand{\img}{\mathrm{i}}

\newcommand{\sdt}{\!\cdot\!}
\newcommand{\tr}{\textrm{tr}}

\newcommand{\lra}{\leftrightarrow}

\newcommand{\eps}{\epsilon}
\newcommand{\la}{\lambda}
\newcommand{\w}{\omega}

\newcommand{\cB}{{\mathcal B}}
\newcommand{\cI}{{\mathcal I}}
\newcommand{\cL}{{\mathcal L}}

\newcommand{\cP}{{\mathcal P}}
\newcommand{\cR}{{\mathcal R}}
\newcommand{\cY}{{\mathcal Y}}

\newcommand{\bn}{\bar{n}}

\newcommand{\bC}{\bar C}
\newcommand{\bnP}{\overline {\mathcal P}}
\newcommand{\bT}{\overline{T}}

\newcommand{\hp}{\hat{p}}

\newcommand{\tB}{\widetilde{B}}
\newcommand{\tM}{\widetilde{M}}

\newcommand{\Dslash}{D\!\!\!\!\slash}
\newcommand{\nslash}{n\!\!\!\slash}
\newcommand{\bnslash}{\bar{n}\!\!\!\slash}

\newcommand{\GeV}{\,\mathrm{GeV}}
\newcommand{\TeV}{\,\mathrm{TeV}}

\newcommand{\nn}{\nonumber}

\newcommand{\lqcd}{\Lambda_\mathrm{QCD}}
\newcommand{\alem}{\alpha_\mathrm{em}}

\newcommand{\lb}{ {\tilde{b}} }         % label b
\newcommand{\lp}{ {\tilde{p}} }         % label p
\newcommand{\lw}{ {\tilde{\omega}} }    % label omega

\newcommand{\soft}{{s}}             % label for soft things
\newcommand{\parton}{\mathrm{part}} % label for \sigma partonic
\newcommand{\thr}{\mathrm{thr}}     % label for threshold
\newcommand{\jthr}{\mathrm{2jthr}}  % label for dijet threshold
\newcommand{\hemiin}{\mathrm{ihemi}}
\newcommand{\jhemiin}{{\cone\mathrm{2j}}}
\newcommand{\hemiout}{\mathrm{hemi}}

\newcommand{\cone}{{\sphericalangle}}

\newcommand{\Obs}{O}

\newcommand{\Ecm}{E_\mathrm{cm}}

\newcommand{\cut}{\mathrm{cut}}
\newcommand{\tree}{\mathrm{tree}}
\newcommand{\oneloop}{\mathrm{1loop}}
\newcommand{\cusp}{\mathrm{cusp}}
\renewcommand{\max}{\mathrm{max}}
\newcommand{\incl}{\mathrm{incl}}
\renewcommand{\det}{\mathrm{det}}
\newcommand{\MJJ}{M_{J\!J}}
\newcommand{\mJJ}{m_{J\!J}}

\newcommand{\zero}{{(0)}}
\newcommand{\one}{{(1)}}

\newcommand{\SCETa}{\ensuremath{{\rm SCET}_{\rm I}}\xspace}
\newcommand{\SCETb}{\ensuremath{{\rm SCET}_{\rm II}}\xspace}

\allowdisplaybreaks[2]

\definecolor{mygreen}{rgb}{0,0.65,0}
\definecolor{myblue}{rgb}{0,0,1}
\definecolor{myorange}{rgb}{1,0.5,0}
\definecolor{mydarkblue}{rgb}{0,0,0.6}
\definecolor{myred}{rgb}{1,0,0}

\newcommand{\beamc}[1]{\textcolor{mygreen}{#1}}

\newcommand{\softc}[1]{\textcolor{myorange}{#1}}
\newcommand{\hardc}[1]{\textcolor{mydarkblue}{#1}}

\color{black}

\begin{document}

%%%%%%%%%%%%%%%%%%%%%%%%%%%%%%%%%%%%%%%%%%%%%%%%%%%%%%%%%%%%%%%%%%%%%%%%%%%%%%%%
% Title page
%%%%%%%%%%%%%%%%%%%%%%%%%%%%%%%%%%%%%%%%%%%%%%%%%%%%%%%%%%%%%%%%%%%%%%%%%%%%%%%%

\preprint{\vbox{\hbox{arXiv:0910.0467}\hbox{MIT--CTP 4064}\hbox{October 2, 2009}}}

\title{Factorization at the LHC: From PDFs to Initial State Jets}

\author{Iain W.~Stewart}
\affiliation{Center for Theoretical Physics, Massachusetts Institute of
  Technology, Cambridge, MA 02139\vspace{2ex}}

\author{Frank J.~Tackmann}
\affiliation{Center for Theoretical Physics, Massachusetts Institute of
  Technology, Cambridge, MA 02139\vspace{2ex}}

\author{Wouter J.~Waalewijn\vspace{1ex}}
\affiliation{Center for Theoretical Physics, Massachusetts Institute of
  Technology, Cambridge, MA 02139\vspace{2ex}}

%%%%%%%%%%%%%%%%%%%%%%%%%%%%%%%%%%%%%%%%%%%%%%%%%%%%%%%%%%%%%%%%%%%%%%%%%%%%%%%%
\begin{abstract}

  We study proton-(anti)proton collisions at the LHC or Tevatron in the presence
  of experimental restrictions on the hadronic final state and for generic
  parton momentum fractions.  At the scale $Q$ of the hard interaction,
  factorization does not yield standard parton distribution functions (PDFs) for
  the initial state. The measurement restricting the hadronic final state
  introduces a new scale $\mu_B \ll Q$ and probes the proton prior to the hard
  collision. This corresponds to evaluating the PDFs at the scale $\mu_B$. After
  the proton is probed, the incoming hard parton is contained in an
  initial-state jet, and the hard collision occurs between partons inside these
  jets rather than inside protons. The proper description of such initial-state
  jets requires ``beam functions''.  At the scale $\mu_B$, the beam function
  factorizes into a convolution of calculable Wilson coefficients and PDFs.
  Below $\mu_B$, the initial-state evolution is described by the usual PDF
  evolution which changes $x$, while above $\mu_B$ it is governed by a different
  renormalization group evolution that sums double logarithms of $\mu_B/Q$ and
  leaves $x$ fixed.  As an example, we prove a factorization theorem for
  ``isolated Drell-Yan'', $pp\to X\ell^+\ell^-$ where $X$ is restricted to have
  no central jets. We comment on the extension to cases where the hadronic final
  state contains a certain number of isolated central jets.

\end{abstract}
%%%%%%%%%%%%%%%%%%%%%%%%%%%%%%%%%%%%%%%%%%%%%%%%%%%%%%%%%%%%%%%%%%%%%%%%%%%%%%%%

\maketitle

%%%%%%%%%%%%%%%%%%%%%%%%%%%%%%%%%%%%%%%%%%%%%%%%%%%%%%%%%%%%%%%%%%%%%%%%%%%%%%%%
\section{Introduction}
\label{sec:intro}
%%%%%%%%%%%%%%%%%%%%%%%%%%%%%%%%%%%%%%%%%%%%%%%%%%%%%%%%%%%%%%%%%%%%%%%%%%%%%%%%

Factorization is one of the most basic concepts for
understanding data from the Tevatron at Fermilab and the CERN Large Hadron
Collider (LHC).  For a review of factorization see Ref.~\cite{Collins:1989gx}.
Typically, factorization is viewed as the statement that the
cross section can be computed through a product of probability functions, namely
parton distribution functions (PDFs), describing the probability to extract a
quark or gluon from the protons in the initial state, a perturbative cross
section for the hard scattering, and a probabilistic description of the final
state by a parton shower Monte Carlo or otherwise.  This factorization is of key
importance in the program to search for new physics, as new physics is primarily
a short-distance modification of the hard scattering that must be distinguished
from the array of QCD interactions in the initial and final states.
Factorization is also necessary for controlling QCD effects. For example, the
momentum distributions of the colliding partons in the protons are
nonperturbative, but factorization can imply that these are described by
universal distributions which have been measured in earlier experiments.

As the primary goal of the experiments at the LHC or Tevatron is to probe the
physics of the hard interaction, measurements often impose restrictions on the
hadronic final state, requiring a certain number of hard leptons or jets in the
final state~\cite{:1999fr,:1999fq, Bayatian:2006zz,Ball:2007zza}.  For example,
a typical new physics search looking for missing transverse energy may also
require a minimum number of jets with $p_T$ above some threshold.  To identify
the new physics and determine the masses of new-physics particles, one has to
reconstruct decay chains with a certain number of jets and leptons in the final
state.

Any theoretical prediction for $pp$ or $p\bar p$ collisions, whether analytic or
via Monte Carlo generators, depends on factorization. However, for the majority
of processes of interest at hadron colliders where one distinguishes properties
of the hadronic final state, so far no rigorous field-theoretic
derivation of a factorization theorem to all orders in perturbation theory
exists.
The most well-known factorization theorem is
%%%
\begin{align} \label{eq:sigff}
\df\sigma = \sum_{i,j}\df\sigma^\parton_{ij}
  \otimes f_i(\xi_a)\otimes f_j(\xi_b)
\,,\end{align}
%%%
where $f_i$ and $f_j$ are the standard PDFs for partons $i,j=\{g,u,\bar
u,d,\ldots\}$ carrying momentum fractions $\xi_a$ and $\xi_b$ (which we use as
our PDF $x$-variables), and $\df\sigma^\parton_{ij}$ is the partonic cross
section to scatter $i$ and $j$ calculated in fixed-order perturbation theory.
In \eq{sigff}, the hadronic final state is treated as fully inclusive.  Hence,
in the presence of experimental restrictions that make a process less inclusive,
\eq{sigff} is a priori not applicable. At best, an additional resummation of
large phase-space logarithms must be carried out by a further factorization of
$\df\sigma^\parton_{ij}$, while at worst, additional nonperturbative information
beyond that contained in the PDFs is required or there is no factorization.

Factorization theorems for threshold resummation in hadron-hadron collisions are
a well-studied case where \eq{sigff} can be extended to sum large phase-space
logarithms~\cite{Sterman:1986aj, Catani:1989ne, Kidonakis:1998bk,
Bonciani:1998vc, Laenen:1998qw, Catani:2003zt, Idilbi:2006dg, Becher:2007ty, Chiu:2008vv}.
The corresponding formalism however requires the limit $x\to 1$, and hence is
not directly relevant at the LHC, where the cross section for most measurements
is dominated by the region $x$ far from one~\cite{Campbell:2006wx}.

Our goal is to study factorization for a situation where the hard interaction
occurs between partons with generic momentum fractions, away from the limit
$x\to 1$, and where the hadronic final state is measured and restricted by
constraints on certain kinematic variables.  These restrictions allow one to
probe more details about the final state and may be used experimentally to
isolate central hard jets or leptons or to control backgrounds.

\begin{figure}[t!]
\includegraphics[scale=0.5]{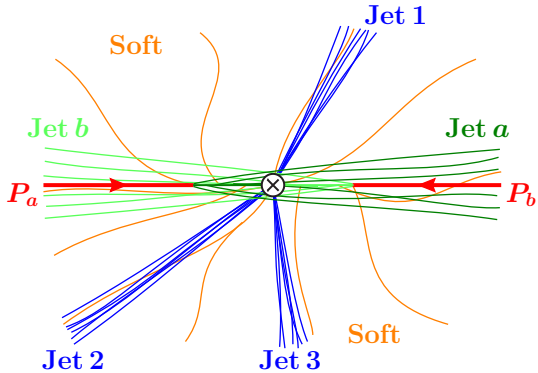}
\caption{A typical event with jet production at the LHC.}
\label{fig:3jetLHC}%
\end{figure}

A typical event at the LHC with three high-$p_T$ jets is illustrated in
\fig{3jetLHC}.  There are several complications one has to face when trying to
derive a factorization theorem in this situation. First, experimentally the number and
properties of the final-state jets are determined with a jet algorithm. Second,
to enhance the ratio of signal over background, the experimental analyses have
to apply kinematic selection cuts.  Third, in addition to the jets produced by
the hard interaction, there is soft radiation everywhere (which is part of what is
sometimes called the ``underlying event''). Fourth, a (large)
fraction of the total energy in the final state is deposited near the beam axes
at high rapidities.  An important component of this radiation can contribute to
measurements, and when it does, it cannot be neglected in the factorization. In
this paper we focus on the last three items. Methods for including
jet algorithms in factorization have been studied in Refs.~\cite{Kidonakis:1998bk, Trott:2006bk,
Bauer:2008jx}

To allow a clean theoretical description, the observables used to constrain the
events must be chosen carefully such that they are infrared safe and sensitive
to emissions everywhere in phase space.  Observables satisfying these criteria
for hadron colliders have been classified and studied in
Refs.~\cite{Banfi:2004nk, Banfi:2005mt}, and are referred to as global event
shapes.  (Issues related to non-global observables have been discussed for
example in Refs.~\cite{Dasgupta:2001sh, Berger:2001ns, Appleby:2003sj,
  Forshaw:2006fk}.)  For our analysis we use a very simple example of such an
observable, constructed as follows. We define two hemispheres, $a$ and $b$,
orthogonal to the beam axis and two unit lightlike vectors $n_a$ and $n_b$ along
the beam axis pointing into each hemisphere. Taking the beam axis along the $z$
direction, hemisphere $a$ is defined as $z > 0$ with $n_a^\mu = (1,0,0,1)$, and
hemisphere $b$ as $z < 0$ with $n_b^\mu = (1,0,0,-1)$. We now divide the total
momentum $p_X$ of the hadronic final state into the contributions from particles
in each hemisphere, $p_X = p_{X_a} + p_{X_b}$.  Next, we remove the momenta
$p_J$ of all jets (defined by an appropriate jet algorithm) in each hemisphere.
Of the remaining hemisphere momenta, we measure the components $B_a^+$ and
$B_b^+$ defined by
%%%
\begin{equation} \label{eq:Badef}
  B_a^+ = n_a \sdt \Bigl(p_{X_a} - \sum_{J\in a} p_J \Bigr)
\,,\end{equation}
%%%
and analogously for $B_b^+$. Because of the dot product with $n_a$ or $n_b$,
energetic particles near the beam axes only give small contributions to $B_a^+$
or $B_b^+$.  In particular, any contributions from particles at very large
rapidities outside the detector reach, including the remnant of unscattered
partons in the proton, are negligible.  All observed particles contribute either
to $B_a^+$, $B_b^+$, or a jet momentum, so we are ensured that we cover all of
phase space.  Demanding that $B_{a,b}^+$ are small restricts the radiation
between central jets, only allowing highly energetic particles either within
these jets or inside jets along the beam directions labeled ``Jet a'' and ``Jet
b'' in \fig{3jetLHC}. Hence, measuring and constraining $B_{a,b}^+$ provides a
theoretically clean method to control the remaining particles in the hadronic
final state. This ensures that observables based on the large momenta of hard
jets or leptons are clean, safe from uncontrolled hadronic effects.

In this paper, we consider the simplest situation where the above setup can be
realized, allowing us to explore the implications of restrictions on the
hadronic final state.  We prove a factorization theorem for Drell-Yan production
$pp\to X\ell^+\ell^-$ where $X$ is allowed to have hard jets close to the beam,
but no hard central jets. We call this ``isolated Drell-Yan''.  Our proof of
factorization uses the soft-collinear effective theory
(SCET)~\cite{Bauer:2000ew, Bauer:2000yr, Bauer:2001ct, Bauer:2001yt} plus
additional arguments to rule out possible Glauber effects based in part on
Refs.~\cite{Collins:1988ig, Aybat:2008ct}. Although we focus our discussion on
Drell-Yan, our factorization theorem applies to processes $pp\to XL$, were the
lepton pair is replaced by other non-strongly interacting particles, such as
Higgs or $Z'$ decaying non-hadronically.  Though our analysis is only rigorous
for $pp\to XL$, we also briefly discuss what the extended factorization formula
may look like for processes with additional identified jets in the final state.

Our main result is to show that process-independent ``beam functions'',
$B_i(t,x)$ with $i=\{g,u, \bar u,d, \ldots\}$, are required to properly describe
the initial state. For the usual PDFs in Drell-Yan production appearing in
\eq{sigff}, the hadronic final state $X$ is treated fully inclusively, and the
effects of initial- and final-state soft radiation cancel
out~\cite{Collins:1989gx}. With restrictions on $X$, the effects of soft
radiation can no longer cancel. Generically, by restricting $X$ one performs an
indirect measurement of the proton prior to the hard collision.  At this point,
the proton is resolved into a colliding hard parton inside a cloud of collinear
and soft radiation. The proper description of this initial-state jet is
given by a beam function in conjunction with an appropriate soft function
describing the soft radiation in the event.

One might worry that the collision of partons inside initial-state jets rather
than partons inside protons could drastically change the physical picture.
Although the changes are not as dramatic, they have important implications.  The
beam function can be computed in an operator product expansion, giving
%%%
\begin{equation} \label{eq:Bisf}
  B_i(t,\xi,\mu_B) = \delta(t) \, f_i(\xi, \mu_B) + \mathcal{O}[\alpha_s(\mu_B)]
\,,\end{equation}
%%%
where $\mu_B$ is an intermediate perturbative scale and $t$ is an invariant-mass
variable closely related to the off-shellness of the colliding parton
(and the Mandelstam variable $t$).  Thus, the beam functions
reduce to standard PDFs at leading order.  For what we call the gluon beam
function, this was already found in Ref.~\cite{Fleming:2006cd}, where the same
matrix element of gluon fields appeared in their computation of $\gamma\,p\to
J/\psi X$ using SCET.

Equation~\eqref{eq:Bisf} implies that the momentum fractions $\xi_{a,b}$ are
determined by PDFs evaluated at the scale $\mu_B\ll Q$, which is parametrically
smaller than the scale $Q$ of the partonic hard interaction. The renormalization
group evolution (RGE) for the initial state now proceeds in two stages. For
scales $\mu < \mu_B$, the RGE is given by the standard PDF
evolution~\cite{Gribov:1972ri, Georgi:1951sr, Gross:1974cs, Altarelli:1977zs,
  Dokshitzer:1977sg}, which sums single logarithms, mixes the PDFs, and
redistributes the momentum fractions in the proton to lower $x$ values. For
scales $\mu > \mu_B$, the jet-like structure of the initial state becomes
relevant and its evolution is properly described by the RGE of the beam
function.  In contrast to the PDF, the evolution of the beam function is
independent of $x$, does not involve any mixing between parton species, and sums
Sudakov double logarithms.  In addition to the change in evolution, the
transition from PDFs to beam functions at the scale $\mu_B$ also involves
explicit $\alpha_s(\mu_B)$ corrections as indicated in \eq{Bisf}. These include
mixing effects, such as a gluon from the proton pair-producing a quark that goes
on to initiate the hard interaction and an antiquark that is radiated into the
final state.  For our observables such fluctuations are not fully accounted for
by the PDF evolution. These beam effects must be taken into account, which can
be done by perturbative calculations.  The standard PDFs are still sufficient to
describe the nonperturbative information required for the initial state.

One should ask whether the description of the initial state by beam functions,
as well as their interplay with the soft radiation, are properly captured by
current Monte Carlo event generators used to simulate events at the LHC and
Tevatron, such as Pythia~\cite{Sjostrand:2006za, Sjostrand:2007gs} and
Herwig~\cite{Corcella:2000bw, Bahr:2008pv}.  In these programs the corresponding
effects should be described at leading order by the initial-state parton shower
in conjunction with models for the underlying event~\cite{Sjostrand:2004pf,
  Sjostrand:2004ef, Butterworth:1996zw, Bahr:2008dy}.  The experimental
implications and reliability of these QCD Monte Carlo models have been studied
extensively~\cite{Affolder:2001xt, Acosta:2004wqa, Kar:2008zza}.  We will see
that the initial-state parton shower is in fact closer to factorization with
beam functions than to the inclusive factorization formula in \eq{sigff}. In
particular, the physical picture of off-shell partons that arises from the
factorization with beam functions has a nice correspondence with the picture
adopted for initial-state parton showers a long time ago~\cite{Sjostrand:1985xi,
  Bengtsson:1986gz}. There are also differences. Our analysis is based solely on
QCD soft-collinear factorization, whereas the initial-state parton shower is
partly based on the picture arising from small-$x$ physics or semihard
QCD~\cite{Gribov:1984tu}.  For the parton distributions our formalism applies in
a situation that is intermediate between the case of very small $x$, where a
resummation of $\ln x$ becomes important, and the case $x\to 1$, where threshold
resummation in $\ln(1-x)$ becomes important.  Numerically, our results apply for
the dominant region of $x$ values that are of interest at the LHC.
Experimentally, measurements of the isolated Drell-Yan cross section provide a
simple obsevable that can rigorously test the accuracy of the initial-state
shower in Monte Carlo programs, by contrasting it with the analytic results
reported here.

In \sec{overview}, we discuss our main results and explain various aspects of
the factorization with beam functions. The goal of this section is to give a
thorough discussion of the physical picture behind our results which is
nontechnical and accessible to non-expert readers.  In \sec{beamfunction}, we
elaborate on the field-theoretic definition and properties of the beam functions
and their relation to the PDFs. We quote explicit results for the quark beam
function at one loop, the derivation of which will be given in a separate
publication~\cite{Stewart:2010qs}.  In \sec{factorization}, we derive in detail the factorization
theorem for isolated $pp\to XL$ using SCET, and apply it to the case of
Drell-Yan.  Readers not interested in the technical details can freely skip this
section.  Plots of the isolated Drell-Yan cross section are given in
\sec{summary}. We conclude in \sec{conclusions}.

%%%%%%%%%%%%%%%%%%%%%%%%%%%%%%%%%%%%%%%%%%%%%%%%%%%%%%%%%%%%%%%%%%%%%%%%%%%%%%%%
\section{Factorization with Beam Functions}
\label{sec:overview}
%%%%%%%%%%%%%%%%%%%%%%%%%%%%%%%%%%%%%%%%%%%%%%%%%%%%%%%%%%%%%%%%%%%%%%%%%%%%%%%%

This section provides an extensive discussion of how factorization with beam
functions works, including the necessary kinematic definitions for the variables
that constrain the hadronic final state. In the interest of avoiding technical
details, we only discuss the physics contained in the factorization theorems.
Readers interested in the field-theoretic definitions for the beam functions are
referred to \sec{beamfunction}, while those interested in the derivation of the
factorization theorem in SCET and explicit definitions for all its ingredients
are referred to \sec{factorization}.

In \subsec{DYfact}, we review the factorization theorems for inclusive Drell-Yan
and threshold Drell-Yan, and then explain the factorization theorem for our
isolated Drell-Yan process. We use a simple setup where measurements on the
final-state hadrons use hemispheres orthogonal to the beam.  These observables
are generalized in \subsec{generalobs} to uniformly account for measurements
that sample over a wide variety of boosts between the hadronic and partonic
center-of-mass frames.  We explain the relation between beam functions and
parton distribution functions in \subsec{BtoF}. We compare the beam-function
renormalization group evolution to initial-state parton showers in
\subsec{showercomparison}.  In \subsec{fixedorder}, we show how the various
pieces in the factorization theorem arise from the point of view of a
fixed-order calculation.  In \subsec{RGE}, we compare the structure of large
logarithms and their resummation for the different factorization theorems. This
yields an independent argument for the necessity of beam functions and provides
a road map for incorporating beam functions in other isolated processes. Finally
in \subsec{Jets}, we comment on the extension of the factorization with beam
functions to the case where one has two or more isolated jets in the final
state.

\begin{figure*}[ht!]
\subfigure[\hspace{1ex}Inclusive Drell-Yan production.]{%
\includegraphics[width=0.33\textwidth]{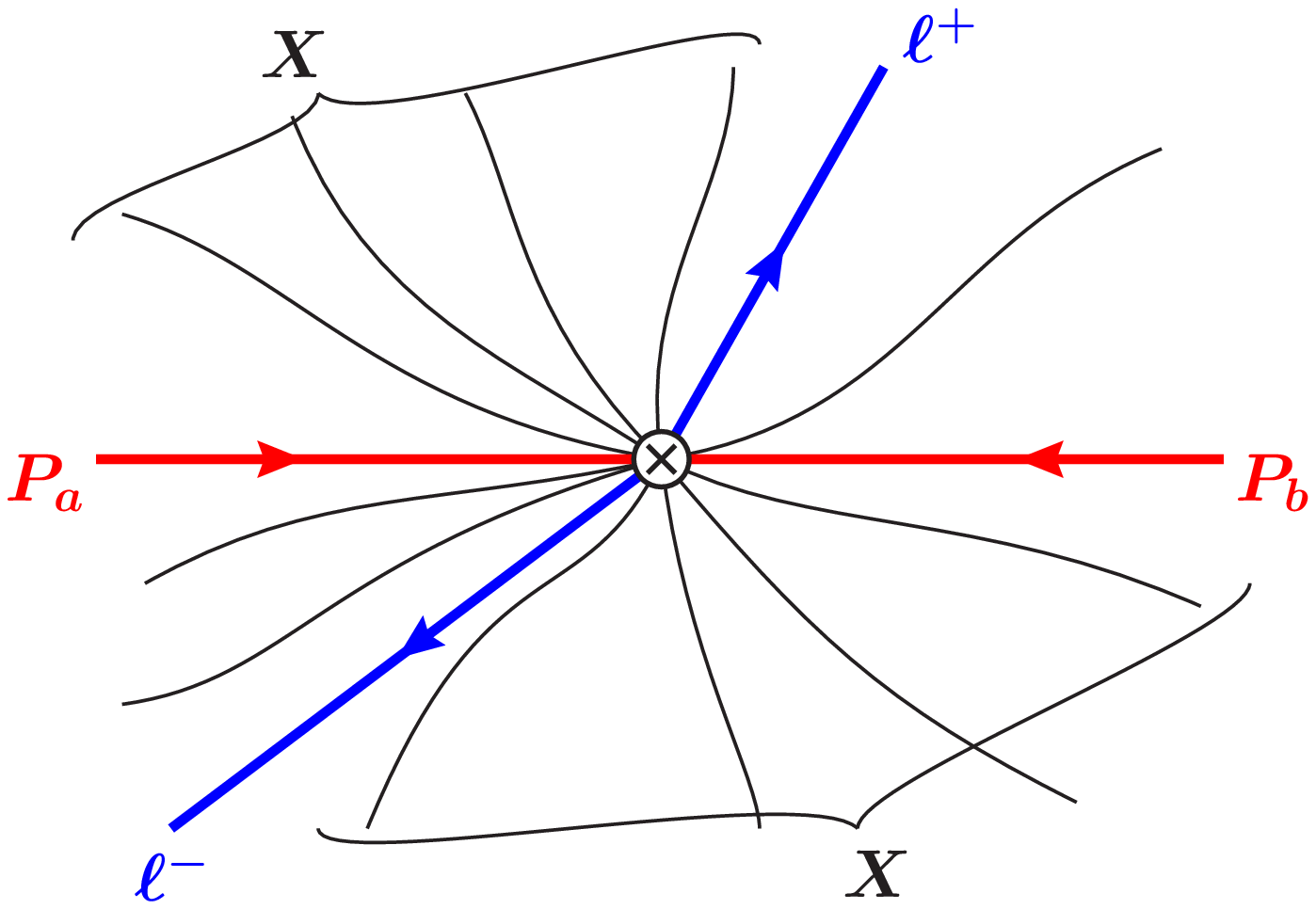}%
\label{fig:drellyan_inclusive}%
}\hfill%
\subfigure[\hspace{1ex}Drell-Yan near threshold.]{%
\includegraphics[width=0.33\textwidth]{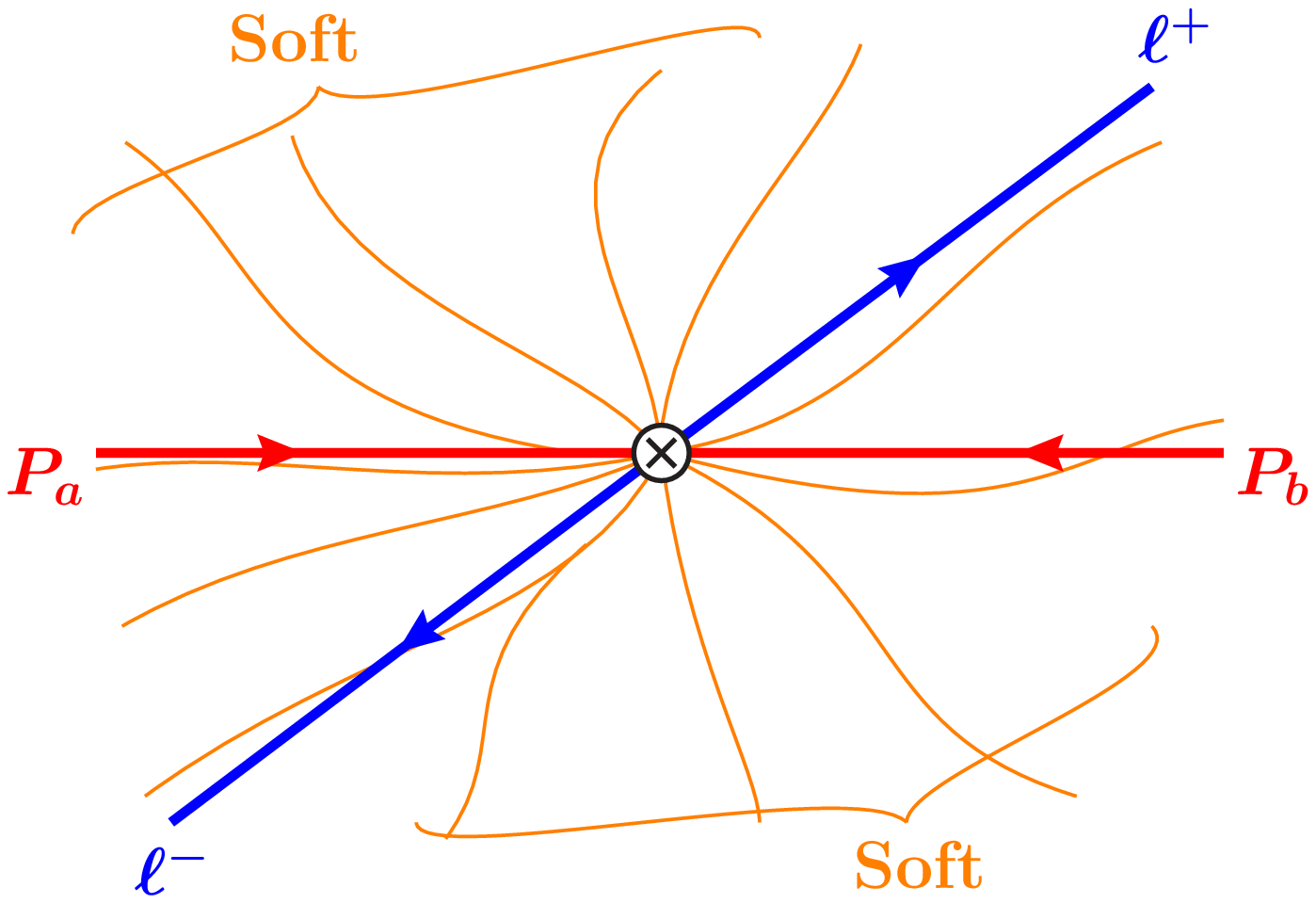}%
\label{fig:drellyan_threshold}%
}\hfill%
\subfigure[\hspace{1ex}Isolated Drell-Yan.]{%
\includegraphics[width=0.33\textwidth]{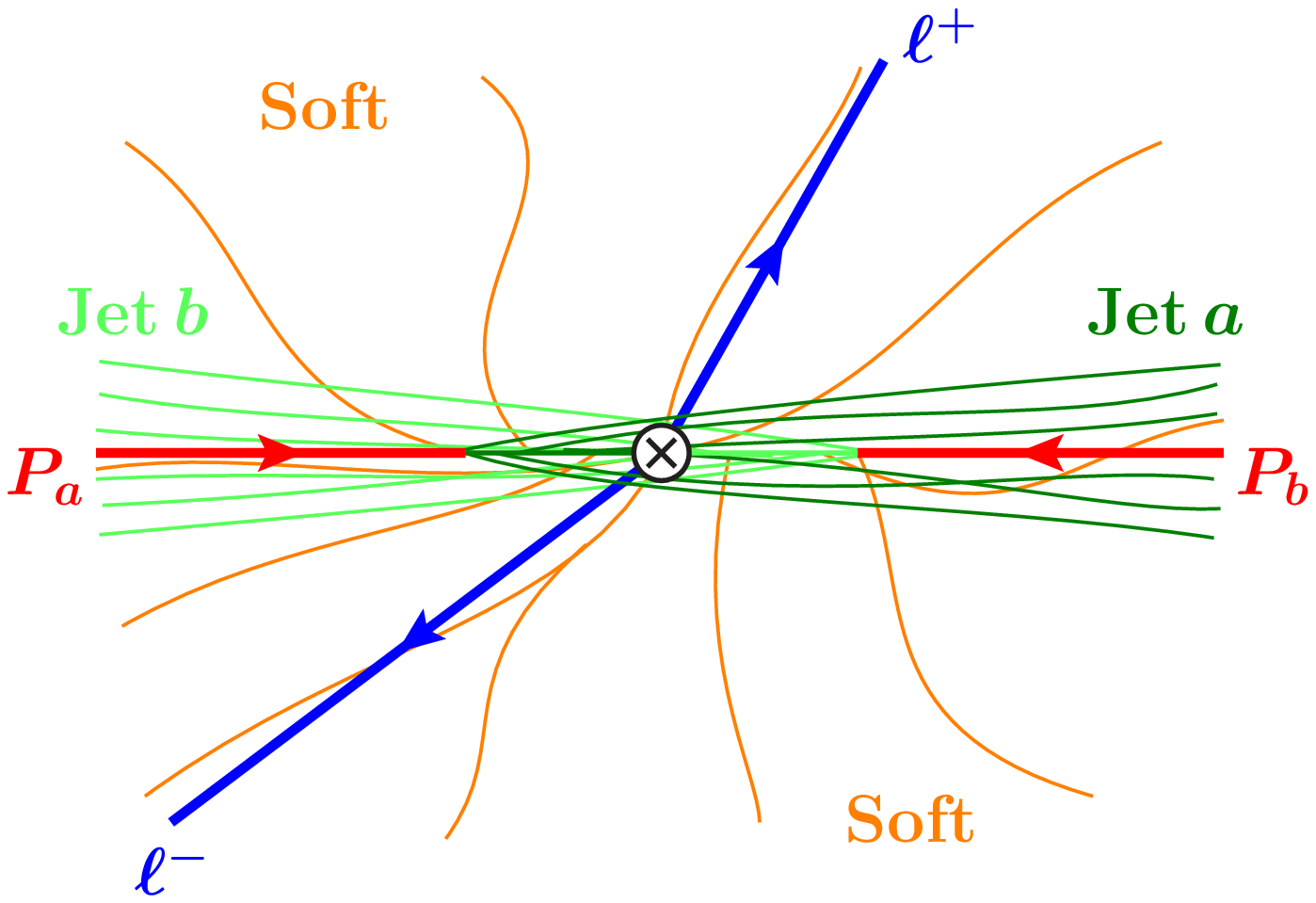}%
\label{fig:drellyan_excl}%
}%
\\
\hspace{0.33\textwidth}\hfill%
\subfigure[\hspace{1ex}Dijet production near threshold.]{%
\includegraphics[width=0.33\textwidth]{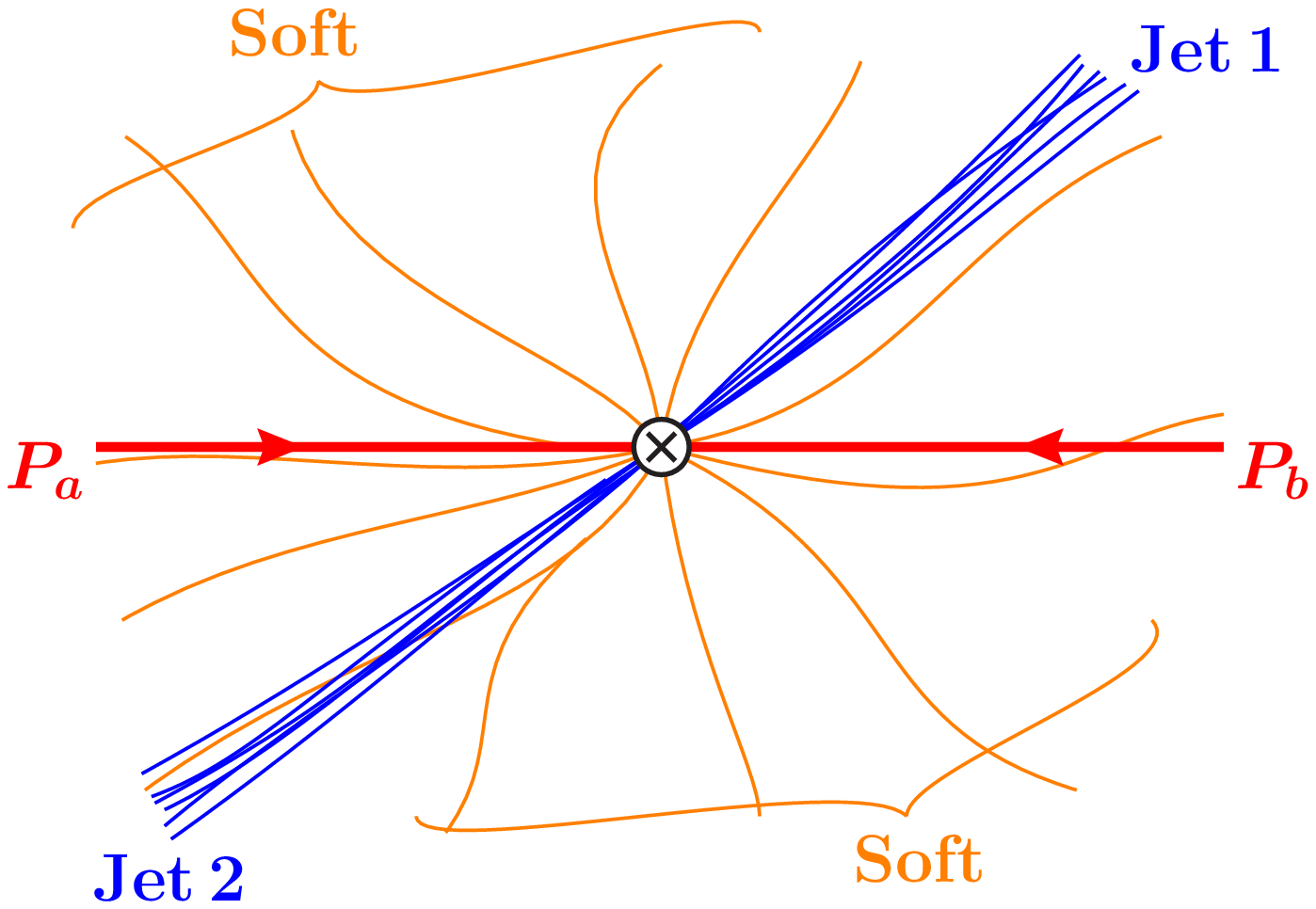}%
\label{fig:2jet_threshold}%
}\hfill%
\subfigure[\hspace{1ex}Isolated dijet production.]{%
\includegraphics[width=0.33\textwidth]{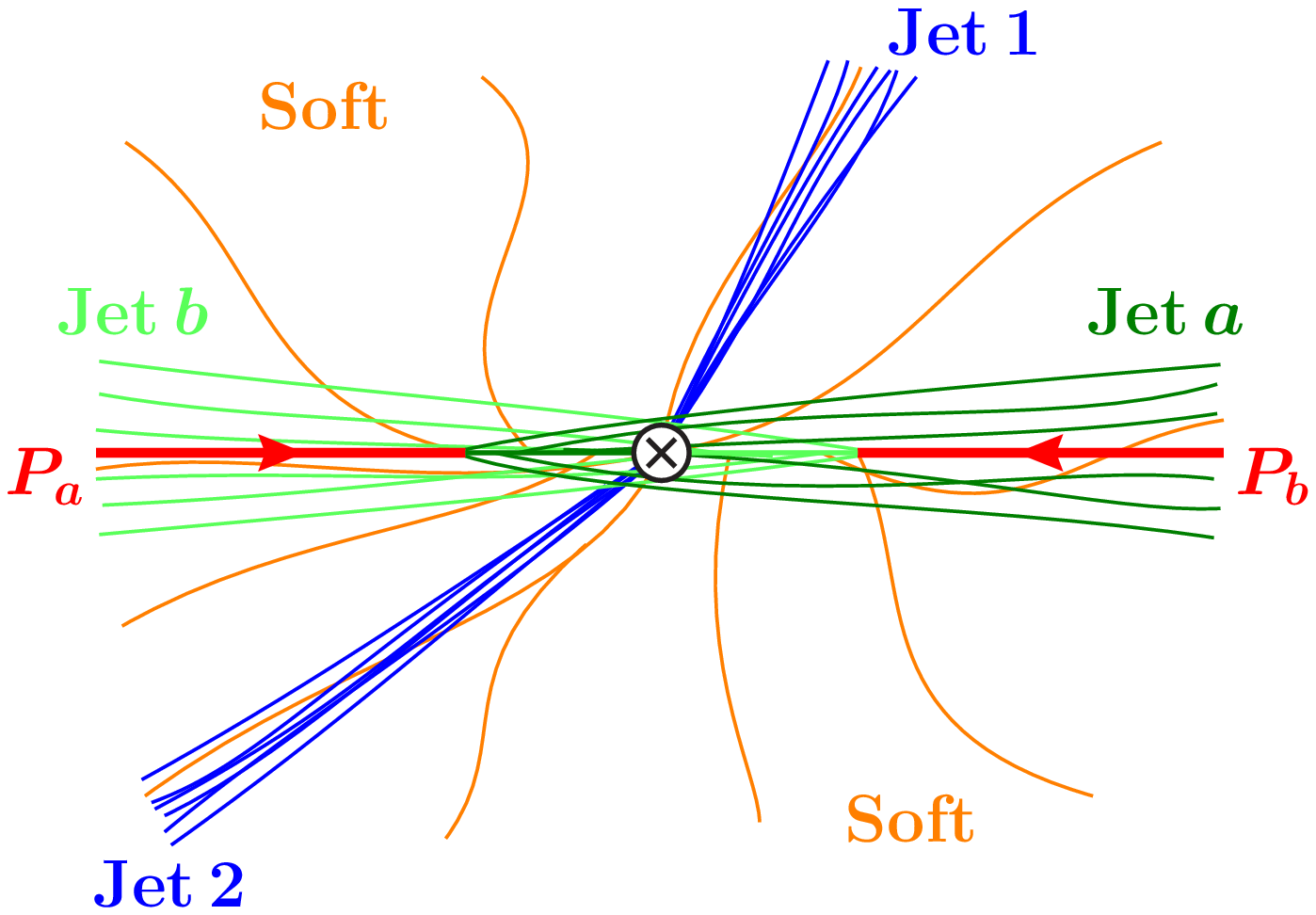}%
\label{fig:2jet_excl}%
}%
\caption{\label{fig:pp} Different final-state configurations for $pp$
  collisions. The top row corresponds to Drell-Yan factorization theorems for
  the (a) inclusive, (b) threshold, and (c) isolated cases. The bottom row
  shows the corresponding pictures with the lepton pair replaced by dijets.}
\end{figure*}

%===============================================================================
\subsection{Drell-Yan Factorization Theorems}
\label{subsec:DYfact}
%===============================================================================

To describe the Drell-Yan process $pp\to X\ell^+\ell^-$ or $p\bar p\to
X\ell^+\ell^-$, we take
%%%
\begin{equation}
P_a^\mu + P_b^\mu = p_X^\mu + q^\mu
\,,\end{equation}
%%%
where $P_{a,b}^\mu$ are the incoming (anti)proton momenta,
$\Ecm = \sqrt{(P_a+P_b)^2}$ is the total center-of-mass energy,
and $q^\mu$ is the total momentum of the $\ell^+\ell^-$ pair. We also define
%%%
\begin{align} \label{eq:DYvars}
\tau &= \frac{q^2}{\Ecm^2}
\,,&
Y &= \frac{1}{2} \ln\frac{P_b \cdot q}{P_a \cdot q}
\,,\nn\\
x_a &= \sqrt{\tau} e^Y
\,,&
x_b &= \sqrt{\tau} e^{-Y}
\,,\end{align}
%%%
where $Y$ is the total rapidity of the leptons with respect to the beam axis,
and $x_a$ and $x_b$ are in one-to-one correspondence with $\tau$ and $Y$.
Their kinematic limits are
%%%
\begin{align} \label{eq:Ylimit}
0 &\leq \tau \leq 1
\,,&
2 \abs{Y} &\leq - \ln\tau
\,,\nn\\
\tau &\leq x_a \leq 1
\,,&
\tau &\leq x_b \leq 1
\,.\end{align}
%%%
The invariant mass of the hadronic final state is bounded by
%%%
\begin{equation} \label{eq:mXincl}
m_X^2 = p_X^2 \leq \Ecm^2(1 - \sqrt{\tau})^2
\,.\end{equation}
%%%
In Drell-Yan
%%%
\begin{equation}
Q = \sqrt{q^2} \gg \lqcd
\end{equation}
%%%
plays the role of the hard interaction scale. In
general, for factorization to be valid at some leading level of approximation
with a perturbative computation of the hard scattering, the measured observable
must be infrared safe and insensitive to the details of the hadronic final
state.

For inclusive Drell-Yan, illustrated in \fig{drellyan_inclusive}, one sums over
all hadronic final states $X$ allowed by \eq{mXincl} without imposing any cuts.
Hence, the measurement is insensitive to any details of $X$ because one sums
over all possibilities. In this situation
there is a rigorous derivation of the classic factorization
theorem~\cite{Bodwin:1984hc, Collins:1985ue, Collins:1988ig}
%%%
\begin{align} \label{eq:DYincl}
\frac{1}{\sigma_0}\, \frac{\df\sigma}{\df q^2 \df Y}
 &= \sum_{i,j} \int\!\frac{\df \xi_a}{\xi_a}\, \frac{\df \xi_b}{\xi_b}\,
  H^\incl_{ij}\Bigl(\frac{x_a}{\xi_a},\frac{x_b}{\xi_b},q^2, \mu \Bigr)\,
\nn\\ &\quad\times
  f_i(\xi_a, \mu)\,f_j(\xi_b, \mu)\Bigl[1 + \ORd{\frac{\lqcd}{Q}}\Bigr]
\,,
\end{align}
%%%
where $\sigma_0=4\pi\alem^2/(3N_c \Ecm^2 q^2)$, and the integration limits are
$x_{a}\le \xi_{a} \le 1$ and $x_{b}\le \xi_{b} \le 1$. The sum is over partons
$i,j=\{g,u,\bar u,d,\ldots\}$, and $f_i(\xi_a)$ is the parton distribution
function for finding parton $i$ inside the proton with light-cone momentum
fraction $\xi_a$ along the proton direction. Note that $\xi_{a,b}$
are partonic variables, whereas $x_{a,b}$ are leptonic, and the two are only
equal at tree level. The inclusive hard function $H^\incl_{ij}$ can be computed in fixed-order
perturbative QCD as the partonic cross section to scatter partons $i$ and
$j$ [corresponding to $\df\sigma_{ij}^\parton$ in \eq{sigff}] and is known to
two loops~\cite{Altarelli:1979ub, Hamberg:1990np, Harlander:2002wh,
Anastasiou:2003yy, Anastasiou:2003ds}.

For threshold Drell-Yan, one imposes strong restrictions to only allow soft
hadronic final states with $m_X\ll Q$, as illustrated in
\fig{drellyan_threshold}. Using \eq{mXincl}, this can be ensured by forcing
$(1-\sqrt{\tau})^2 \ll \tau$, so that one is close to the threshold $\tau \to
1$. In this case, there are large double logarithms that are not accounted for
by the parton distributions.  Furthermore, since
%%%
\begin{align} \label{eq:threshold}
1 \geq \xi_{a,b} \geq x_{a,b} \geq \tau \to 1
\,,\end{align}
%%%
a single parton in each proton carries almost all of the energy, $\xi_{a,b}\to
1$. The partonic analog of $\tau$ is the variable
%%%
\begin{align} \label{eq:DYz}
  z = \frac{q^2}{\xi_a \xi_b \Ecm^2} = \frac{\tau}{\xi_a\xi_b} \leq 1
\,,\end{align}
%%%
and $\tau\to 1$ implies the partonic threshold limit $z\to 1$.
As \eq{Ylimit} forces $Y\to 0$ for $\tau\to 1$, it is convenient to
integrate over $Y$ and consider the $\tau \to 1$ limit for $\df\sigma/\df q^2$.
The relevant factorization theorem in this limit is~\cite{Sterman:1986aj, Catani:1989ne}
%%%
\begin{align} \label{eq:DYendpt}
&\frac{1}{\sigma_0}\, \frac{\df\sigma}{\df q^2}
= \sum_{ij} H_{ij}(q^2, \mu) \int\!\frac{\df\xi_a}{\xi_a}\,\frac{\df\xi_b}{\xi_b}\,f_i(\xi_a, \mu)\, f_{j}(\xi_b, \mu)
\nn\\ &\quad \times
 Q\, S_\thr \Bigl[Q\Bigl(1-\frac{\tau}{\xi_a \xi_b}\Bigr), \mu \Bigr]
\Bigl[1 + \ORd{\frac{\lqcd}{Q}, 1-\tau} \Bigr]
\,,\end{align}
%%%
where we view \eq{DYendpt} as a hadronic factorization theorem in its own
right, rather than simply a refactorization of $H_{ij}^{\rm incl}$ in
\eq{DYincl}.  This Drell-Yan threshold limit has been studied
extensively~\cite{Magnea:1990qg, Korchemsky:1992xv, Catani:1996yz,
  Belitsky:1998tc, Moch:2005ky, Idilbi:2006dg, Becher:2007ty}.  Factorization
theorems of this type are the basis for the resummation of large logarithms in
near-threshold situations.  In contrast to \eq{DYincl}, the sum in \eq{DYendpt}
only includes the dominant $q\bar q$ terms for various flavors, $ij=\{u\bar u,
\bar u u, d\bar d, \ldots \}$.  Other combinations are power-suppressed and only
appear at $\ord{1-\tau}$ or higher.  The threshold hard function $H_{ij} \sim
\abs{C_i C^*_j}$ is given by the square of Wilson coefficients in SCET, and can
be computed from the timelike quark form factor. The threshold Drell-Yan soft
function $S_\thr$ is defined by a matrix element of Wilson lines and contains
both perturbative and nonperturbative physics.  If it is treated purely in
perturbation theory at the soft scale $Q(1-\tau)$, there are in principle
additional power corrections of $\mathcal{O}[\lqcd/Q(1-\tau)]$ in
\eq{DYendpt}~\cite{Korchemsky:1996iq}.

Our goal is to describe the isolated Drell-Yan process shown in
\fig{drellyan_excl}.  Here, the colliding partons in the hard interaction are
far from threshold as in the inclusive case, but we impose a constraint that
does not allow central jets.  Soft radiation still occurs everywhere, including
the central region.  Away from threshold, the hard interaction only carries away
a fraction of the total energy in the collision. The majority of the remaining
energy stays near the beam. The colliding partons emit collinear radiation along
the beams that can be observed in the final state, shown by the green lines
labeled ``Jet $a$'' and ``Jet $b$'' in \fig{drellyan_excl}. This radiation
cannot be neglected in the factorization theorem and necessitates the beam
functions.  In the threshold case, these jets are not allowed by the limit $\tau
\to 1$, which forces all available energy into the leptons and leaves only soft
hadronic radiation.%
\footnote{ Note that the proof of factorization for the partonic cross section
  in the partonic threshold limit $z\to 1$ is not sufficient to establish the
  factorization of the hadronic cross section, unless one takes the limit
  $\tau\to 1$.  The hadronic factorization theorem assumes that all real
  radiation is soft with only virtual hard radiation in the hard function. The
  weaker limit $z\to 1$ still allows the incoming partons to emit energetic real
  radiation that cannot be described by the threshold soft function.  Only the
  $\tau\to 1$ limit forces the radiation to be soft. This point is not related
  to whether or not the threshold terms happen to dominate numerically away from
  $\tau \to 1$ due to the shape of the PDFs or other reasons.}
In the inclusive case there are no restrictions on additional hard
emissions, in which case initial-state radiation is included in the partonic
cross section in $H_{ij}^\incl$.

Also shown in \fig{drellyan_excl} is the fact that the leptons in isolated
Drell-Yan need not be back-to-back, though they are still back-to-back in the
transverse plane [see \subsec{kinematics}]. In this regard, isolated Drell-Yan
is in-between the threshold case, where the leptons are fully back-to-back with
$Y\approx 0$, and the inclusive case, where they are unrestricted.

In \figs{2jet_threshold}{2jet_excl} we show analogs of threshold Drell-Yan and
isolated Drell-Yan where the leptons are replaced by final-state jets. We will
discuss the extension to jets in \subsec{Jets} below.

To formulate isolated Drell-Yan we must first discuss how to veto hard emissions
in the central region. For this purpose, it is important to use an observable
that covers the full phase space. Jet algorithms are good tools to identify
jets, but not necessarily to veto them. Imagine we use a jet algorithm and
require that it does not find any jets in the central region. Although this
procedure covers the full phase space, the restrictions it imposes on the final
state depend in detail on the algorithm and its criteria to decide if something
is considered a jet or not.  It is very hard to incorporate such restrictions
into explicit theoretical calculations, and in particular into a rigorous
factorization theorem. Even if possible in principle, the resulting beam and
soft functions would be very complicated objects, and it would be difficult to
systematically resum the large logarithms arising at higher orders from the
phase-space restrictions.  Therefore, to achieve the best theoretical precision,
it is important to implement the central jet veto using an inclusive kinematic
variable.  This allows us to derive a factorization theorem with analytically
manageable ingredients, which can then be used to sum large phase-space
logarithms.

\begin{figure}[t!]
\includegraphics[scale=0.5]{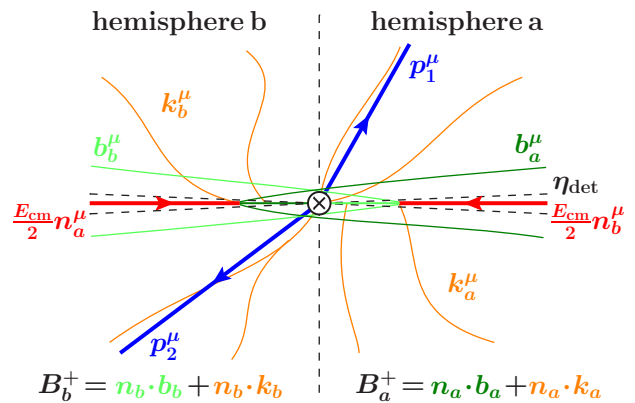}
\caption{Definition of hemispheres and kinematic variables for isolated
Drell-Yan. }
\label{fig:DYkin}
\end{figure}

We will consider a simple kinematic variable that fulfills the above criteria,
leaving the discussion of more sophisticated generalizations to the next
subsection.  The key variables for the isolated Drell-Yan process are shown in
\fig{DYkin}. The proton momenta $P_a^\mu$ and $P_b^\mu$ are used to define
lightlike vectors $n_a^\mu$ and $n_b^\mu$,
%%%
\begin{equation}
 P_a^\mu = \frac{\Ecm}{2}\, n_a^\mu
 \,,\qquad
 P_b^\mu = \frac{\Ecm}{2}\, n_b^\mu
\,,\end{equation}
%%%
where the protons are massless and $n_a^2=0$, $n_b^2=0$, and $n_a\sdt n_b =2$.
Using the beam axis, we define two hemispheres $a$ and $b$ opposite to the
incoming protons. We then divide up the total hadronic momentum as
%%%
\begin{align}
p_X^\mu = B_a^\mu + B_b^\mu
\,,\end{align}
%%%
where $B_a^\mu = p_{X_a}^\mu$ and $B_b^\mu = p_{X_b}^\mu$ are the total final-state
hadronic momenta in hemispheres $a$ and $b$. Of these, we consider the components
%%%
\begin{align}
B_a^+ &= n_a \sdt B_a = B_a^0 (1 + \tanh y_a)\, e^{-2y_a}
\,,\nn\\
B_b^+ &= n_b \sdt B_b = B_b^0 (1 + \tanh y_b)\, e^{-2y_b}
\,,\end{align}
%%%
where $B_{a,b}^0$ are the energy components and $y_{a,b}$ are the total
rapidities of $B_{a,b}^\mu$ with respect to the forward direction $n_{a,b}$ for
each hemisphere.  Here, $\lim_{y\to \infty}(1+\tanh y) = 2$ and $1+\tanh y \geq
1.8$ for $y \geq 1$, so $B_{a,b}^+$ scale exponentially with the rapidities $y_{a,b}$.

In terms of the measured particle momenta $p_k$ in hemisphere $a$,
%%%
\begin{equation} \label{eq:Baplus_calc}
B_a^+ = \sum_{k \in a} n_a\sdt p_k = \sum_{k \in a} E_k (1 + \tanh\eta_k) e^{-2\eta_k}
\,.\end{equation}
%%%
Here, $E_k$ and $\eta_k$ are the experimentally
measured energy and pseudorapidity with respect to $\vec n_a$, and we neglect
the masses of final-state hadrons. An analogous formula applies for $B_b^+$.
Hence, $B_a^+$ and $B_b^+$ receive large
contributions from energetic particles in the central region, while contributions
from particles in the forward region are suppressed. Thus, requiring small
$B_{a,b}^+ \ll Q$ is an effective way to restrict the energetic radiation in
each hemisphere as a smooth function of rapidity, allowing forward jets and
disallowing central jets.  At the same time, soft radiation with energies $\ll
Q$ is measured, but not tightly constrained.

As an example, consider the cut
%%%
\begin{equation} \label{eq:ycut1}
B_{a,b}^+ \leq Q\, e^{-2 y_\cut}
\,.\end{equation}
%%%
This constraint vetoes any events with a combined energy deposit of more than
$Q/2$ per hemisphere in the central rapidity region $\abs{y} \leq y_\cut$.
In the smaller region $\abs{y} \leq y_\cut - 1$, the energy allowed
by \eq{ycut1} is reduced by a factor of $e^2 \simeq 7$, essentially
vetoing any jets there.  In the larger region $\abs{y} \leq y_\cut\!+\! 1$,
it is increased by the same factor, so beyond $y_\cut + 1$ the hadronic final state is
essentially unconstrained.  Thus, a typical experimental value might be $y_\cut
= 2$, which vetoes energetic jets in the central region $\abs{y} \leq 1$. The
precise value of the cut on $B_{a,b}^+$ will of course depend on the
requirements of the experimental analyses.

Note that the variable $B_a^+$ is similar to the total transverse energy in
hemisphere $a$, defined as
%%%
\begin{equation} \label{eq:ET_calc}
E_{Ta} = \sum_{k \in a} \frac{E_k}{\cosh\eta_k} = \sum_{k \in a} E_k (1+\tanh\eta_k)e^{-\eta_k}
\,.\end{equation}
%%%
$B_a^+$ has two advantages over $E_{Ta}$. First, the exponential sensitivity to
rapidity is much stronger for $B_a^+$, which means it provides a stronger
restriction on jets in the central region and at the same time is less sensitive
to jets in the forward region. Second, since $B_a^+$ is a specific four-momentum
component and linear in four-momentum, $(p_1 + p_2)^+ = p_1^+ + p_2^+$, it is
much simpler to work with and to incorporate into the factorization theorem.
It is clear that the isolated Drell-Yan factorization theorem discussed here
can be extended to observables with other exponents, $e^{-a \eta_k}$, much like the angularity
event shapes in $e^+e^-$~\cite{Berger:2003iw}.

One should ask, down to what values can $B_{a,b}^+$ be reliably measured
experimentally? In principle, particles at any rapidity contribute to
$B_{a,b}^+$, but the detectors only have coverage up to a maximum
pseudorapidity $\eta_\det$, as indicated in \fig{DYkin}. For the hadron
calorimeters at the LHC $\eta_\det \simeq 5$ and at the Tevatron
$\eta_\det\simeq 4$.  In the hadronic center-of-mass frame, the
unscattered partons inside the proton have plus components of
$\ord{\lqcd^2/\Ecm}$, so any contributions from the unmeasured proton
remnants are always negligible. The question then is, what is the maximal
contribution to $B_{a,b}^+$ from initial-state radiation that is missed as it is
outside the detector? In the extreme scenario where all proton energy
is deposited right outside $\eta_\det$, we would have $B_{a,b}^+ = 14
\TeV e^{-10} = 0.6 \GeV$ at the LHC and $B_{a,b}^+ = 2\TeV e^{-8} = 0.7 \GeV$ at
the Tevatron. In more realistic scenarios, the contribution
from such radiation is suppressed by at least another factor of $10$ or more.
Therefore, the finite detector range is clearly not an issue for measuring
values $B_{a,b}^+ \gtrsim 2\,{\rm GeV}$, and the relevant limitation will be
the experimental resolution in $B_{a,b}^+$.

The factorization theorem for isolated Drell-Yan, which we prove in
\sec{factorization}, reads
%%%
\begin{align} \label{eq:DYbeam}
 &\frac{1}{\sigma_0} \frac{\df\sigma}{\df q^2 \df Y \df B_a^+\df B_b^+}
 = \sum_{ij} H_{ij}(q^2, \mu) \int\!\df k_a^+\, \df k_b^+\,
\\  &\quad \times
 q^2 B_i[\w_a(B_a^+ -k_a^+), x_a, \mu] B_j[\w_b(B_b^+ -k_b^+),x_b, \mu]
\nn\\\nn &\quad \times
 S_\hemiin(k_a^+,k_b^+, \mu)
\biggl[1 + \ORd{\frac{\lqcd}{Q}, \frac{\w_{a,b} B_{a,b}^+}{Q^2}} \biggr]
.\end{align}
%%%
The physical interpretation of \eq{DYbeam} is that we take partons $i$ and $j$
out of the initial-state jets $B_{i}$, $B_j$ and hard-scatter them to final
state particles with $H_{ij}$, while including $S_\hemiin$ to describe the
accompanying soft radiation.  The hard function $H_{ij}$ is identical to the one
in the threshold factorization theorem in \eq{DYendpt}, and the sum in
\eq{DYbeam} is again only over $ij=\{u\bar u, \bar u u, d\bar d, \ldots\}$. The
quark and antiquark beam functions $B_q$ and $B_{\bar q}$ describe the effects
of the incoming jets and have replaced the PDFs. The variables $\w_{a,b} =
x_{a,b}\Ecm$.  The hard partons are taken from initial-state jets rather than
protons, so unlike in the threshold case the gluon PDF now contributes via the
beam functions. We will see how this works in more detail in \subsec{BtoF}.
Finally, $S_\hemiin$ is the initial-state hemisphere soft function.

The kinematic variables in \eq{DYbeam} are displayed in \fig{DYkin}. The soft
function depends on the momenta $k_a^+ =n_a\sdt k_a$ and $k_b^+ =n_b\sdt k_b$ of
soft particles in hemispheres $a$ and $b$, respectively. Much like PDFs, the
beam functions $B_i(t_a, x_a, \mu)$ and $B_j(t_b, x_b, \mu)$ depend on the
momentum fractions $x_a$ and $x_b$ of the active partons $i$ and $j$
participating in the hard collision.  In addition, they depend on invariant-mass
variables
%%%
\begin{equation}
t_a = \w_a b_a^+\geq 0
\,,\qquad
t_b = \w_b\, b_b^+ \geq 0
\,,\end{equation}
%%%
where $\w_{a,b} = x_{a,b}\Ecm$ are the hard momentum components and $b_a^+ =
n_a\sdt b_a$. The momentum $b_a^\mu$ is defined as the total momentum of the
energetic particles radiated into hemisphere
$a$, as shown in \fig{DYkin}, and similarly for $b_b^+$.
(The kinematics are shown in more detail in \fig{beam_kinematics}.)
Before the hard interaction, the momentum of the active quark can be written as
%%%
\begin{equation}
\w_a\,\frac{n_a^\mu}{2} - b_a^+\, \frac{n_b^\mu}{2} - b_{a\perp}^\mu
\,.\end{equation}
%%%
The first term is its hard momentum along the proton direction, and the last two
terms are from the momentum it lost to radiation, where $b_{a\perp}^2 = -
\vec{b}_{aT}^2$ contains the transverse components. The quark's spacelike
invariant mass is $-\w_a b_a^+ - \vec{b}_{aT}^2 = -t_a - \vec{b}_{aT}^2$. The
beam function $B_i$ for hemisphere $a$ depends on $t_a=\w_a b_a^+ = x_a\Ecm
b_a^+$, which is the negative of the quark's transverse virtuality. (When the
distinction is unimportant we will usually refer to $t$ simply as the quark's
virtuality.) By momentum conservation $b_a^+ = B_a^+ - k_a^+$, leading to the
convolution of the beam and soft functions as shown in \eq{DYbeam}. Physically,
the reason we have to subtract the soft momentum from $B_a^+$ is that the beam
function only properly describes the collinear radiation, while the soft
radiation must be described by the soft function. An analogous
discussion applies to $B_j$ and $t_b$ for hemisphere $b$. The convolutions in
the factorization theorem thus encode the cross talk between the soft radiation
and energetic collinear radiation from the beams.

By measuring and constraining $B^+_a$ we essentially measure the virtuality of
the hard parton in the initial state. As the proton cannot contain partons with
virtualities larger than $\lqcd^2$, the initial state at that point must be
described as an incoming jet containing the hard off-shell parton. This is the
reason why beam functions describing these initial-state jets must appear in
\eq{DYbeam}. It also follows that since $t \gg \lqcd^2$ we can calculate the
beam functions perturbatively in terms of PDFs, which we discuss further in
\subsec{BtoF}.

It is convenient to consider a cumulant cross section, including all events with
$B_{a,b}^+$ up to some specified value, as in \eq{ycut1}. Integrating
\eq{DYbeam} over $0 \leq B_{a,b}^+ \leq B^+_\max $ we obtain
%%%
\begin{align} \label{eq:intDYbeam}
 &\frac{1}{\sigma_0} \frac{\df\sigma}{\df q^2 \df Y}(B_\max^+)
 = \sum_{ij} H_{ij}(q^2, \mu) \int\!\df k_a^+\, \df k_b^+\,
\\  &\quad \times
\tB_i[\w_a(B^+_\max -k_a^+), x_a, \mu] \tB_j[\w_b(B^+_\max -k_b^+), x_b, \mu]
\nn \\ \nn &\quad \times
 S_\hemiin(k_a^+,k_b^+, \mu)
\biggl[1 + \ORd{{\frac{\lqcd}{Q}, \frac{\omega_{a,b}B^+_\max}{Q^2}}} \biggr]
,\end{align}
%%%
where the soft function $S_\hemiin$ is the same as in \eq{DYbeam}, and we
defined the integrated beam function
%%%
\begin{equation} \label{eq:tB_def}
\tB_i(t_\max, x, \mu) = \int\! \df t\, B_i(t, x, \mu)\,\theta(t_\max - t)
\,.\end{equation}
%%%
The cut $B_{a,b}^+ \leq B^+_\max$ implies the limit $b_{a,b}^+ \leq B^+_\max -
k_{a,b}^+$ and $t_{a,b} \leq \w_{a,b} (B^+_\max - k_{a,b}^+)$, leading to the
convolutions in \eq{intDYbeam}.

The factorization theorem \eq{DYbeam} and its integrated version \eq{intDYbeam}
are valid in the limit $t_{a,b}/Q^2 \simeq B_{a,b}^+/Q \equiv \lambda^2 \ll 1$,
and receive power corrections of $\ord{\lambda^2}$. Thus, for $B^+_\max = Q
e^{-2y_\cut}$ with $y_\cut = 1$, we expect the power corrections not to exceed
$e^{-2} \sim 10\%$. This is not a fundamental limitation, because the power corrections
can be computed in SCET if necessary. If the soft function is treated purely
perturbatively, there are additional power corrections of
$\ord{\lqcd/B_{a,b}^+}$, which account for soft singularities as $B_{a,b}^+ \to 0$.
The variables $B_{a,b}^+$ are infrared safe with respect to collinear splittings~\cite{Sterman:1978bj}.

The hard function receives perturbative $\alpha_s$ corrections at the hard scale
$\mu_H\simeq Q$, the beam functions have $\alpha_s$ corrections at the
intermediate beam scale $\mu_B^2 \simeq t_\max \simeq Q B^+_\max$, and the soft
function at the soft scale $\mu_S \simeq B^+_\max$. For example, for $Q \simeq
1\TeV$ and $y_\cut = 2$ we have $\mu_B \simeq 140\GeV$ and $\mu_S \simeq
20\GeV$.  Even with a very small $Q \simeq 100\GeV$, perhaps for Higgs
production, $\mu_B \simeq 14\GeV$ and $\mu_S \simeq 2\GeV$ are still
perturbative (although at this point nonperturbative contributions $\sim
\lqcd/\mu_S$ to the soft function might no longer be small and may be
incorporated with the methods in Refs.~\cite{Hoang:2007vb,Ligeti:2008ac}). In
fixed-order perturbation theory, the cross section contains large single and
double logarithms, $\ln(B^+_\max/Q) \simeq -4$ and $\ln^2(B^+_\max/Q) \simeq
16$, invalidating a fixed-order perturbative expansion. The factorization
theorem allows us to systematically resum these logarithms to all orders in
perturbation theory, which is discussed in more detail in \subsec{RGE}.

The factorization theorem \eq{DYbeam} also applies to other non-hadronic final
states such as $Z' \to \ell^+\ell^-$, or Higgs production with $H\to
\gamma\gamma$ or $H\to Z Z^*\to 4\ell$. In each case, $q^2$ and $Y$ are the
total non-hadronic invariant mass and rapidity, and central jets are vetoed with
a cut on $B_{a,b}^+$. The only dependence on the process is in the hard
function, which must be replaced appropriately and can be taken directly from
the corresponding threshold factorization theorem. One may also consider $W$
production with $W\to \ell\bar\nu$, with an appropriate replacement of $q^2$ and
$Y$ with the charged lepton's rapidity.  For a light Higgs with $Q\sim m_H$, the
isolated Drell-Yan factorization theorem applies to Higgs production through
gluon fusion $gg\to H$ and Higgs-strahlung $q \bar{q} \to VH$, which are the
dominant production channels at the LHC and Tevatron, respectively.%
\footnote{ In vector-boson fusion and associated production $gg\to t \bar{t} H$,
  the situation is more complicated and one has to explicitly consider the
  process $pp\to X jj H$ with two forward (top) jets.}  For a generic process
$pp\to XL$, the sum over $ij = \{gg, u\bar u, \bar u u, d\bar d,\ldots\}$
includes a gluon-gluon contribution, but still no cross terms between different
parton types, and there will be two independent soft functions $S^{q\bar
  q}_\hemiin$ and $S^{gg}_\hemiin$.  [As shown in \sec{factorization}, only the
$q\bar q$ soft function contributes to isolated Drell-Yan, so the labels were
omitted in \eq{DYbeam}.] Indeed, the gluon-gluon contribution involving the
gluon beam and soft functions, $B_g$ and $S^{gg}_\hemiin$, gives the dominant
contribution in the case of Higgs production.

With the above physical picture, we can understand why the gluon beam function
appeared in $\gamma\,p\to J/\psi X$ in the analysis of
Ref.~\cite{Fleming:2006cd} in the limit where $E_{J/\psi} \to E_\gamma$. Taking
$p_X$ as the total momentum of final-state hadrons other than the $J/\psi$, one has
$n\sdt p_X \sim \Ecm(1 - E_{J/\psi}/E_\gamma)$, where $n$ is the proton
direction. For $E_{J/\psi}$ close to $E_\gamma$, energetic radiation in the final state is
restricted to a jet close to the $n$ direction. Just as for our
$B_{a,b}^+$, the measurement of $E_{J/\psi}$ probes the radiation emitted by the
colliding gluon in the initial state. Thus, the proton is broken apart prior to the
hard collision, and the gluon beam function is required to describe the initial
state.

%===============================================================================
\subsection{Generalized Observables}
\label{subsec:generalobs}
%===============================================================================

The factorization theorem in \eq{DYbeam} applies for $t_a \ll q^2$ and $t_b \ll
q^2$. This includes the situation where in the hadronic center-of-mass frame
there is a numerically significant asymmetry $\w_a = x_a\Ecm > \w_b = x_b\Ecm$.
This means that the boost between the hadronic and partonic center-of-mass frames,
given by the leptonic $Y = \ln\sqrt{\w_a/\w_b} = \ln\sqrt{x_a/x_b}$, is
significantly different from zero. We explore the implications of this here.

If there is no hierarchy, $\w_a \approx \w_b \approx \sqrt{\w_a \w_b} = Q$,
corresponding to $Y\approx 0$, we can define a simple variable to constrain both
hemispheres simultaneously,
%%%
\begin{equation}
\widehat{B} = \frac{B_a^+ + B_b^+}{Q}
\,.\end{equation}
%%%
From \eq{DYbeam}, this gives
%%%
\begin{align}
\frac{1}{\sigma_0} \frac{\df\sigma}{\df q^2 \df Y \df \widehat B}
&= \sum_{ij} H_{ij}(q^2, \mu) \int\!\df t_a\, \df t_b\,
\nn\\ &\quad \times
B_i(t_a, x_a, \mu)\, B_j(t_b, x_b, \mu)
\nn\\ &\quad \times
Q\, S_B\Bigl(Q\widehat{B} - \frac{t_a}{\w_a}- \frac{t_b}{\w_b}, \mu\Bigr)
\,,\end{align}
%%%
where the soft function is defined as
%%%
\begin{equation} \label{eq:SB}
S_B(k^+, \mu)
= \!\int\!\df k_a^+ \df k_b^+\, S_\hemiin(k_a^+, k_b^+, \mu )\,
\delta(k^+ - k_a^+ - k_b^+)
\,.\end{equation}
%%%
The advantage of using $\widehat B$ is that the soft function now only depends
on the single variable $k^+ = k_a^+ + k_b^+$, much like the soft function for
thrust in $e^+e^-$ collisions.

\begin{figure}[t!]
\includegraphics[scale=0.5]{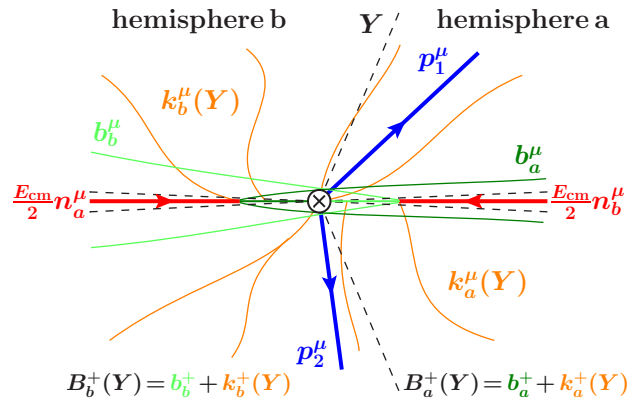}
\caption{Generalized definition of hemispheres. The total rapidity of
the leptons is $Y$, $b_{a,b}^+ = n_{a,b}\sdt b_{a,b}$, and $k_{a,b}^+(Y) = n_{a,b}\sdt k_{a,b}(Y)$.}
\label{fig:DYkin_boosted}
\end{figure}

If we have a hierarchy $\w_b < Q < \w_a$, the final state has a substantial
boost in the $n_a$ direction, as shown in \fig{DYkin_boosted}. In this case, the
energetic radiation will generically be much closer to the beam axis in
hemisphere $a$ than in hemisphere $b$. To take this into account, it is natural
to impose different cuts on $B_a^+$ and $B_b^+$. Using the boost-invariant
combinations $\w_a B_a^+/q^2$ and $\w_b B_b^+/q^2$ to define the cut, we obtain
%%%
\begin{equation} \label{eq:ycut2}
\frac{\w_a B_a^+}{q^2} = \frac{B_a^+}{\w_b} \leq e^{-2y_\cut}
\,,\qquad
\frac{\w_b B_b^+}{q^2} = \frac{B_b^+}{\w_a} \leq e^{-2y_\cut}
\,,\end{equation}
%%%
so $B_a^+$ has a tighter constraint than $B_b^+$, as desired.  If we simply
replace $\widehat{B}$ by $B_a^+/\w_b + B_b^+/\w_a$, the soft function analogous
to $S_B$ in \eq{SB} will depend on the combination $(\w_ak_a^+ + \w_b k_b^+)/Q^2$.

However, we should also adjust the hemispheres themselves to take into account
the significant boost of the partonic center-of-mass frame. We therefore define
a generalized hemisphere $a$ as $y > Y$ and hemisphere $b$ as $y < Y$, as shown
in \fig{DYkin_boosted}. The corresponding total hemisphere momenta are denoted as
$B_{a,b}^+(Y)$ and the soft hemisphere momenta as $k_{a,b}^+(Y)$. The original
definitions in \fig{DYkin} correspond to $B_{a,b}^+(0) \equiv B_{a,b}^+$ and
$k_{a,b}^+(0) \equiv k_{a,b}^+$. The generalization of $\widehat B$ is given by
the boost-invariant combination
%%%
\begin{equation} \label{eq:tauB}
\tau_B = \frac{\w_a B_a^+(Y) + \w_b B_b^+(Y)}{q^2}
\,.\end{equation}
%%%
With the generalized definition of the hemispheres, $B_{a,b}^+(Y)$ and $\w_{a,b}$ transform
under a boost by $y$ in the $n_a$ direction as
%%%
\begin{align}
B_a^+(Y) &\to B_a^{+\prime}(Y + y) = e^{-y} B_a^+(Y)
\,,\nn\\
B_b^+(Y) &\to B_b^{+\prime}(Y + y) = e^{y} B_b^+(Y)
\,,\nn\\
\w_a &\to \w_a' = e^{y} \w_a
\,,\nn\\
\w_b &\to \w_b' = e^{-y} \w_b
\,.\end{align}
%%%
Thus, boosting by $y = -Y$ from the hadronic to the partonic
center-of-mass frame gives
%%%
\begin{equation} \label{eq:tauB_partoniccms}
\tau_B
= \frac{\w_a' B_a^{+\prime}(0) + \w_b' B_b^{+\prime}(0)}{q^2}
= \frac{B_a^{+\prime}(0) + B_b^{+\prime}(0)}{Q}
\,.\end{equation}
%%%
In the partonic center-of-mass frame we have $\w_a' = \w_b' = Q$, so there is no hierarchy. Correspondingly, the generalized hemispheres in this frame are again perpendicular to the beam axis, so \eq{tauB_partoniccms} has the same form as $\widehat B$.

Note that for $e^+e^- \to$ jets, one can use the thrust axis to define two
hemispheres with $n_{a,b}$ analogous to our case. In the 2-jet limit, thrust is
then given by $1 - T = (Q\, n_a\sdt p_{X_a} + Q\, n_b\sdt p_{X_b})/2Q^2$.
Hence, we can think of $\tau_B$ as the analog of thrust for incoming jets. For
this reason we will call $\tau_B$ the ``beam thrust''.

In analogy to \eqs{ycut1}{ycut2}, we define the cutoff on $\tau_B$ by
%%%
\begin{equation} \label{eq:tauBycut}
\tau_B \leq e^{-2 y_B^\cut}
\,.\end{equation}
%%%
For $\tau_B \to 0$ or equivalently $y_B^\cut \to \infty$ the jets along the beam axes
become pencil-like, while for
generic $y_B^\cut$ we allow energetic particles up to rapidities $y \lesssim y_B^\cut$
(with $y$ measured in the partonic center-of-mass frame).

The beam functions are boost-invariant along the beam axis, so the different
hemisphere definitions do not affect them. The soft function is boost-invariant
up to the hemisphere definition, which defines its arguments $k_{a,b}^+$. Hence,
boosting by $-Y$ we have $S_\hemiin[e^Y k_a^+, e^{-Y} k_b^+; Y] =
S_\hemiin[k_a^+, k_b^+; 0] = S_\hemiin(k_a^+, k_b^+)$, where the third argument
denotes the definition of the hemispheres. This implies that the soft function for
$\tau_B$ is the same as in \eq{SB}. The factorization theorem for $\tau_B$
following from \eq{DYbeam} is
%%%
\begin{align} \label{eq:dsigma_tauB}
\frac{1}{\sigma_0} \frac{\df\sigma}{\df q^2 \df Y \df \tau_B}
&= \sum_{ij} H_{ij}(q^2, \mu) \int\!\df t_a\, \df t_b\,
\nn\\ &\quad \times
B_i(t_a, x_a, \mu)\, B_j(t_b, x_b, \mu)
\nn\\ &\quad \times
Q\,S_B\Bigl(Q\,\tau_B - \frac{t_a + t_b}{Q}, \mu\Bigr)
\,.\end{align}
%%%
Integrating over $0 \leq \tau_B \leq \exp(-2y_B^\cut)$ we obtain
%%%
\begin{equation} \label{eq:sigma_yBcut}
\frac{\df\sigma}{\df q^2 \df Y}(y_B^\cut)
= \int_0^{\exp(-2y_B^\cut)}\!\df \tau_B\,  \frac{\df\sigma}{\df q^2 \df Y\df\tau_B}
\,.\end{equation}
%%%
We will use \eqs{dsigma_tauB}{sigma_yBcut} to show plots of our results in \sec{summary}.

%===============================================================================
\subsection{Relating Beam Functions and PDFs}
\label{subsec:BtoF}
%===============================================================================

\begin{figure*}[t!]
\includegraphics[scale=0.75]{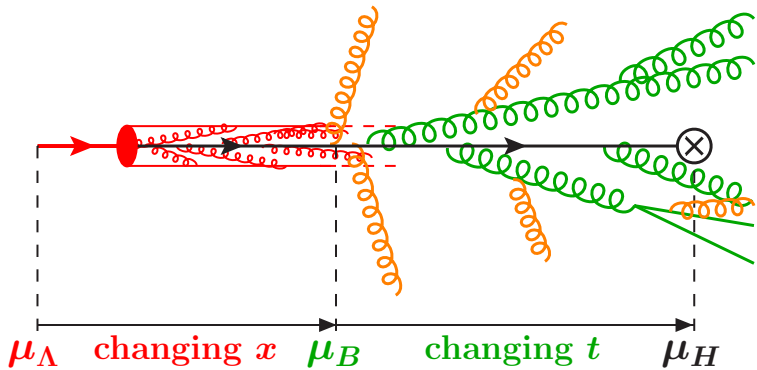}
\caption{Evolution of the initial state. Starting from the low scale
  $\mu_\Lambda$, the incoming proton is described by the $x$-dependent evolution of the PDFs, which
  redistributes the total momentum of the proton between its constituents. At
  the scale $\mu_B$, the proton is probed by measuring the radiation in the final state
  and breaks apart.
  This is the scale where the PDFs are evaluated and the $x$-dependent evolution stops.
  Above $\mu_B$, the proton has ceased to exist, and the initial state behaves
  like an incoming jet, whose evolution is governed by the virtuality $t$ of the
  off-shell spacelike parton that eventually enters the hard interaction at the
  scale $\mu_H$.}
\label{fig:beam}
\end{figure*}

The beam functions can be related to the PDFs by performing an operator product
expansion, because $t_{a,b} \gg \lqcd^2$.\footnote{A detailed discussion of the
  appropriate operator product expansion is given in
  Ref.~\cite{Stewart:2010qs}.} This yields the factorization formula
%%%
\begin{align} \label{eq:B_fact}
B_i(t, x, \mu)
&= \sum_j\!\int_x^1 \frac{\df\xi}{\xi}\, \cI_{ij}\Bigl(t,\frac{x}{\xi},\mu \Bigr) f_j(\xi, \mu)
\nn\\ &\quad\times
\biggl[1 + \ORd{\frac{\lqcd^2}{t}}\biggr]
,\end{align}
%%%
where we sum over partons $j=\{g, u,\bar u, d, \ldots\}$, $\cI_{ij}$ are
perturbatively calculable Wilson coefficients, and $f_j$ is the standard PDF
for parton $j$. The $\ord{\lqcd^2/t}$ power corrections in
\eq{B_fact} involve proton structure functions at subleading twist. Further
mathematical details on \eq{B_fact} are discussed in
\sec{beamfunction}, whereas here we focus on the physical ramifications.

The interpretation of \eq{B_fact} is illustrated in \fig{beam}. At a
hadronic scale $\mu_\Lambda\sim 1\GeV$, the initial conditions for the
PDFs $f_j$ can be specified, and one has the standard DGLAP evolution up to the
scale $\mu_B$,
%%%
\begin{align} \label{eq:fevolution}
\mu \frac{\df}{\df\mu} f_j(\xi, \mu)
= \sum_{j'} \int\!\frac{\df\xi'}{\xi'}\, P_{jj'}\Bigl(\frac{\xi}{\xi'}, \mu\Bigr) f_{j'}(\xi', \mu)
\,.\end{align}
%%%
The anomalous dimensions $P_{jj'}$ are the standard QCD splitting
functions for quarks, antiquarks, and gluons (including the color factors and coupling constant).
Equation~\eqref{eq:B_fact} applies at the scale
$\mu=\mu_B$, since this is the scale at which a measurement on the proton is
performed by observing the soft and collinear radiation
contributing to $B_{a,b}^+$.  At this scale, a parton $j$ with momentum fraction
$\xi$ is taken out of the incoming proton according to the probability
distribution $f_j(\xi, \mu)$.  As the parton continues to propagate and evolve
with $\mu>\mu_B$, it is modified by virtual radiation and by the emission of
real radiation, which forms a jet. The evolution in this region no longer depends on
$\xi$, but instead on the virtuality $t$. This evolution occurs with
fixed $x$ and fixed parton type $i$, via the beam function RGE
%%%
\begin{align} \label{eq:Bevolution}
\mu \frac{\df}{\df\mu} B_i(t, x, \mu) = \int\!\df t'\, \gamma^i_B(t - t', \mu)\, B_i(t',x, \mu)
\,.\end{align}
%%%
This result for initial-state jet evolution has the same structure as the
evolution for final-state jets.  In fact, the anomalous dimension $\gamma^q_B$ is
identical to that for the quark jet function to all orders in perturbation
theory~\cite{Stewart:2010qs}. We discuss this correspondence further in
\sec{beamfunction}.

The effect of initial-state real and virtual radiation is described by the
perturbatively calculable Wilson coefficients $\cI_{ij}(t, x/\xi, \mu)$ at the
scale $\mu=\mu_B$. They encode several physical effects.  The virtual loop
corrections contribute to the $\cI_{ii}$ and modify the effective strength of
the various partons.  If the radiation is real, it has physical timelike
momentum. Hence, it pushes the active parton in the jet off shell with spacelike
virtuality $-t < 0$ and reduces its light-cone momentum fraction from
$\xi$ to $x$.

In addition, the real radiation can change the identity of the colliding parton,
giving rise to the sum over $j$ in \eq{B_fact}. For example, an incoming quark
can radiate an energetic gluon which enters the hard interaction, while the
quark itself goes into the final state.  This gives a contribution of the quark
PDF to the gluon beam function through $\cI_{gq}$.  Similarly, an incoming gluon
can pair-produce, with the quark participating in the hard interaction and the
antiquark going into the final state.  This gives a contribution of the gluon
PDF to the quark beam function through $\cI_{qg}$. There are also of course real
radiation contributions to the diagonal terms, $\cI_{qq}$ and $\cI_{gg}$, where
the parton in the PDF and the parton participating in the hard interaction have
the same identity.

At lowest order in perturbation theory, the parton taken out of the proton
directly enters the hard interaction without emitting radiation,
%%%
\begin{align}
\cI_{ij}^\tree\Bigl(t,\frac{x}{\xi},\mu \Bigr)
  = \delta_{ij}\, \delta(t)\, \delta\Bigl(1 - \frac{x}{\xi}\Bigr)
\,.\end{align}
%%%
Thus at tree level, the beam function reduces to the PDF
%%%
\begin{equation} \label{eq:Bi_tree}
B^\tree_i(t, x, \mu) = \delta(t)\, f_i(x,\mu)
\,.\end{equation}
%%%
Beyond tree level, $\cI_{ij}(t, x/\xi, \mu)$ can be determined perturbatively as
discussed in more detail in \sec{beamfunction}, where we give precise field-theoretic
definitions of the beam functions and quote the one-loop results for
$\cI_{qq}$ and $\cI_{qg}$.

Interestingly, in the threshold factorization theorem \eq{DYendpt}, cross terms
between quark and gluon PDFs are power suppressed, so the gluon PDF does not
contribute at leading order.  In the inclusive case \eq{DYincl}, such cross
terms are leading order in the power counting. For isolated Drell-Yan, there are
no cross terms between quark and gluon beam functions, but there are
leading-order cross terms between different PDFs, which appear via the
contributions of different PDFs to a given beam function in \eq{B_fact}. Thus,
the isolated case is again in-between the inclusive and threshold cases.

%===============================================================================
\subsection{Comparison with Initial-State Parton Shower}
\label{subsec:showercomparison}
%===============================================================================

The physical situation associated with the beam evolution has an interesting
correspondence with that of initial-state parton showers. As pictured in the
region between $\mu_B$ and $\mu_H$ in \fig{beam}, the parton in the beam
function evolves forward in time while emitting a shower of radiation into the
final state governed by the anomalous dimension $\gamma_B^i(t - t',\mu)$ in \eq{Bevolution}.
This equation has no parton mixing. Each emission by the radiating parton
increases the magnitude of its spacelike virtuality
$-t<0$, pushing it further off-shell in a spacelike direction. At the time the
parton is annihilated in the hard collision, it has evolved to some $t$ with $\abs{t}\ll q^2$,
so the large momentum transfer $q^2$ guarantees that no partons in the final state
are spacelike. This description agrees quite well with
the physical picture associated with the evolution of the primary parton in an
initial-state parton shower, as summarized in Ref.~\cite{Sjostrand:2006za}.

Differences in the description arise when one considers the initial-state
parton shower in more detail (for simplicity we focus on the so-called longitudinal
evolution). The shower is based on the evolution equation for the PDFs in
\eq{fevolution}. An evolution forward in time is not practical because of the
lack of prior knowledge of the scale of the hard interaction, so the
shower uses backward evolution starting at a given partonic hard scale
$Q$~\cite{Sjostrand:1985xi}. Knowing the identity of the final parton $i$, the
shower evolves based on the probability $\df\mathcal{P}_i/\df t$ that parton
$i$ is unresolved into parton $j$ via the splitting $j\to ij'$ at an earlier
(lower) scale $t$. The evolution equation is~\cite{Sjostrand:2006za}
%%%
\begin{align} \label{eq:ishower}
\frac{\df \mathcal{P}_i(x, t_\max, t)}{\df t}
&= \biggl[\sum_{jj'} \int_x^{z_\max}\!\frac{\df z}{z}\,
  P_{j\to ij'}(z, t)\, \frac{f_j(x/z,t)}{f_i(x,t)} \biggr]
\nn\\* & \quad \times
\frac{1}{t}\, \mathcal{P}_i(x,t_\max,t)
\,,\end{align}
%%%
where $\mathcal{P}_i(x,t_\max,t)$ is the shower Sudakov exponential, which is interpreted
as the probability for no emissions to occur between the initial value $t_\max$ and $t$.
The evolution variable $t$, which determines the scale of the splitting, is
usually chosen as the virtuality or transverse momentum of the parton.

The mixing of partons in the PDF evolution influences the shower. In particular,
the evolution kernel depends on the PDF $f_j(x/z, t)$, which determines the number density of
partons of type $j$ at the scale $t$, and inversely on the PDF $f_i(x, t)$.
Thus, unlike in the beam evolution in \eq{Bevolution}, the shower evolution in
\eq{ishower} still knows the identity of the initial-state hadron. Double
logarithms in the initial-state parton shower are generated in $q\to qg$ and $g\to gg$
splittings because of the soft-gluon singularity $\sim 1/(1-z)$ in the splitting
functions. This singularity is regulated~\cite{Sjostrand:2006za} by the upper
cutoff $z_\max = x/(x+x_\eps)$, where $x_\epsilon$ provides a lower cutoff on
the gluon energy in the rest frame of the hard scattering, $E_g \geq
x_\eps\gamma\Ecm/2 \simeq 2\GeV$ (where $\gamma$ is the boost factor of the hard
scattering). Hence, one logarithm, $\ln x_\eps$, is generated by the $z$
integration, and one logarithm, $\ln t$, by the collinear $1/t$ singularity.  In contrast,
the beam function contains double logarithms $\ln^2 t$
similar to a final-state parton shower, where the $z$ integration yields a kernel
$\sim(\ln t)/t$ that produces a double logarithm $\ln^2 t$ via the $t$
evolution.

The above comparison is very rough. For example, the influence of soft radiation
on both the shower and on the isolated factorization theorem was not compared
and is likely to be important. Furthermore, the goal of the shower is to provide
a universal method for populating fully exclusive final states, while the beam
function applies for a more inclusive situation with a particular measurement.
Note that just the presence of mixing in the initial-state parton shower and absence of
mixing in the beam-function evolution does not imply an inconsistency. For example,
it is well known that the final-state parton shower reproduces the correct
double logarithms for $e^+e^-$ event shapes~\cite{Catani:1992ua}, even
though there is no parton mixing in the evolution of the corresponding hard,
jet, and soft functions.  In the future it would be interesting to test in
detail the correspondence between the double logarithms generated by the initial-state
parton shower and those predicted by our factorization theorem for the isolated
Drell-Yan process.

\begin{figure*}[t!]
\vspace{-3ex}
\hfill\subfigure[]{ \label{fig:DYBtree}
\parbox{0.4\textwidth}{%
\begin{equation*}
\begin{aligned}
\parbox[c]{23.3ex}{\includegraphics[width=23.3ex]{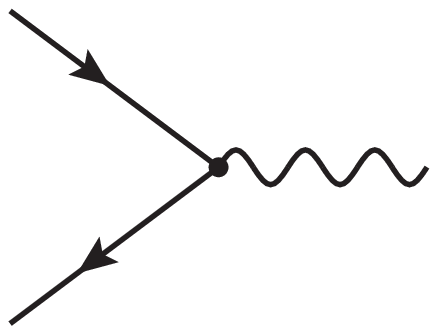}}
&=
\parbox[c]{23.3ex}{\includegraphics[width=23.3ex]{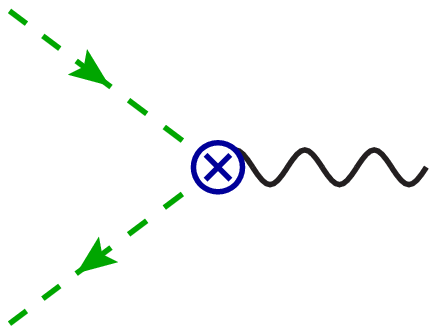}}
\\ &\qquad
\hardc{H_{q\bar q}^\zero}\,\beamc{B_q^\zero\,B_{\bar q}^\zero}\,\softc{S_{q\bar q}^\zero}
\end{aligned}
\end{equation*}}}%
\hfill\hfill\subfigure[]{ \label{fig:DYBglue}
\parbox{0.4\textwidth}{%
\begin{equation*}
\begin{aligned}
\parbox[c]{23.3ex}{\includegraphics[width=23.3ex]{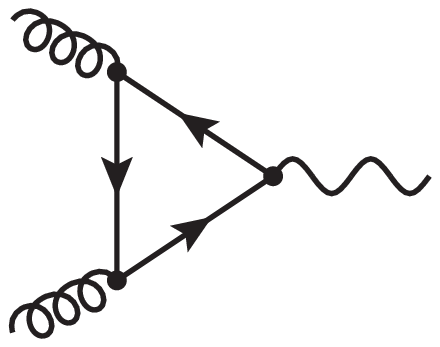}}
&=
\parbox[c]{23.3ex}{\includegraphics[width=23.3ex]{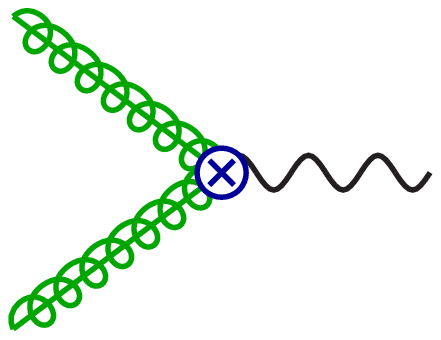}}
\\ &\qquad
\hardc{H_{gg}^\one}\,\beamc{B_g^\zero\,B_g^\zero}\,\softc{S_{gg}^\zero}
\end{aligned}
\end{equation*}}%
}\hspace*{\fill}%

\vspace{-3ex}
\subfigure[]{\label{fig:DYBvertex} \parbox{\textwidth}{%
\begin{equation*}
\begin{aligned}
\parbox[c]{23.3ex}{\includegraphics[width=23.3ex]{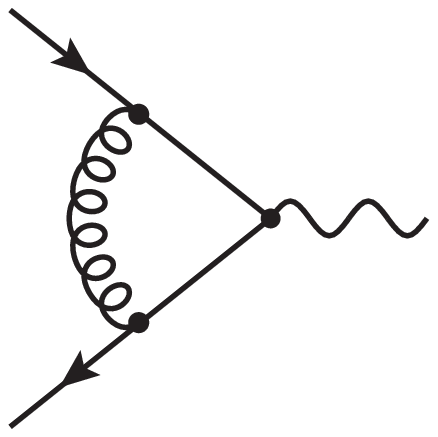}}
&=
\parbox[c]{23.3ex}{\includegraphics[width=23.3ex]{figs/feyn/qqbar_Z_0}}
\mspace{-18mu}&&+
\parbox[c]{23.3ex}{\includegraphics[width=23.3ex]{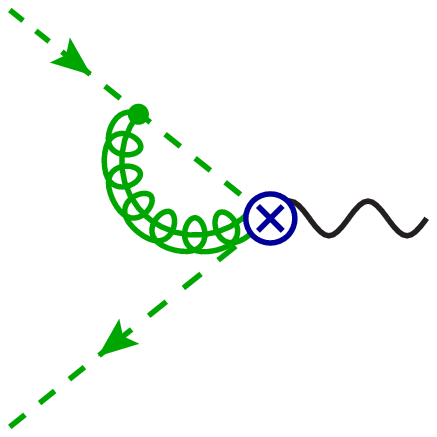}}
\mspace{-18mu}&&+
\parbox[c]{23.3ex}{\includegraphics[width=23.3ex]{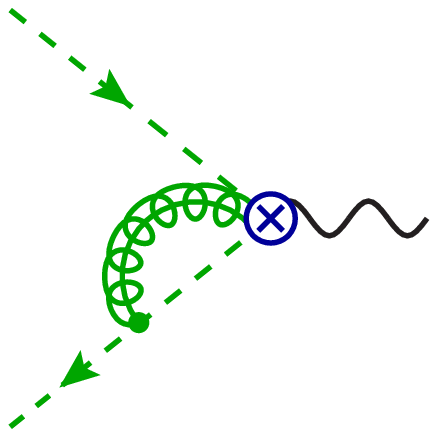}}
\mspace{-18mu}&&+
\parbox[c]{23.3ex}{\includegraphics[width=23.3ex]{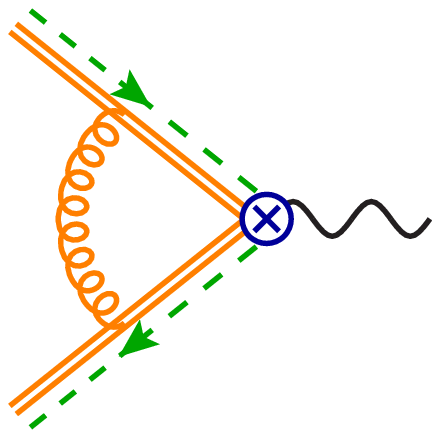}}
\\ &\qquad
\hardc{H_{q\bar q}^\one}\,\beamc{B_q^\zero\,B_{\bar q}^\zero}\,\softc{S_{q\bar q}^\zero}
&&\qquad
\hardc{H_{q\bar q}^\zero}\,\beamc{B_q^\one\,B_{\bar q}^\zero}\,\softc{S_{q\bar q}^\zero}
&&\qquad
\hardc{H_{q\bar q}^\zero}\,\beamc{B_q^\zero\,B_{\bar q}^\one}\,\softc{S_{q\bar q}^\zero}
&&\qquad
\hardc{H_{q\bar q}^\zero}\,\beamc{B_q^\zero\,B_{\bar q}^\zero}\,\softc{S_{q\bar q}^\one}
\end{aligned}
\end{equation*}}}

\vspace{-3ex}
\hfill\subfigure[]{\label{fig:DYBbeamqg}
\parbox{0.8\textwidth}{%
\begin{equation*}
\begin{aligned}
\parbox[c]{23.3ex}{\includegraphics[width=23.3ex]{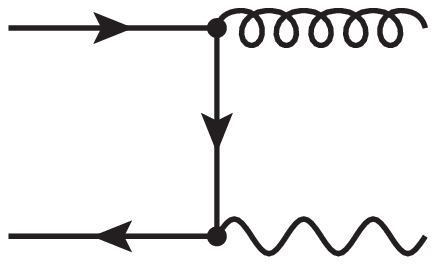}}
&=
\parbox[c]{23.3ex}{\includegraphics[width=23.3ex]{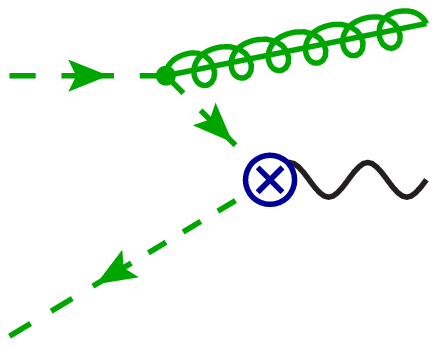}}
\mspace{-18mu}&&+
\parbox[c]{23.3ex}{\includegraphics[width=23.3ex]{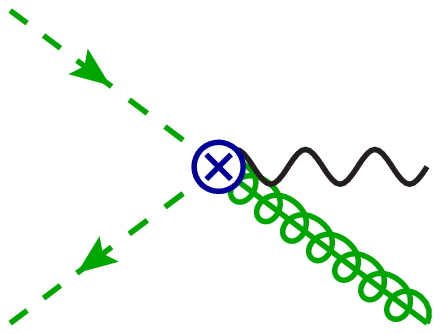}}
\mspace{-18mu}&&+
\parbox[c]{23.3ex}{\includegraphics[width=23.3ex]{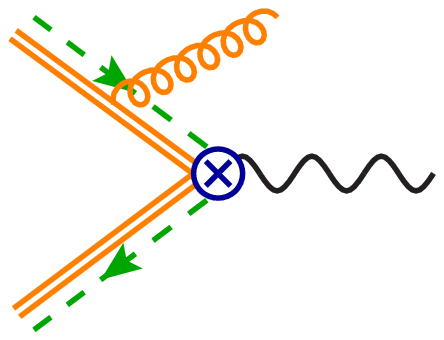}}
\\ & \qquad
\hardc{H_{q\bar q}^\zero}\,\beamc{B_q^\one\,B_{\bar q}^\zero}\,\softc{S_{q\bar q}^\zero}
&&\qquad \hardc{H_{q\bar q}^\zero}\,\beamc{B_q^\zero\,B_{\bar q}^\one}\,\softc{S_{q\bar q}^\zero}
&&\qquad \hardc{H_{q\bar q}^\zero}\,\beamc{B_q^\zero\,B_{\bar q}^\zero}\,\softc{S_{q\bar q}^\one}
\end{aligned}
\end{equation*}}}\hspace*{\fill}

\vspace{-3ex}
\hfill\subfigure[]{\label{fig:DYBbeamgq}
\parbox{0.4\textwidth}{%
\begin{equation*}
\begin{aligned}
\parbox[c]{23.3ex}{\includegraphics[width=23.3ex]{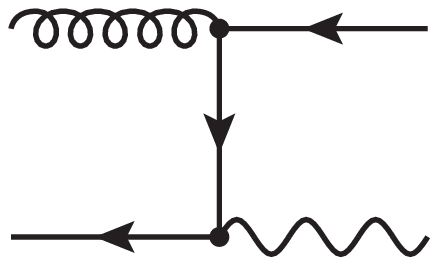}}
&=
\parbox[c]{23.3ex}{\includegraphics[width=23.3ex]{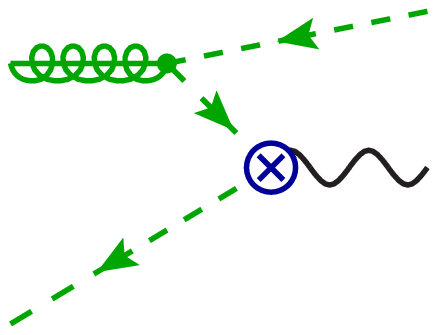}}
\\ & \qquad
\hardc{H_{q\bar q}^\zero}\,\beamc{B_q^\one\,B_{\bar q}^\zero}\,\softc{S_{q\bar q}^\zero}
\end{aligned}
\end{equation*}}}%
\hfill\hfill\subfigure[]{\label{fig:DYBschannel}
\parbox{0.4\textwidth}{%
\begin{equation*}
\parbox[c]{23.3ex}{\includegraphics[width=23.3ex]{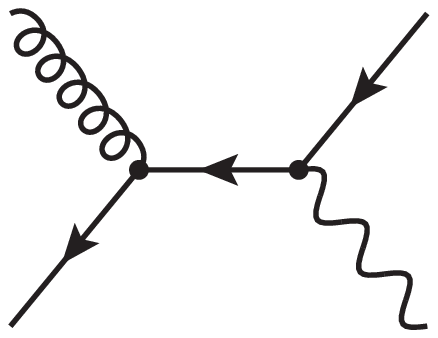}}
= \text{power correction}
\end{equation*}}}\hspace*{\fill}
\caption{\label{fig:DYBgraphs} Factorization for isolated Drell-Yan in pictures.
  The left-hand side of each equality are graphs in QCD, while the right-hand
  side shows the sum of the corresponding SCET diagrams. Dashed lines are collinear quarks, and springs with
  a line through them are collinear gluons. The double lines denote soft Wilson lines, and the gluons attached to them are soft. }
\end{figure*}

%===============================================================================
\subsection{Relation to Fixed-Order Calculation}
\label{subsec:fixedorder}
%===============================================================================

The factorization theorem for the cross section in \eq{DYbeam} and the
factorization for the beam function in \eq{B_fact} together allow us to describe
in more detail how various Feynman diagrams that would appear in a fixed-order
calculation contribute to the cross section in our kinematic region. Various
examples are shown in \fig{DYBgraphs}.

In \fig{DYBtree}, we have the tree-level $q\bar{q}$
annihilation producing a $\gamma$ or $Z$, which involves the
tree-level $\ord{\alpha_s^0}$ hard function, beam functions, and soft function,
denoted by a superscript $\zero$ in the figure. In \fig{DYBglue}, initial-state
gluons couple to a quark loop (e.g. a top quark), which subsequently annihilates
into a $\gamma$, $Z$, or Higgs. The quarks in this loop are far off
shell, so they can be integrated out and appear as one-loop corrections,
$H_{gg}^\one$, to the hard coefficient in the factorization theorem. Other
possibilities for this graph are power suppressed.

The situation for the vertex correction in \fig{DYBvertex} is more involved. If
the gluon in the loop is hard, all particles in the loop are far off shell and
can be integrated out, giving the one-loop hard function $H_{q\bar q}^\one$
shown as the first term on the right-hand side. In the second term, the gluon is
collinear to the incoming quark beam and gives a virtual one-loop contribution
to the quark beam function, $B_q^\one$. The third term is the analog of the
second, but now with the gluon collinear to the incoming antiquark. Finally in
the fourth term, the gluon is soft, communicating between the incoming collinear
beams. Here, the eikonal approximation holds for describing the quark
propagators. The generalization of this to all orders in $\alpha_s$ leads to the
fact that the soft function is a matrix element of Wilson lines. Although a
single loop graph contributes in several different places in the factorization
theorem, all of these contributions have a precise separation in SCET. We
will use this separation in \sec{factorization} to prove the isolated Drell-Yan
factorization theorem.

An interesting contribution occurs in \fig{DYBbeamqg}, where a gluon is radiated
into the final state. Because of the kinematic restrictions in isolated
Drell-Yan, this gluon can only be collinear to the incoming quark, collinear to
the incoming antiquark, or soft, and these three possibilities are represented
by the diagrams on the right-hand side of the equality. In the first case, we
have a real-emission correction to the quark beam function, $B_q^\one$.  In the
second case, the intermediate quark is far off shell and can be integrated out,
and the gluon collinear to the antiquark arises from a collinear Wilson line
contribution in $B_{\bar q}^\one$.  The third case gives a real-emission
correction to the soft function, $S^\one_{q\bar q}$.  The full-theory graph in
\fig{DYBbeamqg} has a $t$-channel singularity.  An important fact about the
isolated Drell-Yan factorization theorem is that it fully captures the dominant
parts of this singularity, and allows a simple framework for a resummation of
higher order $\alpha_s$ corrections enhanced by large double logarithms due to
this singularity.  For threshold Drell-Yan, the kinematic restrictions are
stronger and only allow the third graph with soft initial-state radiation.  In
inclusive Drell-Yan, the gluon is treated as hard, and the graph in
\fig{DYBbeamqg} only corrects $H_{q\bar q}^\incl$, without providing a framework
for summing the large double logarithms that appear when we make a global
measurement of the radiation in each hemisphere defined by the beams.

\begin{figure*}[ht!]
\includegraphics[scale=0.75]{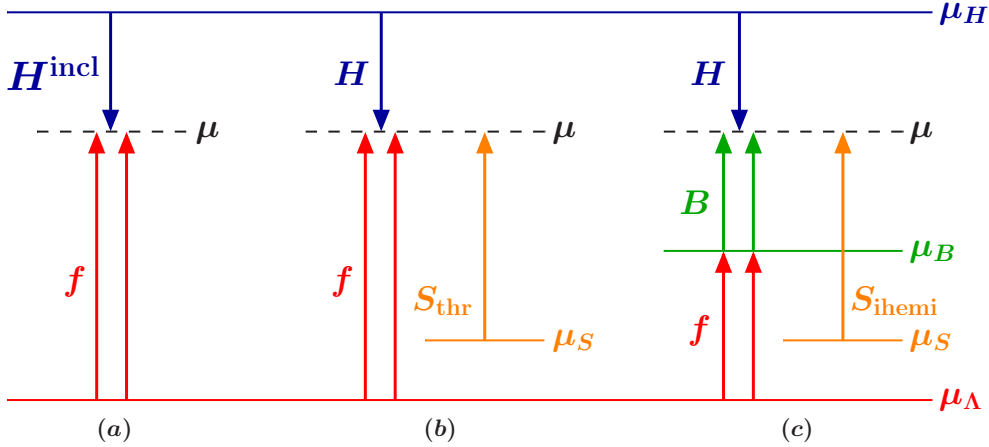}
\caption{RGE running for different Drell-Yan scenarios. Case (a) corresponds to
  the inclusive case. Case (b) corresponds to the threshold case, where the kinematics
  forces all hadrons in the final state to be soft.  Case (c) corresponds to the
  isolated case. Here, the PDFs freeze out at the intermediate beam scale
  $\mu_B$, above which they are replaced by beam functions.}
\label{fig:running_drellyan}
\end{figure*}

The situation is a bit simpler for \figs{DYBbeamgq}{DYBschannel}. In
\fig{DYBbeamgq}, the incoming collinear gluon from the PDF pair-produces a quark
and antiquark both collinear to this beam direction, and the quark enters the
hard interaction.  Therefore, this is a one-loop correction to the quark beam
function, $B_q^\one$, proportional to the gluon PDF $f_g$. The beam functions
again allow us to resum the possibly large logarithms due to this $t$-channel
singularity.  Other possibilities for the final-state antiquark in
\fig{DYBbeamgq} lead to power-suppressed contributions. Similarly, the
$s$-channel graph in \fig{DYBschannel}, which has the same initial and final
states as \fig{DYBbeamgq}, has no leading-power contribution and only
contributes to \eq{DYbeam} in the power-suppressed terms.  The same is also true
for Drell-Yan in the threshold region. Only inclusive Drell-Yan receives a
leading-order hard contribution from the $s$-channel graph, which is then
treated as of the same size as the $t$-channel graphs.

%===============================================================================
\subsection{Renormalization Group Evolution}
\label{subsec:RGE}
%===============================================================================

In this subsection, we discuss and compare the structure of large logarithms in
the cross sections for inclusive, threshold, and isolated Drell-Yan.  These
large logarithms may be summed using the renormalization group evolution of the
individual functions appearing in the factorization theorems. In fact, the
structure of large logarithms in the differential $B_{a,b}^+$ cross section
allows us to infer the necessity of the beam functions in the isolated
factorization theorem.  This procedure provides a method of determining whether
beam functions enter for other observables or processes than those studied here.
The consistency of the RGE was used to provide a similar consistency check in
Ref.~\cite{Fleming:2007qr} when deriving a new factorization theorem for the
invariant-mass distribution of jets initiated by a massive quark in $e^+e^-$
collisions.  In that case, the RGE consistency provided important constraints on
the structure of the factorization theorem at scales below the heavy-quark mass.

In inclusive Drell-Yan, the hard functions $H^\incl_{ij}$ are sensitive to the
scale $\mu_H \simeq Q$ of the hard interaction, and the proton mass defines a
low scale $\mu_\Lambda \simeq 1\GeV \gtrsim \lqcd$ (which is still
large enough so perturbation theory can be applied for the PDF evolution). The measurement of $q^2$
and $Y$ in this case does not introduce additional scales, and thus does not
influence the structure of the logarithms. Thus, we have the hierarchy
$\mu_\Lambda \ll \mu_H$, and the large logarithms are $L =
\ln(\mu_\Lambda/\mu_H)$. Here, only single-logarithmic series, $(\alpha_s L)^k$,
are generated at higher orders in perturbation theory. The logarithms are
factorized as $\ln(\mu/\mu_H) + \ln(\mu_\Lambda/\mu)$ in the factorization
theorem in \eq{DYincl} and may then be resummed.  The general form of the
running is pictured in \fig{running_drellyan}(a). The logarithms
$\ln(\mu_\Lambda/\mu)$ are summed by evolving the PDFs $f_i(\xi_a, \mu)$ and
$f_j(\xi_b, \mu)$ from $\mu_\Lambda$ up to the common scale $\mu$. The inclusive
hard function, $H^\incl(x_a/\xi_a,x_b/\xi_b,q^2,\mu)$, is evolved from $\mu_H$
down to $\mu$, summing the logarithms $\ln(\mu/\mu_H)$. The choice of $\mu$ is
arbitrary.  Taking $\mu\simeq\mu_H$ corresponds to only running the PDFs up,
while for $\mu\simeq \mu_\Lambda$ only $H^\incl$ runs down. The equivalence of
these two choices implies that $H^\incl$ must be convoluted with the two PDFs
and exhibit a factorized structure for logarithms in the $a$ and $b$ variables.

For threshold Drell-Yan, the kinematic restrictions only allow soft radiation in
the final state. This induces additional large logarithms $\ln(1-\tau)$.  These
can be written in terms of a ratio of scales $\ln(\mu_S/\mu_H)$, where the soft
scale $\mu_S \simeq Q (1-\tau)$ is another important scale in the analysis. The
logarithms $L = \ln(\mu_S/\mu_H)$ appear as double-logarithmic series $(\alpha_s
L^2)^k$ in the cross section.  In the threshold factorization theorem in
\eq{DYendpt}, these double logarithms can be summed by evolving the PDFs and the
threshold soft and hard functions, $S_\thr$ and $H$, to a common scale $\mu$, as
shown in \fig{running_drellyan}(b).  Since $\xi_{a,b}\to 1$, the logarithms
$\ln(1-\xi_a)$ and $\ln(1-\xi_b)$ are also large. The RGE for the PDFs must be
expanded, and the result sums a double-logarithmic series of $\ln^2(1-\xi)$
terms.  The threshold soft function sums double logarithms $\ln^2(\mu/\mu_S)$
between $\mu_S$ and $\mu$, while the threshold hard function sums double
logarithms $\ln^2(\mu/\mu_H)$ between $\mu_H$ and $\mu$.  The evolution
equations are
%%%
\begin{align}\label{eq:RGEendpt}
\mu\,\frac{\df}{\df\mu} H(q^2, \mu)
&= \gamma_H(q^2,\mu)\, H(q^2,\mu)
\,,\nn\\
\mu\,\frac{\df}{\df\mu} f_i(\xi,\mu)
&= \int\!\frac{\df\xi'}{\xi'}\, P^{\rm expanded}_{ii}\Bigl(\frac{\xi}{\xi'}, \mu\Bigr)\, f_i(\xi',\mu)
\,,\\\nn
\mu\, \frac{\df}{\df\mu} S_\thr(k,\mu)
&= \int\!\df k_s'\, \gamma_{S_\thr}(k - k',\mu)\, S_\thr(k',\mu)
\,.\end{align}
%%%
The consistency of the RGE at the scale $\mu$ shown in \fig{running_drellyan}(b)
implies that the double logarithms in $f_i$, $f_j$, and $S_\thr$ combine in such
a way that the RGE of the convolution $f_i f_j\otimes S_\thr$ is identical to
that of $H$, and hence only depends on $q^2$.

For isolated Drell-Yan, the kinematic restrictions allow both soft and collinear
initial-state radiation, and induce an invariant-mass scale for each beam
function, $\mu_B^2 \simeq x_a \Ecm B_{a}^+ $ and $\mu_B^2 \simeq x_b \Ecm B_b^+
$, and a soft scale $\mu_S \simeq B_{a,b}^+ $. For simplicity, we use a common
scale $\mu_B$ for both beam functions in our discussion here. (Since the
evolution of the two beam functions is independent, one can just as easily
implement two independent beam scales.) As we saw in \subsec{DYfact}, at
partonic center-of-mass energies of a hundred GeV to a few TeV there is a large
hierarchy between the different scales, $\mu_\Lambda \ll \mu_S \ll \mu_B \ll
\mu_H$, and correspondingly large double and single logarithms of the ratios of
these scales.  The RGE running for this case is shown in
\fig{running_drellyan}(c).  Here, the PDFs are not restricted to their
endpoints, so their evolution is given by \eq{fevolution}, which involves the
unexpanded and nondiagonal $P_{ij}(\xi/\xi')$ and sums single logarithms,
$(\alpha_s L)^k$. For each $f_j$ this evolution joins at $\mu = \mu_B$ with the
Wilson coefficients $\cI_{ij}$ in the beam function factorization $B_i =
\cI_{ij}\otimes f_j$ of \eq{B_fact}. The $\cI_{ij}$ cancel the $\xi$-dependent
evolution of $f_j$, and turn it into the $t$-dependent evolution of $B_i$, which
sums a double-logarithmic series. The objects meeting at the common scale $\mu$
in \fig{running_drellyan}(c) are the hard function, which is identical to the
threshold case in \eq{RGEendpt}, and the beam and soft functions,
%%%
\begin{align}
\mu\,\frac{\df}{\df\mu} B_i(t, x, \mu)
&= \int\!\df t'\, \gamma_B^i(t - t',\mu)\, B_i(t',x,\mu)
\,,\nn\\
\mu\,\frac{\df}{\df\mu} S_\hemiin(k_a^+, k_b^+,\mu)
&= \int\!\df k_a'\,\df k_b'\, S_\hemiin(k_a', k_b',\mu)
\nn\\ & \quad\times
\gamma_{S_\hemiin}(k_a^+ - k_a', k_b^+ - k_b',\mu)\,
\,.\end{align}
%%%
The consistency of the RGE at $\mu$ now implies that the double-logarithmic running
in the different variables for $B_i$, $B_j$, and $S_\hemiin$ cancels such that the
convolution $B_i B_j \otimes S_\hemiin$ has an RGE identical to $H$, which only depends
on $q^2$.  (A detailed discussion of this consistency can be found in
Ref.~\cite{Fleming:2007xt} for the analogous case of two jet functions and the
final-state hemisphere soft function, $J J\otimes S_\hemiout$, and in
Ref.~\cite{Stewart:2010qs} for the case discussed here.) It is important that this
cancellation would not be possible if we tried to replace $B_i$ by $f_i$ in the
isolated factorization theorem.  Given the type of double logarithms in
the cross section, the single logarithms summed by the PDFs at generic $x$
cannot combine with the double logarithms in $S_\hemiin$ to
give a result in agreement with the double logarithms in $H$. Thus, the structure
of double logarithms necessitates the presence of beam functions in the
isolated factorization theorem.

By the same argument we can conclude that for all processes involving a
threshold-type hard function $H$ with double logarithms, and with $x_{a,b}$ away
from one, the description of the initial-state radiation will require beam
functions $B_i$.  This includes all situations where $H$ is the square of Wilson
coefficients of SCET operators, $H=\sum_k\abs{C_k}^2$ (for example when the
energetic partons in the hard collision all have distinct collinear directions).
In particular, the theoretical description of any threshold process with $x\to
1$ can be extended to a factorization theorem for the respective isolated case
with $x$ away from one. This is achieved by adding variables $B_{a,b}^+$,
replacing the PDFs by beam functions, and replacing the threshold soft function
by an appropriate soft function for the isolated case.

Thus, beam functions are quite prevalent for cross sections that one may wish to
study at the LHC. In situations where the hadronic final state is constrained
with variables that are more complicated than $B_{a,b}^+$, one
generically expects to find different beam functions and different soft
functions encoding these constraints.  This extension is analogous to how
the choice of jet algorithm modifies the definition of the jet and soft functions
for central jets produced by the hard collision~\cite{Bauer:2008jx}.  Even with
this generalization, the beam and soft functions will both
sum double-logarithmic series, and we expect that the factorization relating the beam
function to the PDFs will carry through, just with different coefficients
$\cI_{ij}$.

%===============================================================================
\subsection{Extension to Final-State Jets}
\label{subsec:Jets}
%===============================================================================

\begin{figure*}[t!]
\includegraphics[scale=0.75]{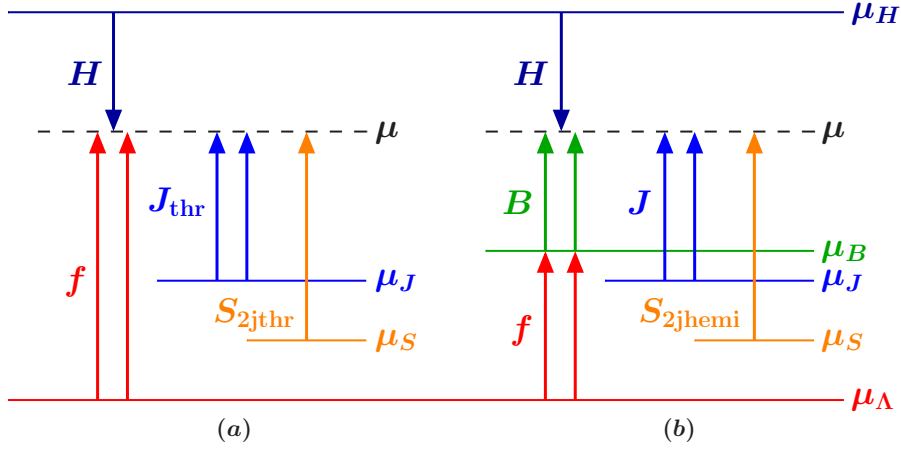}
\caption{RGE running for dijet production for (a) the threshold situation and (b)
  the isolated situation.}
\label{fig:running_2jet}
\end{figure*}

In \fig{pp}, where we show the types of hadronic final states for inclusive, endpoint,
and isolated Drell-Yan,  we also included analogs where the lepton
pair is replaced by two jets. Figure~\ref{fig:2jet_threshold} shows the
threshold dijet production process studied in Ref.~\cite{Kidonakis:1998bk},
which is the generalization of the Drell-Yan threshold process in
\fig{drellyan_threshold}.  Figure~\ref{fig:2jet_excl} shows the isolated dijet
production process, which is the generalization of isolated Drell-Yan in
\fig{drellyan_excl}. The goal of this subsection is to give a rough idea of how
the isolated factorization theorem will look when it is extended to include
final-state jets. Recall that our proof of factorization is only for the
Drell-Yan case. We stress that the factorization formula for the isolated dijet
case discussed here expresses our expectations and has not been rigorously
derived.

Final-state jets are identified by a jet algorithm
as more-or-less isolated groups of energetic particles within a
cone%
\footnote{Here $\phi$ and $\eta$ denote the azimuthal angle and
  pseudorapidity. Although our notation corresponds to a cone algorithm, at the level
  of our discussion we may equally well substitute a $k_T$ algorithm.}
of some radius $R = [(\Delta \phi)^2 + (\Delta \eta)^2]^{1/2}$.  For a dijet event, the
jet algorithm allows us to define the total jet momenta $P_1^\mu$ and $P_2^\mu$.%
\footnote{We take $P_{1,2}^\mu$ to be the momenta of the two hardest jets found by
the jet algorithm. We will see below that the final-state restrictions considered here
eliminate the possibility of having other hard central jets.}
Given these, we let $y_1$ and $y_2$ be the rapidities of the two jets relative to the beam
axis and define $\Delta y = y_1 - y_2$. The invariant masses of the jets are denoted as $M_{1,2}^2 = P_{1,2}^2$.
Two analogs for the $q^2$-variable of Drell-Yan are
%%%
\begin{align}
\MJJ^2 &= (P_1+P_2)^2
\,, \nn\\
\mJJ^2 &= 2 P_1\sdt P_2 = \MJJ^2 - M_1^2 - M_2^2
\,,\end{align}
%%%
where $\MJJ^2$ is the total invariant mass of the two jets, and $\mJJ^2$ is their total invariant mass minus their
individual invariant masses. The corresponding analogs of the Drell-Yan $\tau$-variable are then
%%%
\begin{equation}
\tau_J = \frac{ \MJJ^2}{\Ecm^2}
\,,\qquad
\Delta\tau_J = \frac{\mJJ^2}{\Ecm^2}
\,.\end{equation}
%%%

Identifying the two jets already restricts the hadronic final state, which means
there is no analog of the inclusive Drell-Yan factorization theorem. In the
threshold case, we take the limit $\tau_J\to 1$, which ensures that the final
state consists of two back-to-back jets plus soft radiation and no additional
energetic jets, as shown in \fig{2jet_threshold}. The limit $\Delta\tau_J\to 1$
is even more restrictive, since it also forces the two jets to have very small
invariant masses, essentially behaving like massless particles for the
factorization theorem.

The threshold factorization for $\mJJ^2$ was considered in
Ref.~\cite{Kidonakis:1998bk},%
\footnote{To the best of our knowledge a proof of the decoupling of Glauber
  gluons does not exist for threshold dijet production in hadronic collisions,
  so there is no complete proof of \eq{2jetthreshold}.}%
${}^{,}$\footnote{When replacing $\mJJ^2$ by $\MJJ^2$ in \eq{2jetthreshold}, the
  main difference is that now $\rho_i = 2k_i^0/\MJJ$. In this case, the threshold
  limit $\tau_J\to 1$ alone does not constrain the jet invariant masses,
  $M_i^2$, to be small. Since $M_i^2\sim R^2 \MJJ^2$, they are constrained to be
  small by jet algorithms with $R^2\ll 1$. This induces complications in
  deriving an all-orders factorization theorem, but still suffices to imply that
  the factorization formula in \eq{2jetthreshold} with the replacement
  $\mJJ^2\to \MJJ^2$ will sum the next-to-leading
  logarithms~\cite{Kidonakis:1998bk}.}%
%%%
\begin{align} \label{eq:2jetthreshold}
  & \frac{\df\sigma}{\df\mJJ^2\df(\Delta y)} = \frac{1}{\Ecm^2}
  H^{IL}(\mJJ^2,\Delta y) \int\!
  \df \xi_a\, \df \xi_b\, \df \rho_1\, \df \rho_2\,
\nn\\  & \ \times
f(\xi_a)\, f(\xi_b)\, J_\thr(\mJJ, \rho_1, R)\, J_\thr (\mJJ, \rho_2, R)
\nn\\ & \ \times
  S^{LI}_\jthr\bigl[\mJJ(1\!-\!\Delta\tau_J
   \!-\! (1\!-\!\xi_a)\!-\!(1\!-\!\xi_b)\!-\!\rho_1\!-\!\rho_2),\Delta y\bigr]
\nn\\ & \ \times
  \Bigl[1 + \ORd{\frac{\lqcd}{\mJJ}, R, 1-\Delta\tau_J} \Bigr]
\,,\end{align}
%%%
where one sums over the color basis $IL$, and for simplicity the dependence on
flavor labels and $\mu$ of the various functions have been suppressed.%
\footnote{We also made a redefinition so that the PDFs $f$ depend on light-cone
  momentum fractions rather than fixed energy as in
  Ref.~\cite{Kidonakis:1998bk}, absorbing the difference into the hard functions
  $H_{IL}$.}  The first argument of the soft function $S^{LI}_\jthr$ is the
energy it radiates outside the jet cones. The jet function $J_\thr$ depends on
$\rho_i = 2 k_i^0/\mJJ + M_i^2/\mJJ^2$, where $k_i^0$ is the energy of particles it
radiates outside its cone.  The threshold limit $\Delta \tau_J\to 1$ forces
$\xi_{a,b}\to 1$ and $\rho_{1,2}\to 0$, so we have PDFs, jet functions, and soft
functions that all correspond to the threshold limit and contain double
logarithms.%
\footnote{We follow the SCET definition of soft functions as matrix elements of
  eikonal Wilson lines without subtractions, so $S_\jthr$ has double logarithms.
  In SCET the jet functions have subtractions.}  When these functions are
convoluted they consistently reproduce the double logarithms encoded in the
renormalization group evolution of $H^{IL}$, which is shown in
\fig{running_2jet}(a).  Compared to threshold Drell-Yan as described in
\subsec{RGE}, there are extra convolutions for the jets and a more sophisticated
soft function that is a matrix in color space.

To extend the isolated Drell-Yan factorization theorem to the dijet case,
we need to define analogs of the $B_{a,b}^+$ variables in \subsec{DYfact} that
can constrain the final state in an appropriate manner. We
first define lightlike vectors along each jet direction, $n_1^\mu=(1,\vec{n}_1)$ and
$n_2^\mu=(1,\vec{n}_2)$ with $\vec{n}_{1,2} = \vec{P}_{1,2}/\abs{\vec{P}_{1,2}}$,
and corresponding lightlike vectors in the opposite directions, $\bar{n}_{1,2}^\mu = (1,-\vec{n}_{1,2})$.
Next, the final-state particles are
divided into four categories $\{ a,b,1,2\}$ as follows.
We define $\cR_1$ and $\cR_2$ as all particles that have been grouped into
jets $1$ and $2$ by the jet algorithm.
The remaining particles not grouped into either of $\cR_{1,2}$ are divided
into the two hemispheres $a$ and $b$ as before, which defines $\cR_{a,b}$. We then define plus momenta
%%%
\begin{align} \label{eq:plusmomenta}
  P_1^+ &=  \sum_{k\in \cR_1} n_1 \sdt p_k
\,,&
  P_2^+ &=  \sum_{k\in \cR_2} n_2 \sdt p_k
\,, \nn\\
  B_a^+ &=  \sum_{k\in \cR_a} n_a \sdt p_k
\,,&
  B_b^+ &=  \sum_{k\in \cR_b} n_b \sdt p_k
\,.\end{align}
%%%
This definition of $B_{a,b}^+$ is identical to \eq{Badef}.
Since the jet algorithm is used for the grouping, these variables are infrared safe.
The union of the four categories covers all of phase space, so the measurement of all the momenta
in \eq{plusmomenta} defines a global observable sensitive to all
radiation in the event.  Just as in our Drell-Yan discussion, the
definition of the variables $B_{a,b}^+$ ensures that radiation outside the reach of the
detector can safely be ignored.

For the isolated dijet limit we demand that $P_{1,2}^+ /\MJJ \ll 1$ and $B_{a,b}^+ /\MJJ \ll 1$.
In addition, we constrain $y_1$ and $y_2$ such that the jets lie in the central region sufficiently
separated from the beam directions. The condition on $P_{1,2}^+$ ensures that the jet regions $\cR_{1,2}$ only contain energetic radiation along the direction of their jet plus soft radiation. The condition on $B_{a,b}^+$ has a similar effect as for isolated Drell-Yan. It ensures that there is only soft and no energetic radiation in the central region apart from the two jets. Thus, we have exactly two isolated central jets.

Since each category $\{ a,b,1,2\}$ predominantly contains all the corresponding
collinear particles, this division of phase space mainly affects how the soft
radiation is associated to each jet.  In analogy to isolated Drell-Yan, we
divide the total soft momentum as $k = k_a + k_b + k_1 + k_2$, where each $k_i$
is the total momentum of soft particles in $\cR_i$, and we define $k_i^+ =
n_i\sdt k_i$.  The corresponding isolated dijet soft function,
$S^{LI}_\jhemiin(k_a^+, k_b^+, k_1^+, k_2^+, y_1, y_2)$, depends on all four
directions $n_1$, $n_2$, $n_a$, $n_b$ and hence on the rapidities $y_1$ and
$y_2$. It now contains both incoming and outgoing soft Wilson lines and is a
matrix in color space, where we can use the same color basis $\{LI\}$ as in the
threshold case. The soft function itself of course differs from the threshold
case.

The total jet momenta can now be written as
%%%
\begin{equation}
P_1^\mu = \w_1 \frac{n_1^\mu}{2} + q_1^+ \frac{\bar{n}_1^\mu}{2} + q_\perp^\mu + k_1^\mu
\,,\end{equation}
%%%
and similarly for $P_2^\mu$. The first three terms on the right-hand side are the contributions from the
energetic radiation in the jet, with $\w_{1,2} \sim \MJJ$ and $q_1^+, q_\perp \ll \MJJ$. Expanding in the small components of the beam and jet momenta, the hard momentum components in the beams and jets have to satisfy
%%%
\begin{equation}
\w_a \frac{n_a^\mu}{2} + \w_b \frac{n_b^\mu}{2} = \w_1 \frac{n_1^\mu}{2} + \w_2 \frac{n_2^\mu}{2}
\,.\end{equation}
%%%
Thus, in this limit the two jets are massless and back-to-back in
the transverse plane, but need not be back-to-back in three dimensions.
In terms of $y_{1,2}$ and $\MJJ^2 = \w_a \w_b = \w_1\w_2 n_1\sdt n_2/2$, we then have
%%%
\begin{align}
\w_a &= \MJJ\, e^{(y_1 + y_2)/2} \equiv x_a\Ecm
\,,\nn\\
\w_b &= \MJJ\, e^{-(y_1 + y_2)/2} \equiv x_b\Ecm
\,,\nn\\
\w_{1,2} &= \MJJ\,\frac{\cosh y_{1,2}}{\cosh(\Delta y/2)}
\,.\end{align}
%%%

The collinear radiation in the jets is described by jet functions that depend on the invariant-mass variables
$\w_1 q_1^+$ and $\w_2 q_2^+$. By momentum conservation we have $q_{1,2}^+ = P_{1,2}^+ - k_{1,2}^+$, so
the jet functions are convoluted with the soft function through $k_{1,2}^+$. The subtraction of $k_{1,2}^+$
is necessary to remove the plus momentum of soft particles in the jet, since
the momentum distribution of these particles is properly described by the soft
function not by the jet function.  Just as for isolated Drell-Yan, the collinear initial-state radiation in the beams is described by beam functions, which depend on the invariant-mass variable $t_a = \w_a (B_a^+-k_a^+)$ and
momentum fraction $x_a$, and similarly for hemisphere $b$. Since the jets are well-separated from the beams,
removing the particles in $\cR_{1,2}$ from the hemispheres $\cR_{a,b}$ mainly affects the soft radiation and not the energetic partons collinear to the beams. Therefore up to power corrections, we expect the same inclusive beam functions as before.

From the above discussion it is natural to suppose that the factorization theorem for isolated dijet production will be
%%%
\begin{align} \label{eq:2jetisolated}
&\frac{\df\sigma}{\df \MJJ^2 \df y_1 \df y_2 \df B_a^+ \df B_b^+ \df P_1^+ \df P_2^+}
\\ &\qquad
= H^{IL}(\MJJ^2,y_1, y_2) \int \!\df k_a^+\, \df k_b^+\, \df k_1^+\, \df k_2^+
\nn\\ &\qquad\quad \times
J_\cone[\w_1(P_1^+ -k_1^+)]\, J_\cone[\w_2(P_2^+ - k_2^+)]
\nn\\ &\qquad\quad \times
B[\w_a(B_a^+ - k_a^+), x_a]\, B[\w_b(B_b^+ - k_b^+), x_b]
\nn\\ &\qquad\quad \times
S^{LI}_\jhemiin(k_a^+, k_b^+, k_1^+, k_2^+, y_1, y_2)
\nn\\\nn &\qquad\quad \times
\biggl\{1 + \mathcal{O}\biggl[\frac{\lqcd}{\MJJ}, \frac{\w_{a,b} B_{a,b}^+}{\MJJ^2}, \frac{\w_{1,2} P_{1,2}^+}{\MJJ^2} \biggr] \biggr\}
\,,\end{align}
%%%
where we again suppressed flavor labels and $\mu$ dependence.  The hard function
$H^{IL}(\MJJ,y_i)$ is precisely the threshold hard function, and we sum over the
same color basis $\{IL\}$.  The subscript $\cone$ on the soft and jet functions
denotes the fact that their plus momenta depend on the regions $\cR_i$, which in
turn depend on $y_i$.

The consistency of the RGE for the isolated dijet factorization theorem, shown
in \fig{running_2jet}(b), again provides important constraints on its structure.
Each of the functions $J_\cone$, $B$, and $S_\jhemiin^{LI}$ includes a series of
double logarithms, and when these functions are convoluted over the $k_i^+$
variables at a common scale $\mu$, these different series have to collapse to
precisely the double-logarithmic series of the hard function $H^{IL}$. The RGE
for the hard function $H^{IL}$ is a matrix equation in color space, but has no
convolutions of kinematic variables. We expect that this equivalence will occur
in the same manner as it does for the isolated Drell-Yan case.

Key missing ingredients in providing a rigorous derivation of \eq{2jetisolated}
include i) providing a mathematically rigorous treatment of the separation of
jets and beams in the factorization, and ii) determining the role of Glauber
degrees of freedom, that in principle may couple the final-state jets and spoil
factorization.  It should be evident that if such a proof becomes available, it
will be straightforward to generalize the above discussion to the case where we
produce $N$ isolated jets rather than just two.

%%%%%%%%%%%%%%%%%%%%%%%%%%%%%%%%%%%%%%%%%%%%%%%%%%%%%%%%%%%%%%%%%%%%%%%%%%%%%%%%
\section{The Beam Function}
\label{sec:beamfunction}
%%%%%%%%%%%%%%%%%%%%%%%%%%%%%%%%%%%%%%%%%%%%%%%%%%%%%%%%%%%%%%%%%%%%%%%%%%%%%%%%

In this section, we discuss the properties of the beam function in more detail.
We present its definition and relation to the standard PDF, as well as its
renormalization group evolution. We will display explicit results for the quark
beam function at one loop (leaving a detailed derivation to a dedicated
publication~\cite{Stewart:2010qs}).  The comparison of effects in the beam functions
and PDFs are illustrated with plots.

The quark, antiquark, and gluon beam functions are defined in SCET as
%%%
\begin{widetext}
\begin{align} \label{eq:B_def}
B_q(\w b^+, \w/P^-, \mu)
&= \frac{\theta(\w)}{\w} \int\! \frac{\df y^-}{4\pi}\, e^{\img b^+ y^-/2}
   \MAe{p_n(P^-)}{\bar\chi_n\Bigl(y^-\frac{n}{2}\Bigr) \delta(\w-\bnP_n)\,
   \frac{\bnslash}{2} \chi_n(0)}{p_n(P^-)}
\,, \nn\\
B_{\bar{q}}(\w b^+, \w/P^-, \mu)
&= \frac{\theta(\w)}{\w} \int\! \frac{\df y^-}{4\pi}\,e^{\img b^+ y^-/2}
   \MAe{p_n(P^-)}{\tr_\mathrm{spin}\Bigl[\frac{\bnslash}{2} \chi_n\Bigl(y^-\frac{n}{2}\Bigr)\,
   \delta(\w-\bnP_n)\, \bar\chi_n(0)\Bigr]}{p_n(P^-)}
\,, \nn\\
B_g(\w b^+,\w/P^-, \mu)
&= -\theta(\w) \int\! \frac{\df y^-}{4\pi}\,e^{\img b^+ y^-/2}
   \MAe{p_n(P^-)}{\cB_{n\perp\mu}^c\Bigl(y^- \frac{n}{2}\Bigr)\,
   \delta(\w-\bnP_n)\, \cB_{n\perp}^{\mu c}(0)}{p_n(P^-)}
\,.\end{align}
%%%
We will briefly explain the relevant notation. (A more detailed overview of SCET
and the definitions of the objects in \eq{B_def} are given in \subsec{SCET}.) As
before, $n^\mu = (1,\vec{n})$ and $\bn^\mu = (1,-\vec{n})$ are
lightlike vectors, $n^2 = \bn^2 = 0$, $n\sdt \bn = 2$, where $\vec{n}$ is a unit
three-vector in the direction of the proton. The proton states
$\ket{p_n(P^-)}$ have lightlike momentum $P^\mu = P^- n^\mu/2$, and the matrix
elements are always implicitly averaged over the proton spin. The SCET fields for
collinear quarks and gluons,
$\chi_n(y)$ and $\cB_{n\perp}^\mu(y)$, are composite fields containing
Wilson lines of collinear gluons [see \eq{chiB}]. Matrix elements with these fields include
so-called zero-bin subtractions~\cite{Manohar:2006nz}, which effectively divide
by a matrix element of Wilson lines~\cite{Lee:2006nr}. At lowest order in the
strong coupling, the fields describe an energetic quark or gluon
moving in the $n$ direction with momentum $p^- n^\mu/2 + k^\mu$ with $k \ll
p^-$. The momentum operator $\bnP_n$ picks out the large light-cone component
$p^-$ of all particles annihilated by $\chi_n$ or $\cB_{n\perp}^\mu$.  Thus, when these
fields annihilate the incoming colliding parton, the
$\delta$ function in \eq{B_def} sets $\w$ equal to the $p^-$ of that
parton. Therefore, $x = \w/P^-$ is the fraction of the proton's light-cone
momentum that is carried by the parton into the hard collision. At the time of the
collision, this parton is propagating in an initial-state jet rather than the
proton, which is encoded by the dependence of the beam functions on the variable
$b^+ = -k^+$. Here, $k^+ = n\sdt k$ is the small component of the incoming collinear
parton's momentum. The variable $t = \w b^+ \sim -p^2$ measures the parton's virtuality,
where $t > 0$, because the parton is spacelike.
As we already saw in \eq{DYbeam}, the beam functions are convoluted
with the soft function through $b^+$.

The beam function definitions in \eq{B_def} can be compared with those of the
standard quark, antiquark, and gluon PDFs in SCET~\cite{Bauer:2002nz},
%%%
\begin{align} \label{eq:f_def_SCET}
f_q(\w'/P^-, \mu)
&= \theta(\w') \MAe{p_n(P^-)}{\bar\chi_n(0)\,
  \delta(\w' - \bnP_n)\,\frac{\bnslash}{2}\chi_n(0)}{p_n(P^-)}
\,,\nn\\
f_{\bar{q}}(\w'/P^-, \mu)
&= \theta(\w') \MAe{p_n(P^-)}{\tr_\mathrm{spin}\Bigl[\frac{\bnslash}{2}
   \chi_n(0)\, \delta(\w' - \bnP_n)\, \bar\chi_n(0) \Bigr]}{p_n(P^-)}
\,,\nn\\
f_g(\w'/P^-, \mu)
&= -\theta(\w')\, \w' \MAe{ p_n(P^-)}{\cB_{n\perp\mu}^c(0)\,
  \delta(\w' - \bnP_n)\, \cB_{n\perp}^{\mu c}(0) }{p_n(P^-)}
\,.\end{align}
%%%
The $f_i$ depend on the analogous light-cone momentum fraction $\xi = \w'/P^-$.
As discussed in the beginning of \subsec{BtoF}, $\xi$ can be interpreted as the
momentum fraction of the hard parton when it is taken out of the proton and due
to perturbative corrections generally differs from $x = \w/P^-$ appearing in the
beam functions.  A more common and equivalent definition of the PDFs is in terms
of QCD fields.  For example, for the quark PDF,
%%%
\begin{equation} \label{eq:f_def_QCD}
f_q(\w'/P^-, \mu) = \theta(\w') \int\! \frac{\df y^+}{4\pi}\,
  e^{-\img\, \w'y^+/2}
  \MAe{p_n(P^-)}{\bar\psi \Bigl(y^+\frac{\bn}{2}\Bigr) \frac{\bnslash}{2}\,
   W_{\bn}\Bigl(y^+\frac{\bn}{2},0\Bigr) \psi(0)}{p_n(P^-)}
\,.\end{equation}
\end{widetext}
%%%
Equation~\eqref{eq:f_def_SCET} is essentially the Fourier transform of
\eq{f_def_QCD}, where the SCET fields are written in momentum space with respect
to $\w'$, while the QCD fields are separated along the $\bn$ direction between
$0$ and $y^+\bn/2$. The lightlike Wilson line $W_{\bn}(y^+\bn/2,0)$ is required
to render the product of the quark fields at different space-time points gauge
invariant, and the corresponding Wilson lines in \eq{f_def_SCET} are those
hidden in the definitions of $\chi_n$ and $\cB_{n\perp}^\mu$.

\begin{figure}[b!]
\subfigure[]{\includegraphics[scale=0.7]{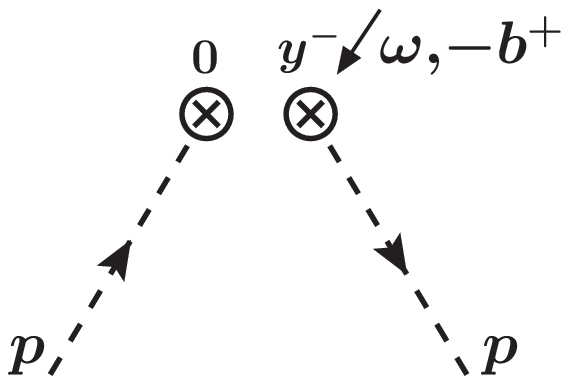}%
\label{fig:Btree}%
}\hfill%
\subfigure[]{\includegraphics[scale=0.7]{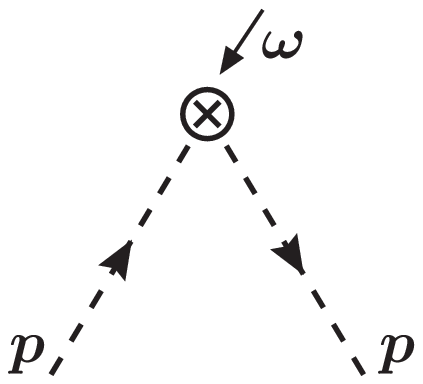}%
\label{fig:ftree}%
}
\caption{Tree-level diagrams for the quark beam function (a) and quark PDF (b). }
\end{figure}

\begin{figure}[b!]
\subfigure[]{\includegraphics[scale=0.7]{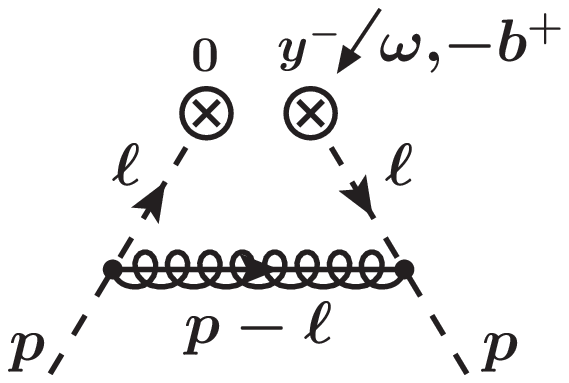}%
\label{fig:Bone_a}%
}\hfill%
\subfigure[]{\includegraphics[scale=0.7]{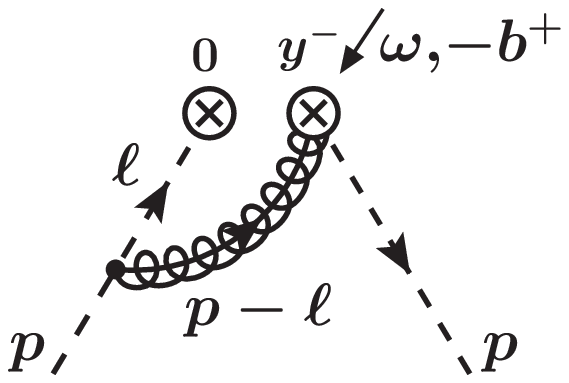}%
\label{fig:Bone_b}%
}
\\
\subfigure[]{\includegraphics[scale=0.7]{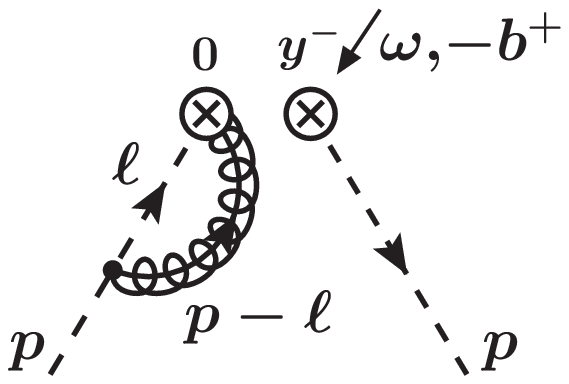}%
\label{fig:Bone_c}%
}\hfill%
\subfigure[]{\includegraphics[scale=0.7]{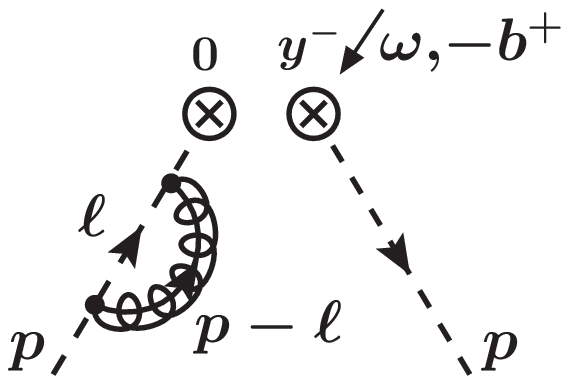}%
\label{fig:Bone_d}%
}\hspace*{\fill}%
\\
% \hfill%
\subfigure[]{\includegraphics[scale=0.7]{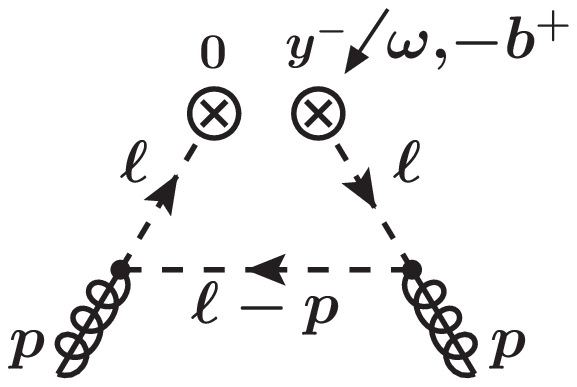}%
\label{fig:Bone_e}%
}\hfill%
\subfigure[]{\includegraphics[scale=0.7]{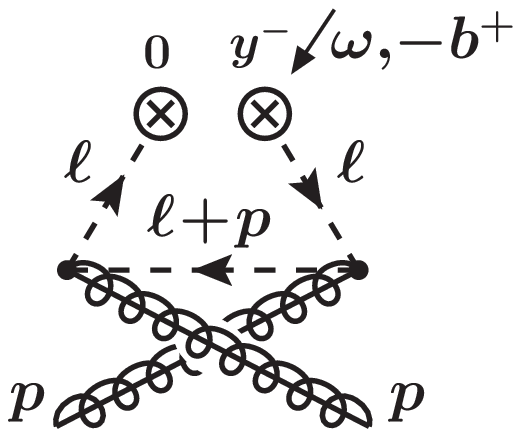}%
\label{fig:Bone_f}%
}\hspace*{\fill}%
\caption{One-loop diagrams for the quark beam function. Graphs (a) and (b) correspond to real gluon emission, while (c) and (d) are virtual corrections. Graphs (e) and (f) determine the contribution of the gluon PDF to the quark beam function.}
\label{fig:Bone}
\end{figure}

In the beam functions in \eq{B_def}, the fields are in addition separated along
the $n$ direction, with a large separation $y^- \gg y^+$ corresponding to the
small momentum $b^+ \ll \w$.  This $y^-$ separation is formulated with a gauge
invariant multipole expansion of fields in SCET. The possible gauge
transformations in the effective theory are divided into global, collinear, and
soft, and it is the coupling to soft gluons and the corresponding soft gauge
transformations that are relevant for making the $y^-$ separation gauge
invariant. The collinear fields in \eq{B_def} are the ones that occur after
making a field redefinition to decouple soft gluons into the soft function, and
the resulting collinear fields no longer transform under soft gauge
transformations.  Hence, the SCET definitions in \eq{B_def} are gauge invariant.

Note that formulating equivalent definitions of the beam functions directly in
QCD is more challenging. It seems to require QCD fields that are simultaneously
separated in the $n$ and $\bn$ directions, and a priori it is not clear how to
obtain an unambiguous gauge-invariant expression in this case, because Wilson
lines connecting the fields along different paths are not equivalent.  For the
beam functions, which one might think of as $b^+$-dependent PDFs, this problem
is solved in SCET, because the effective theory distinguishes the large and
small momentum components with the multipole expansion, resolving the ambiguity.

For $t = \w b^+ \gg \lqcd^2$, or equivalently $y^- \ll \w/\lqcd^2$, the beam
functions can be related to the PDFs by performing an operator product expansion
in $\lqcd^2/t \ll 1$. This leads to the factorized form
%%%
\begin{align} \label{eq:B_fact2}
B_i(t, x, \mu)
&= \sum_j\!\int_x^1 \frac{\df\xi}{\xi}\, \cI_{ij}\Bigl(t,\frac{x}{\xi},\mu \Bigr) f_j(\xi, \mu)
\nn\\ &\quad\times
\biggl[1 + \ORd{\frac{\lqcd^2}{t}}\biggr]
,\end{align}
%%%
where $j = \{g, u, \bar u, d, \ldots\}$ and $\cI_{ij}$ is a perturbatively
calculable Wilson coefficient.  The physical interpretation of this equation was
discussed in \subsec{BtoF}.  For $B_g$, the equivalent of the matching
expression in \eq{B_fact2} was derived in Ref.~\cite{Fleming:2006cd} for the
$\cI_{gg}$ term using a moment-space OPE to match \SCETa onto \SCETb.
Ref.~\cite{Fleming:2006cd} considered this matching at the level of the matrix
element defining $B_g$, without the accompanying physical picture advocated here
that implies that beam functions will occur in a wide variety of interesting
processes. The mixing contributions were missed in their analysis, but the
extension of their proof to the general case is straightforward~\cite{Stewart:2010qs}.

The coefficients $\cI_{ij}(t, x/\xi, \mu)$ can be determined perturbatively by
computing both sides of \eq{B_fact2} with the proton states in the definitions
of $B_i$ and $f_j$ replaced by quark and gluon states. The tree-level diagram
for the quark beam function is shown in \fig{Btree} and for the PDF in
\fig{ftree}. They give
%%%
\begin{align}
B^\tree_{q/q}(t, x, \mu) &= \delta(t)\, \delta(1 - x)
\,, \nn \\*
f^\tree_{q/q}(\xi,\mu) &= \delta(1 - \xi)
\,,\end{align}
%%%
from which we deduce $\cI_{qq}^\tree(t, x/\xi, \mu) = \delta(t) \delta(1 - x/\xi)$.
In general, we have
%%%
\begin{equation}
\cI_{ij}^\tree\Bigl(t, \frac{x}{\xi}, \mu \Bigr) = \delta_{ij}\,
  \delta(t)\, \delta\Bigl(1 - \frac{x}{\xi}\Bigr)
\,,\end{equation}
%%%
so the tree-level beam functions reduce to the PDFs
%%%
\begin{equation}
B^\tree_i(t, x, \mu) = \delta(t)\, f_i(x, \mu)
\,.\end{equation}
%%%

The one-loop coefficients for the quark beam function are determined from the
diagrams in \fig{Bone} together with the corresponding diagrams for the PDFs.
The beam functions and PDFs are renormalized using dimensional regularization
with $\overline{\text{MS}}$.  From the first four diagrams in \fig{Bone} we find
the one-loop correction to the quark-quark coefficient (here $z = x/\xi$)
%%%
\begin{widetext}
\begin{align} \label{eq:Iqq_oneloop}
\cI_{qq}^\oneloop(t, z, \mu)
&= \frac{\alpha_s(\mu)C_F}{2\pi}\, \theta(z)\,
  \biggl\{
  \frac{2}{\mu^2} \biggl[\frac{\theta(t/\mu^2)\ln(t/\mu^2)}{t/\mu^2}\biggr]_+\delta(1-z)
  + \frac{1}{\mu^2} \biggl[\frac{\theta(t/\mu^2)}{t/\mu^2}\biggr]_+
     \biggl[\biggl[\theta(1-z)\frac{1 + z^2}{1-z}\biggr]_+\!\! - \frac{3}{2}\,\delta(1-z)\biggr]
\nn\\ &\quad
  + \delta(t) \biggl[
    \biggl[\frac{\theta(1-z) \ln(1-z)}{1-z}\biggr]_+(1+z^2)
    - \frac{\pi^2}{6}\, \delta(1-z)
    + \theta(1-z)\biggl(1-z - \frac{1 + z^2}{1-z} \ln z\biggr)
    \biggr] \biggr\}
% \nn\\
% &= \frac{\alpha_s(\mu)C_F}{2\pi}\, \theta(z)\,
%  \biggl\{
%  \frac{2}{\mu^2} \biggl[\frac{\theta(t/\mu^2)\ln(t/\mu^2)}{t/\mu^2}\biggr]_+\delta(1-z)
%   + \frac{1}{\mu^2} \biggl[\frac{\theta(t/\mu^2)}{t/\mu^2}\biggr]_+
%      \biggl\{\biggl[\theta(1-z)\frac{1 + z^2}{1-z}\biggr]_+\!\! - \frac{3}{2}\,\delta(1-z)\biggr\}
% \nn\\ &\quad
%  + \delta(t) \biggl\{\biggl[\theta(1-z)\biggl(\frac{1 + z^2}{1-z}
%  \ln\frac{1-z}{z} + 1-z\biggr)\biggr]_+
%    + \Bigl(1 + \frac{\pi^2}{6}\Bigr)\, \delta(1-z) \biggr\} \biggr\}
\,.\end{align}
%%%
The last two diagrams in \fig{Bone} determine the one-loop contribution of the
gluon PDF to the quark beam function,
%%%
\begin{align} \label{eq:Iqg_oneloop}
\cI_{qg}^\oneloop(t,z, \mu)
&= \frac{\alpha_s(\mu)\,T_F}{2\pi}\,
  \theta(z)\, \theta(1 - z)
  \biggl\{
  \biggl[\frac{1}{\mu^2} \biggl[\frac{\theta(t/\mu^2)}{t/\mu^2}\biggr]_+
  + \delta(t)\ln \frac{1 - z}{z}\biggr]\bigl[z^2 + (1-z)^2\bigr]
  + \delta(t)\, 2z(1-z)\biggr\}
\,.\end{align}
%%%
At two loops, $\cI_{q\bar{q}}(t, z, \mu)$ will start to contribute as well. The
plus distributions are defined as
%%%
\begin{equation} \label{eq:plus_def}
\bigl[\theta(x) g(x)\bigr]_+
= \lim_{\eps \to 0} \frac{\df}{\df x} \bigl[\theta(x-\eps)\, G(x) \bigr]
\qquad\text{with}\qquad
G(x) = \int_1^x\!\df x'\, g(x')
\,,\end{equation}
%%%
satisfying the boundary condition $\int_0^1 \df x\, [\theta(x) g(x)]_+ = 0$. In
particular,
%%%
\begin{align}
\int_x^\infty\!\frac{\df z}{z}\, \bigl[\theta(1 - z)g(1 - z)\bigr]_+
   f\Bigl(\frac{x}{z}\Bigr)
&= \int_x^1\!\df z\, g(1 - z)\Bigl[\frac{1}{z}f\Bigl(\frac{x}{z}\Bigr)
   - f(x)\Bigr] + f(x)\, G(1-x)
\,,\nn\\
\frac{1}{\mu^2}\int_{-\infty}^{t_\max}\!\df t\,
  \biggl[\frac{\theta(t/\mu^2)\ln^n(t/\mu^2)}{t/\mu^2}\biggr]_+
&= \frac{1}{n+1}\,\ln^{n+1}\frac{t_\max}{\mu^2}
\,.\end{align}
%%%

The infrared (IR) divergences in the diagrams in \fig{Bone} precisely cancel those in the
PDF calculation as they must, so the matching coefficients in
\eqs{Iqq_oneloop}{Iqg_oneloop} are IR finite and independent of the IR
regulator.  The ultraviolet (UV) divergences in the diagrams determine the one-loop
RGE and anomalous dimension of the quark beam function, which in $\overline{\rm MS}$ are
%%%
\begin{equation} \label{eq:Bq_RGE}
\mu\,\frac{\df}{\df\mu} B_q(t,x, \mu) = \int\!\df t'\,
  \gamma_B^q(t - t', \mu)\, B_q(t', x, \mu)
\,,\qquad
\gamma_B^q(t, \mu)
= -2\Gamma_\cusp[\alpha_s(\mu)] \frac{1}{\mu^2}\biggl[\frac{\theta(t/\mu^2)}{t/\mu^2}\biggr]_+
  + \frac{\alpha_s(\mu)}{4\pi}\, 6C_F\,\delta(t)
\,.\end{equation}
%%%
Here, $\Gamma_\cusp[\alpha_s(\mu)]$ is the cusp anomalous dimension~\cite{Korchemsky:1987wg},
and the coefficient of the plus function is equal to $\Gamma_\cusp$
to all orders in perturbation theory. The non-cusp term of the anomalous dimension is
equal to that for the quark jet function at one-loop, and in Ref.~\cite{Stewart:2010qs} we prove that the
anomalous dimensions for the quark beam and jet functions are identical to
all orders in $\alpha_s$. As stated before, the RGE in \eq{Bq_RGE}
does not change $x$. Also, the mixing graphs in \figs{Bone_e}{Bone_f} have no UV
divergences and hence the gluon beam function does not mix into the quark beam
functions under renormalization. Equation~\eqref{eq:Bq_RGE} leads to the physical picture discussed in
\subsec{BtoF}. The RGE has a solution~\cite{Balzereit:1998yf, Neubert:2004dd, Fleming:2007xt}, which
can be written as~\cite{Ligeti:2008ac}
%%%
\begin{equation}
B_q(t, x, \mu) = \int\!\df t'\, B_q(t - t', x, \mu_0)\, U_B(t', \mu_0, \mu)
\,,\qquad
U_B(t, \mu_0, \mu) = e^{K_B}\,\frac{e^{-\gamma_E\, \eta_B}}{\Gamma(1 + \eta_B)}\,
\biggl\{\frac{\eta_B}{\mu_0^2} \biggl[\frac{\theta(t/\mu_0^2)]}{(t/\mu_0^2)^{1-\eta_B}}\biggr]_+ + \delta(t) \biggr\}
\,,\end{equation}
%%%
with the plus distribution defined according to \eq{plus_def}. Furthermore, $K_B \equiv K_B(\mu_0, \mu)$ and $\eta_B \equiv \eta_B(\mu_0, \mu)$ are
%%%
\begin{equation} \label{eq:KetaB_def}
K_B(\mu_0, \mu)
= \int_{\alpha_s(\mu_0)}^{\alpha_s(\mu)}\!\frac{\df\alpha_s}{\beta(\alpha_s)}\,
\biggl[4\, \Gamma_\cusp(\alpha_s) \int_{\alpha_s(\mu_0)}^{\alpha_s} \frac{\df \alpha_s'}{\beta(\alpha_s')}
   + \gamma_B^q(\alpha_s) \biggr]
\,,\qquad
\eta_B(\mu_0, \mu)
= -2 \int_{\alpha_s(\mu_0)}^{\alpha_s(\mu)}\!\frac{\df\alpha_s}{\beta(\alpha_s)}\, \Gamma_\cusp(\alpha_s)
\,,\end{equation}
%%%
where $\beta(\alpha_s)$ is the QCD $\beta$ function and $\gamma^q_B(\alpha_s)$ is the coefficient of $\delta(t)$ in $\gamma^q_B(t, \mu)$ in \eq{Bq_RGE}.

Using the above results for the quark beam function, we can see explicitly that
when we integrate over $0\leq t\leq t_\max$ to get the beam function
$\tB_q(t_\max, x, \mu)$ in \eq{tB_def}, the result contains double and single
logarithms of $t_\max/\mu^2$,
%%%
\begin{align} \label{eq:tB_log}
\tB_q(t_\max,x, \mu)
&= \theta(t_\max) f_q(x, \mu) + \theta(t_\max) \frac{\alpha_s(\mu)}{2\pi} \biggl\{
  C_F \biggl(\ln^2\frac{t_\max}{\mu^2} - \frac{3}{2} \ln\frac{t_\max}{\mu^2}\biggr) f_q(x, \mu)
\nn\\ &\quad
+ \ln\frac{t_\max}{\mu^2} \int_x^1\!\frac{\df z}{z}\, \biggl\{
C_F \biggl[\theta(1-z)\frac{1 + z^2}{1 - z}\biggr]_+\! f_q\Bigl(\frac{x}{z}, \mu\Bigr)
+ T_F \bigl[z^2 + (1-z)^2\bigr] f_g\Bigl(\frac{x}{z}, \mu\Bigr) \biggr\}
 + \dotsb \biggr\}
\,.\end{align}
\end{widetext}
%%%
The ellipses denote $x$-dependent terms that have no $\ln(t_\max/\mu^2)$.
Equation~\eqref{eq:tB_log} shows that the natural scale for the beam function is
$\mu=\mu_B \sim t_\max$. The logarithms of $t_\max/\mu^2$ are summed by solving
the beam function's RGE in \eq{Bq_RGE}.  From \eq{tB_log} we can see how the
matching coefficients $\cI_{ij}$ convert the PDF running into the beam function
running at one loop.  Expanding \eq{Bq_RGE} to $\ord{\alpha_s}$, the integrated
beam function satisfies
%%%
\begin{align} \label{eq:tB_RGE}
&\mu \frac{\df}{\df\mu} \tB_q(t,x, \mu)
\nn\\ &\qquad
= \frac{\alpha_s(\mu)\,C_F}{\pi}\, \Bigl(\frac{3}{2} - 2\ln\frac{t}{\mu^2}
\Bigr) \tB^\tree_q(t, x, \mu) + \dotsb
\,.\end{align}
%%%
Taking the derivative of \eq{tB_log} with respect to $\mu$, the first term in
curly brackets proportional to $f_q(x, \mu)$ reproduces the overall factor in
\eq{tB_RGE}, while the terms in the second line precisely cancel the $\mu$
dependence of the tree-level term $f_q(x, \mu)$. Thus, even though at tree level
$\tB_q(t, x, \mu) = \delta(t)\,f_q(x, \mu)$, the running of $\tB_q$ does not depend on $x$.

\begin{figure}[b!]
\includegraphics[width=\columnwidth]{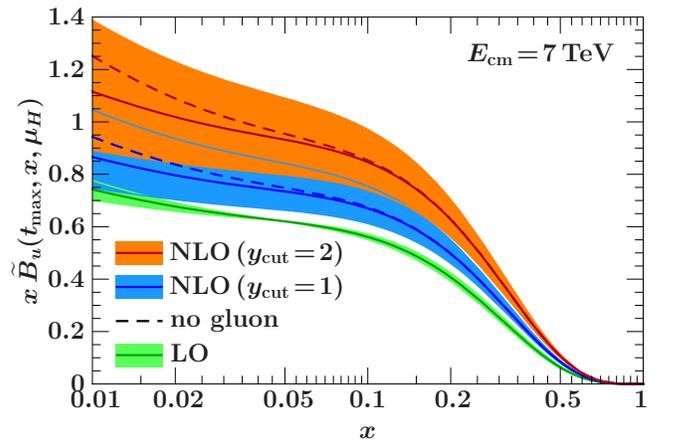}%
\caption{The $u$-quark PDF (lower green band) and $u$-quark beam functions for $y_\cut = 1$ (middle blue band) and $y_\cut = 2$ upper orange band. All functions are evaluated at the same hard scale, with the bands showing the scale variation by factors of two.}
\label{fig:Bu_muH}
\end{figure}

\begin{figure*}[t!]
\includegraphics[width=\columnwidth]{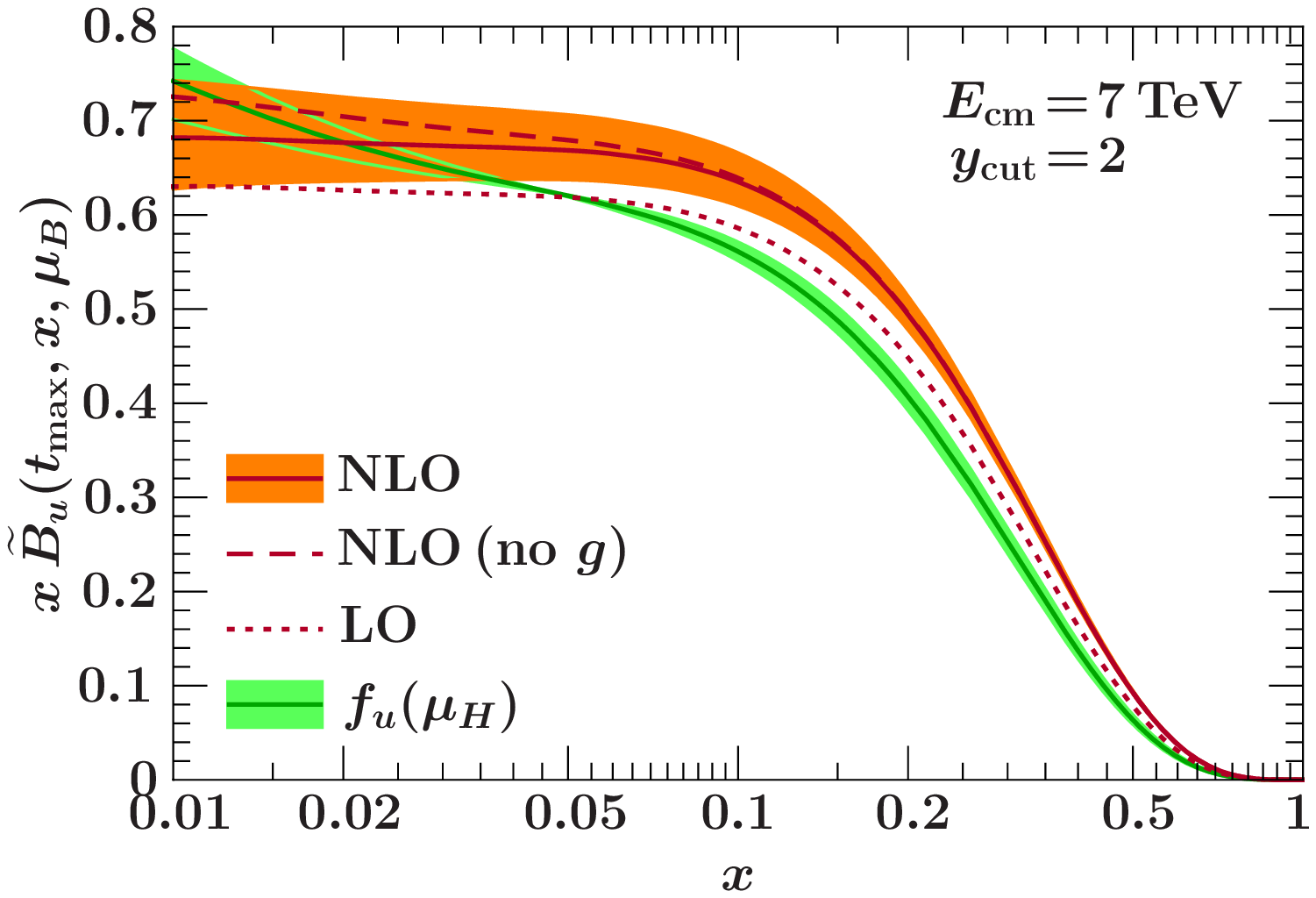}%
\hspace{\columnsep}%
\includegraphics[width=\columnwidth]{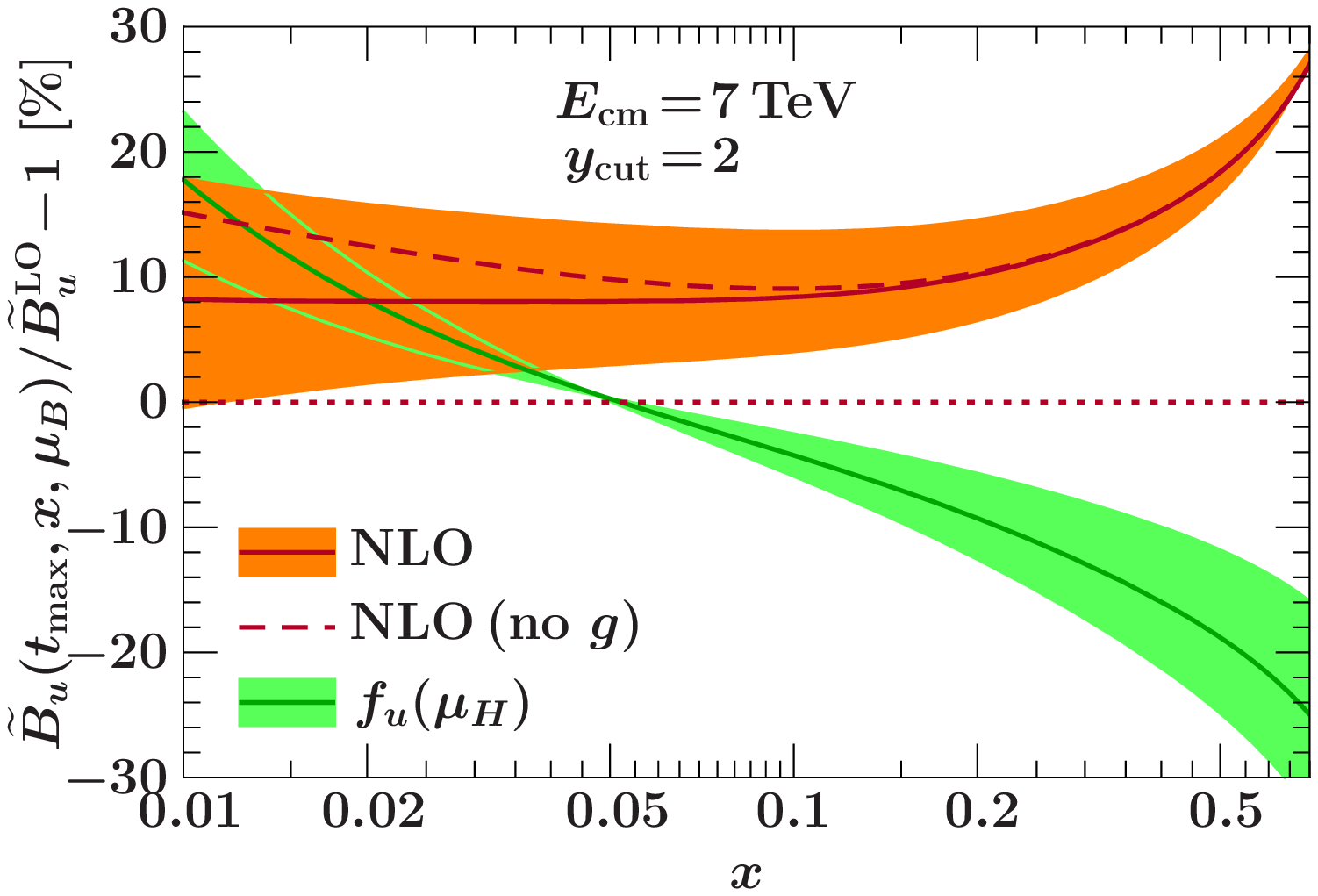}
\includegraphics[width=\columnwidth]{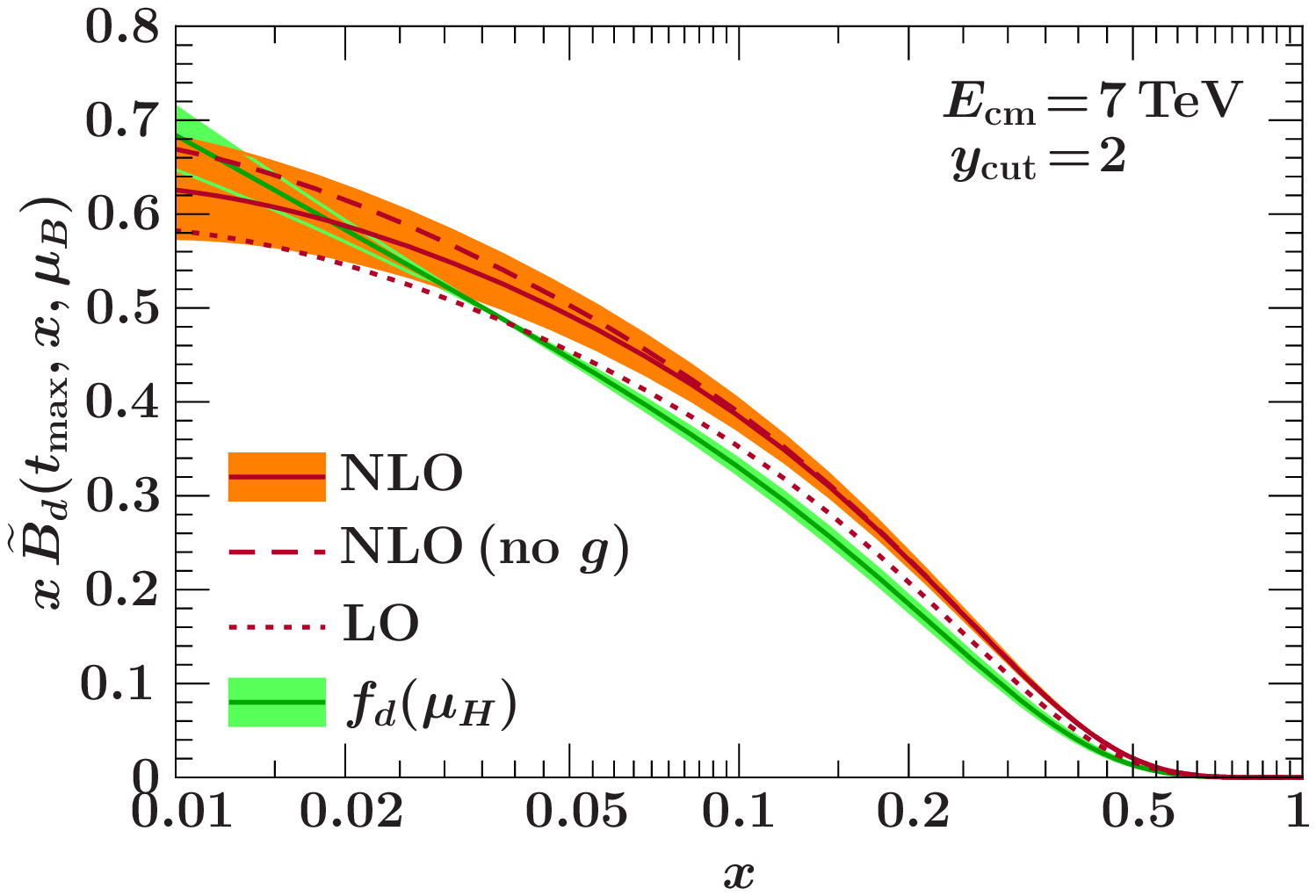}%
\hspace{\columnsep}%
\includegraphics[width=\columnwidth]{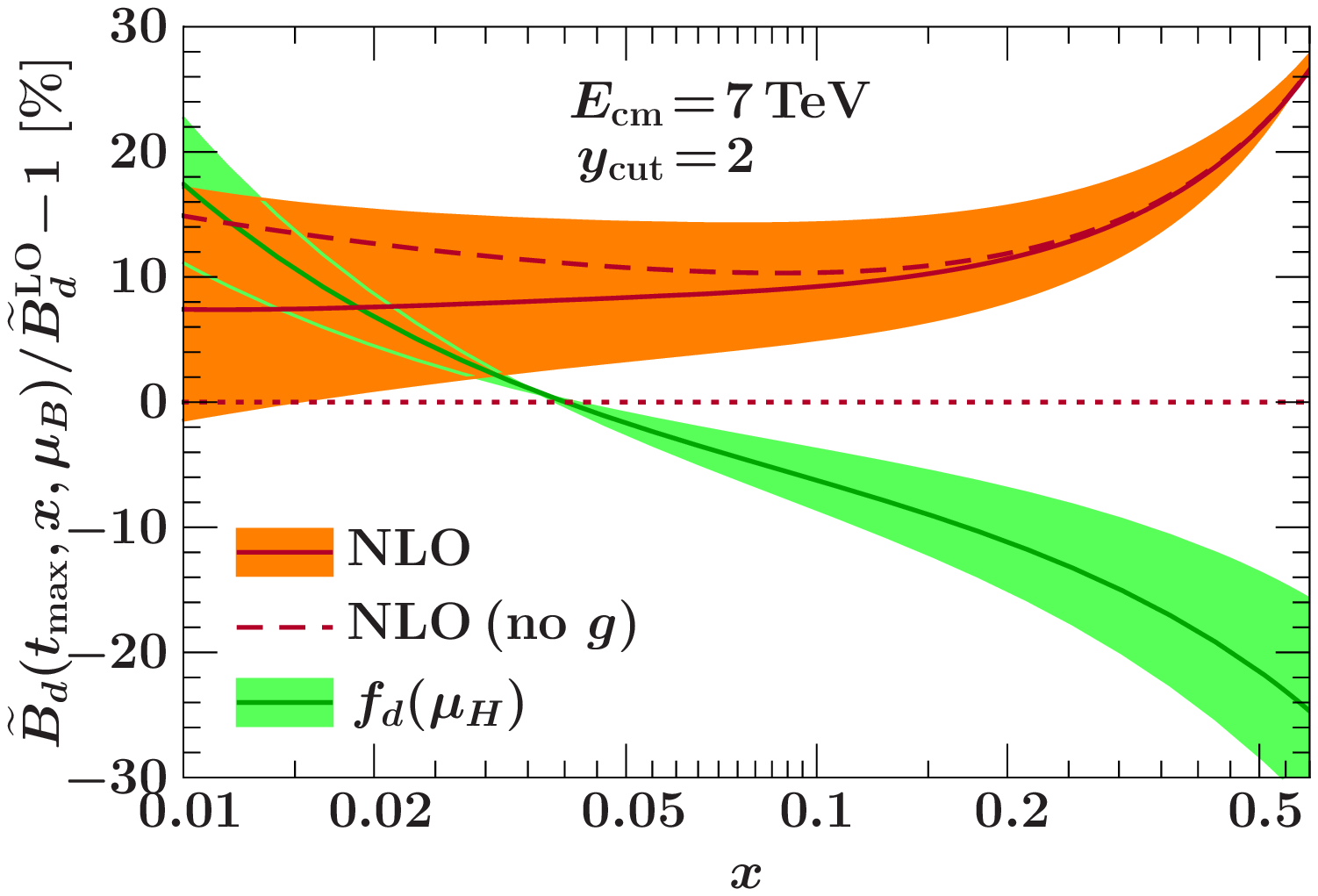}
\includegraphics[width=\columnwidth]{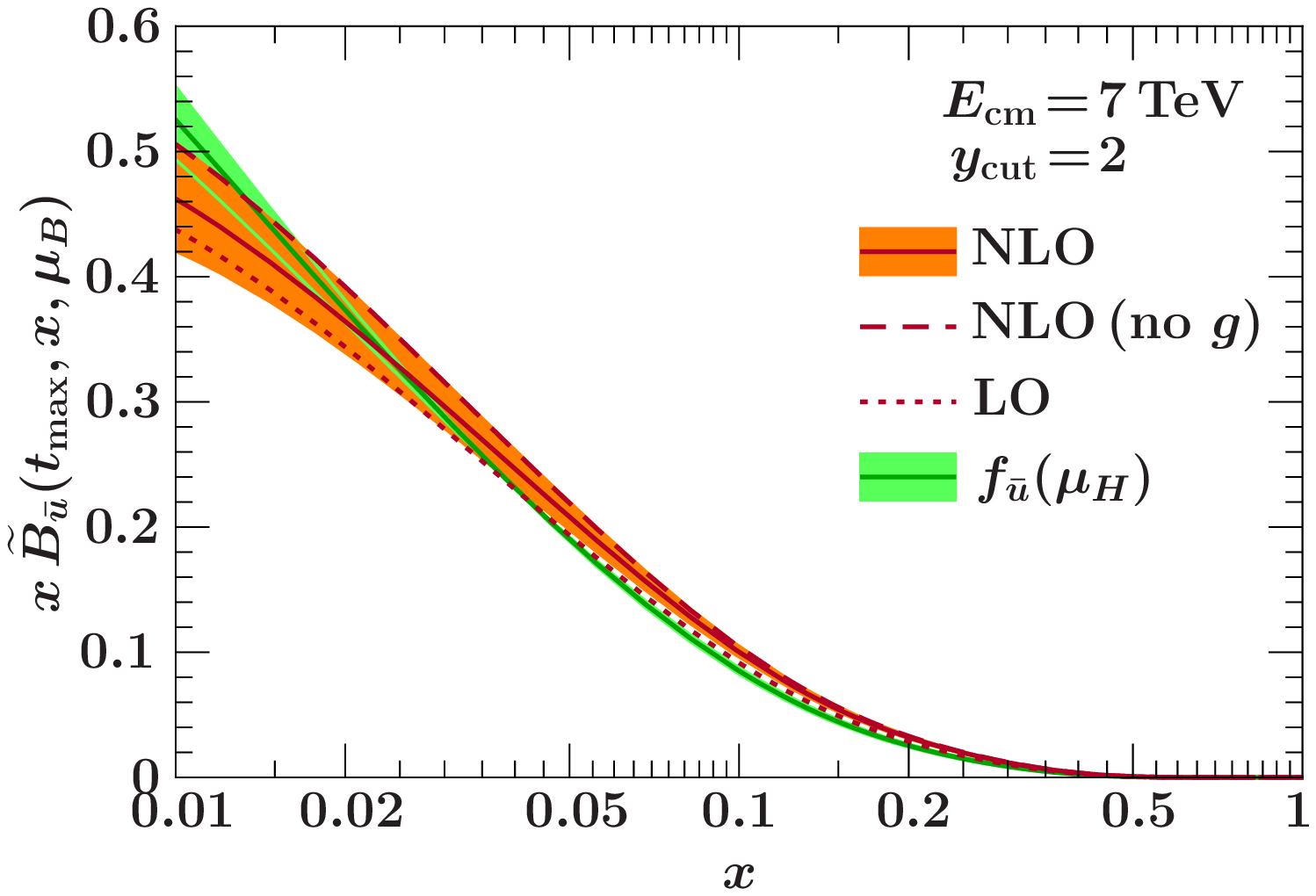}%
\hspace{\columnsep}%
\includegraphics[width=\columnwidth]{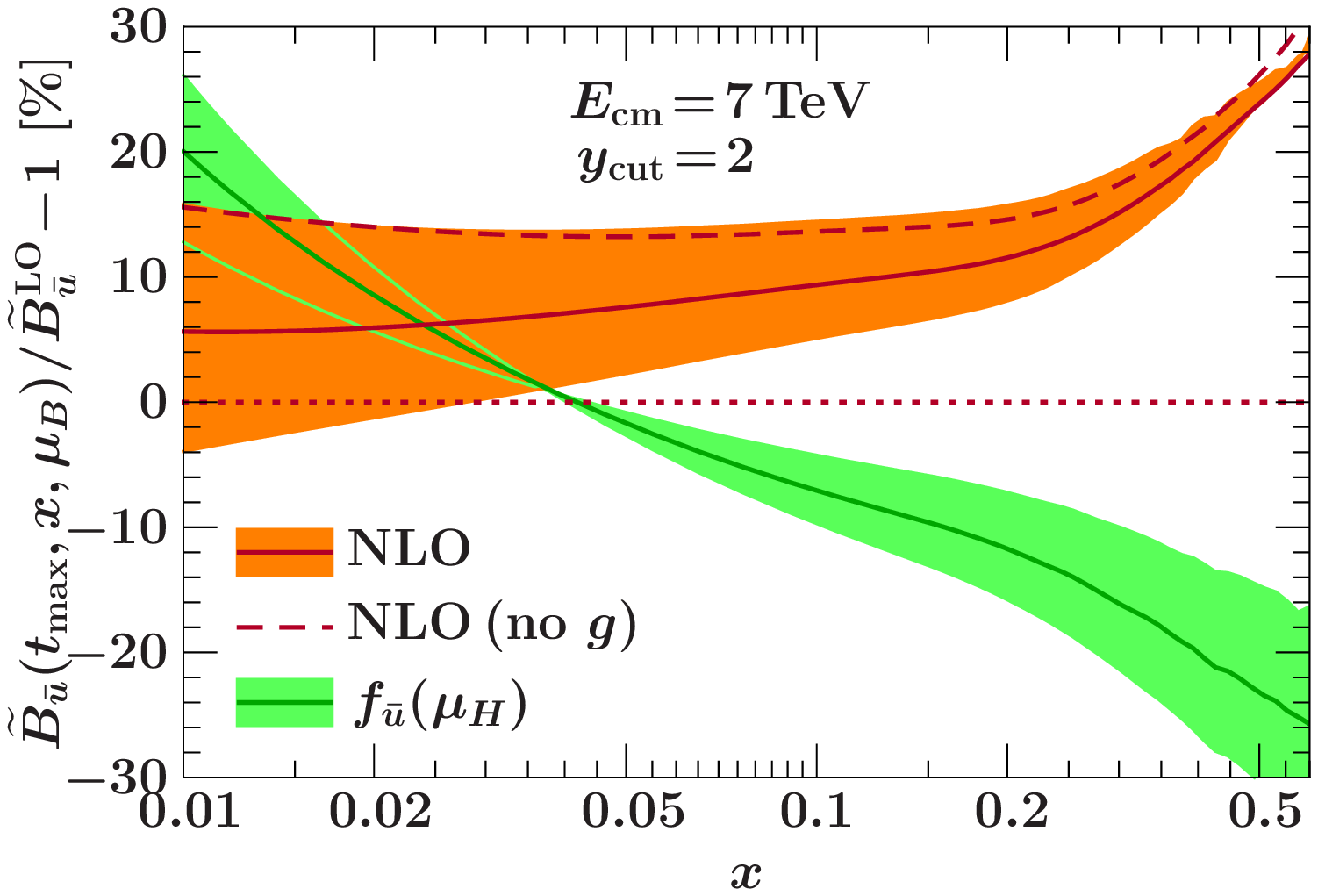}
\caption{The $u$ (top row), $d$ (middle row), and $\bar{u}$ (bottom row) beam functions at the beam scale $\mu_B^2 \simeq t_\max$ for $\Ecm = 7\TeV$ and $t_\max$ with $y_\cut = 2$ [see \eq{smaxchoice}]. The panels on the left show the functions times $x$. The right panels show the relative differences compared to the respective tree-level (LO) beam functions given by the dotted lines on the left. The bands show the scale uncertainties as explained in the text.}
\label{fig:Bq_muB}
\end{figure*}

To illustrate the difference between the beam functions and the PDFs, we may
compare the PDFs and $\tB_q(t_\max, x, \mu)$ for fixed $t_\max$ as a function of
$x$.  For $t_\max$, following the discussion in \subsec{DYfact},  we take
%%%
\begin{equation} \label{eq:smaxchoice}
t_\max = \w B^+_\max = Q^2 e^{-2y_\cut} = (x\Ecm)^2 e^{-2y_\cut}
\,,\end{equation}
%%%
where the hard scale is taken as $Q^2 = (x\Ecm)^2$. By default we use
$\Ecm = 7\TeV$ and $y_\cut = 2$. For the parton distributions we use the NLO
results of MSTW2008~\cite{Martin:2009iq}.

The effect of the large logarithms contained in the beam function is illustrated
in \fig{Bu_muH}, where we show the $u$-quark beam function $x \tB_u(t_\max, x,
\mu_H)$ at NLO for two different values of $t_\max$, along with the $u$-quark
PDF $x f_u(x, \mu_H)$, which is equal to the beam function at tree level.  All
solid central curves are evaluated at the common hard scale $\mu = Q = x\Ecm$, and
the surrounding bands correspond to varying the scale by a factor of $2$.  The
lower green curve and band show the PDF (tree-level beam function).  Including
the one-loop matching corrections the beam functions are given for $y_\cut = 1$
by the middle blue band and for $y_\cut = 2$ by the upper orange band.  Clearly,
with this scale choice, the beam functions receive large $\alpha_s$ corrections
with a dramatically increased scale dependence compared to the PDF, which is
caused by the large logarithms of $t_\max/\mu^2$ in \eq{tB_log}.

To eliminate the large logarithms in the matching, we have to compute the beam
functions at the scale $\mu_B^2 \simeq t_\max$, which we show in \fig{Bq_muB}.
Lets first consider the up-quark beam function $\tB_u$ shown in the top row.
The top left panel shows $x \tB_u(t_\max, x, \mu_B)$. The green (light) band
shows the PDF $x f_u(x, \mu_H)$ evaluated at the hard scale $\mu_H = x\Ecm$,
varying the scale by a factor of two (which for the top left panel is identical
to the corresponding band in \fig{Bu_muH}). The one-loop beam function evaluated
at $\mu_B^2 \simeq t_\max$ is shown by the orange (medium) bands. Here, the
maximum scale variation for $\mu_B^2\in [t_\max/2,2 t_\max]$ is not obtained at
the edges of this region, but is closely approximated by taking $\mu_B
=\{0.7,2.0 \}\,\sqrt{t_\max}$.  The solid line shows the corresponding central
value at $\mu_B = 1.4\, \sqrt{t_\max}$. For comparison, the dotted line shows
the tree-level result, i.e. the PDF at the scale $\mu_B$.  The top right panel
shows the same curves as the top left panel, but normalized to the tree-level
beam function. The plot shows that the beam function can be reliably calculated
at $\mu=\mu_B$.  Using $\mu_B^2 \simeq t_\max$, the shift from the dotted to
solid line is now of reasonable size, and the scale uncertainties are now
similar in size to those of the PDFs.  We also see that evaluating the PDF at
the beam scale rather than the hard scale has a significant effect. The
difference between evaluating the PDF at $\mu_H$ and $\mu_B$ is a $+30\%$
($-20\%$) correction at large (small) $x$, while the $\alpha_s$ corrections from
the beam function at $\mu_B$ are only $\sim 10\%$. (The beam function curves
increase for $x \gtrsim 0.4$ due to the threshold term $[\ln(1-z)/(1-z)]_+$ in
$\cI_{qq}$, but this region only has a small total contribution as can be seen
on the left panel.)  Since the residual $\mu_B$ dependence is only canceled by
other contributions in the factorization theorem, these plots do not determine
the overall size of the $\alpha_s$ corrections, nor their uncertainty. These
questions are addressed by plots of the full cross section in \sec{conclusions}.

In the central and bottom panels of \fig{Bq_muB} we show analogous plots for the
down-quark beam function, $\tB_d$, and the antiup-quark, $\tB_{\bar{u}}$. While
the absolute size of the functions for different flavors in the left panels are
quite different, the relative corrections shown in the right panel are very similar.

\fig{Bq_muB} also contains dashed lines, which show how the solid lines are
modified if we remove the gluon contribution $\cI_{qg}$ from the one-loop
beam function.  The gluon contribution to the quark beam functions
becomes noticeable at $x < 0.1$,
increasing to about $-5\%$ at $x = 0.01$, while for the antiquark beam function
it is important in the entire $x$ range. This is expected, since the antiquark
PDF is much smaller, so the gluon PDF can have a bigger impact. The gluon
contribution is always negative and partially compensates the quark matching
correction. Even at $x = 0.01$ there is no indication that treating the
logarithms $\ln x$ in fixed-order perturbation theory causes any problems. The
contribution from the $\ln z$ term in \eq{Iqq_oneloop} is of similar size as
other contributions and within the perturbative uncertainties. It has roughly
the same size as the gluon contribution.

%%%%%%%%%%%%%%%%%%%%%%%%%%%%%%%%%%%%%%%%%%%%%%%%%%%%%%%%%%%%%%%%%%%%%%%%%%%%%%%%
\section{Isolated Factorization Theorem}
\label{sec:factorization}
%%%%%%%%%%%%%%%%%%%%%%%%%%%%%%%%%%%%%%%%%%%%%%%%%%%%%%%%%%%%%%%%%%%%%%%%%%%%%%%%

In this section, we derive the isolated factorization theorem in \eq{DYbeam}.
Our analysis is based on factorization in SCET, which rigorously and
systematically separates hard, soft and collinear
contributions~\cite{Bauer:2001ct,Bauer:2001yt,Bauer:2002nz}.  We make use of a
setup with \SCETa and \SCETb~\cite{Bauer:2002aj}, carrying out the factorization
in two stages at the scales $Q^2$ and $\omega_{a,b}B_{a,b}^+$ respectively.  We
have an \SCETa analysis to factorize initial-state jets from soft radiation.
The initial-state jets described by beam functions in \SCETa are then matched
onto initial-state PDFs with lower offshellness for the collinear particles in
\SCETb. In this section, we carry out the \SCETa computation, while the matching
onto \SCETb was discussed in \sec{beamfunction}.  Our analysis below uses
similar tools as used in the derivation of the factorization theorem for
hemisphere invariant masses for $e^+e^-\to $ dijets in
Ref.~\cite{Fleming:2007qr}, but differs significantly due to the kinematics, and
the fact that we have initial-state rather than final-state jets and a further
matching onto \SCETb.  The soft dynamics of $e^+e^-\to $ dijets was studied
earlier in SCET in Refs.~\cite{Bauer:2002ie,Bauer:2003di}.  We start with a
brief overview of the necessary SCET ingredients in \subsec{SCET} and describe
the relevant kinematics in \subsec{kinematics}.  We derive the factorization
theorem for isolated $pp\to XL$ in \subsec{facttheorem}, including arguments to
rule out contributions from so-called Glauber degrees of freedom. Finally in
\subsec{DY_final}, we apply the factorization theorem to $pp\to X\ell^+\ell^-$
and quote final results for the beam thrust cross section with one-loop
corrections and logarithmic resummation.

%===============================================================================
\subsection{SCET}
\label{subsec:SCET}
%===============================================================================

Soft-collinear effective theory is an effective field theory of QCD that
describes the interactions of collinear and soft particles~\cite{Bauer:2000ew,
  Bauer:2000yr, Bauer:2001ct, Bauer:2001yt}.  Collinear particles are
characterized by having large energy and small invariant mass. To separate the
large and small momentum components, it is convenient to use light-cone
coordinates. We define two light-cone vectors
%%%
\begin{equation}
n^\mu = (1, \vec{n})
\,,\qquad
\bn^\mu = (1, -\vec{n})
\,,\end{equation}
%%%
with $n^2 = \bn^2 = 0$, $n\sdt\bn = 2$, and $\vec{n}$ is a unit three-vector.
Any four-momentum $p$ can then be decomposed as
%%%
\begin{equation}
p^\mu = \bn\sdt p\,\frac{n^\mu}{2} + n\sdt p\,\frac{\bn^\mu}{2} + p^\mu_{n\perp}
\,.\end{equation}
%%%
Choosing $\vec{n}$ close to the direction of a collinear particle, its momentum
$p$ scales as $(n\sdt p, \bn\sdt p, p_{n\perp}) \sim \bn\sdt p\,(\la^2,1,\la)$,
with $\la \ll 1$ a small parameter. For example, for a jet of collinear
particles in the $\vec{n}$ direction with total momentum $p_X$, $\bn \sdt p_X
\simeq 2E_X$ corresponds to the large energy of the jet, while $n \sdt p_X
\simeq p_X^2/E_X \ll E_X$, so $\la^2 \simeq p_X^2/E_X^2 \ll 1$.

To construct the fields of the effective theory, the momentum is written as
%%%
\begin{equation}
p^\mu = \lp^\mu + k^\mu = \bn\sdt\lp\, \frac{n^\mu}{2} + \lp_{n\perp}^\mu + k^\mu
\,\end{equation}
%%%
where $\bn\cdot\lp \sim Q$ and $\lp_{n\perp} \sim \la Q$ are the large momentum
components, where $Q$ is the scale of the hard interaction, while $k\sim \la^2
Q$ is a small residual momentum. The effective theory expansion is in powers of
the small parameter $\la$.

The SCET fields for $n$-collinear quarks and gluons, $\xi_{n,\lp}(x)$ and
$A_{n,\lp}(x)$, are labeled by the collinear direction $n$ and their large
momentum $\lp$. They are written in position space with respect to the residual
momentum and in momentum space with respect to the large momentum components.
Frequently, we will only keep the label $n$ denoting the collinear direction,
while the momentum labels are summed over and suppressed. Derivatives acting on
the fields pick out the residual momentum dependence, $\img\partial^\mu \sim k
\sim \la^2 Q$. The large label momentum is obtained from the momentum operator
$\cP_n^\mu$, e.g. $\cP_n^\mu\, \xi_{n,\lp} = \lp^\mu\, \xi_{n,\lp}$. If there
are several fields, $\cP_n$ returns the sum of the label momenta of all
$n$-collinear fields. For convenience, we define $\bnP_n = \bn\sdt\cP_n$, which
picks out the large minus component.

Collinear operators are constructed out of products of fields and Wilson lines
that are invariant under collinear gauge
transformations~\cite{Bauer:2000yr,Bauer:2001ct}.  The smallest building blocks
are collinearly gauge-invariant quark and gluon fields, defined as
%%%
\begin{align} \label{eq:chiB}
\chi_{n,\w}(x) &= \Bigl[\delta(\w - \bnP_n)\, W_n^\dagger(x)\, \xi_n(x) \Bigr]
\,,\nn\\
\cB_{n,\w\perp}^\mu(x)
&= \frac{1}{g}\Bigl[\delta(\w + \bnP_n)\, W_n^\dagger(x)\,\img D_{n\perp}^\mu W_n(x)\Bigr]
\,,\end{align}
%%%
where
%%%
\begin{equation}
\img D_{n\perp}^\mu = \cP^\mu_{n\perp} + g A^\mu_{n\perp}
\end{equation}
%%%
is the collinear covariant derivative and
%%%
\begin{equation} \label{eq:Wn}
W_n(x) = \biggl[\sum_\text{perms} \exp\Bigl(\frac{-g}{\bnP_n}\,\bn\sdt A_n(x)\Bigr)\biggr]
\,.\end{equation}
%%%
The label operators in \eqs{chiB}{Wn} only act inside the square brackets. Here,
$W_n(x)$ is a Wilson line of $n$-collinear gluons in label momentum space. It
sums up arbitrary emissions of $n$-collinear gluons from an $n$-collinear quark
or gluon, which are $\ord{1}$ in the power counting. Since $W_n(x)$ is
localized with respect to the residual position $x$, we can treat
$\chi_{n,\w}(x)$ and $\cB_{n,\w}^\mu(x)$ as local quark and gluon fields. The
label momentum $\w$ is treated as a continuous variable, which is why we
use a $\delta$-function operator in \eq{chiB}. It is set equal to the sum of the
minus label momenta of all fields that the $\delta$ function acts on, including
those in the Wilson lines, while the label momenta of the individual fields are
summed over.

In general, the effective theory can contain several collinear sectors, each
containing collinear fields along a different collinear direction. To have a
well-defined power expansion in this case, the different collinear directions
$n_i$ have to be well separated~\cite{Bauer:2002nz},
%%%
\begin{equation} \label{eq:nijsep}
  n_i\sdt n_j \gg \la^2 \qquad\text{for}\qquad i\neq j
\,,\end{equation}
%%%
which is simply the requirement that different collinear sectors are distinct
and do not overlap. For $pp\to X\ell^+\ell^-$, we need two collinear sectors,
$n_a$ and $n_b$, along the directions of the two beams. We use a bar to denote
the conjugate lightlike vector, so $n_i\cdot\bar n_i=2$. As the beams are
back-to-back, we have $n_a \sim \bn_b$, so $n_a\sdt n_b \sim 2$ and \eq{nijsep}
is easily satisfied.

Particles that exchange large momentum of $\ord{Q}$ between collinear particles
moving in different directions have to be off shell by an amount of
$\ord{n_i\cdot n_j Q^2}$. These modes can be integrated out of the theory at the
hard scale $Q$ by matching full QCD onto SCET, which yields the hard function.
The effective theory below the scale $Q$ then splits into several distinct
collinear sectors, where particles in the same collinear sector can still
interact with each other, while at leading order in the power counting particles
from different collinear sectors can only interact by the exchange of soft
particles. This means that before and after the hard interaction takes place,
the jets described by the different collinear sectors evolve independently from
each other with only soft but no hard interactions between them.

The soft degrees of freedom, responsible for the  radiation between collinear
jets, are described in the effective theory by soft%
\footnote{In some situations it is necessary to distinguish two types of soft
  sectors, referred to as soft and ultrasoft in the SCET literature. In this
  paper we only need what are usually called ultrasoft particles, so we will
  simply refer to these as soft.}  quark and gluon fields, $q_\soft(x)$ and
$A_\soft(x)$, which only have residual soft momentum dependence
$\img\partial^\mu \sim \la^2Q$.
They couple to the collinear sectors via the soft covariant derivative
%%%
\begin{equation}
\img D_\soft^\mu = \img \partial^\mu + g A_\soft^\mu
\end{equation}
%%%
acting on the collinear fields. At leading order in $\la$, $n$-collinear particles only couple to the $n\sdt A_\soft$ component of soft gluons, so the leading-order $n$-collinear Lagrangian only depends on $n\sdt D_\soft$. For $n$-collinear quarks~\cite{Bauer:2000yr, Bauer:2001ct}
%%%
\begin{equation}
\cL_n = \bar{\xi}_n \Bigl(\img n\sdt D_\soft + g\,n\sdt A_n + \img\Dslash_{n\perp} W_n \frac{1}{\bnP_n}\, W_n^\dagger\,\img\Dslash_{n\perp} \Bigr)\frac{\bnslash}{2} \xi_n
\,.\end{equation}
%%%
The leading-order $n$-collinear Lagrangian for gluons is given in Ref.~\cite{Bauer:2001yt}.

The coupling of soft gluons to collinear particles can be removed at leading order by defining new collinear fields~\cite{Bauer:2001yt}
%%%
\begin{align} \label{eq:BPS}
\chi^\zero_{n,\w}(x) &= Y_n^\dagger(x)\,\chi_{n,\w}(x)
\,,\\\nn
\cB^{\mu\zero}_{n,\w\perp}(x) &= Y_n^\dagger(x)\,\cB^\mu_{n,\w\perp}(x)\,Y_n(x)
= \cB^{\mu d}_{n,\w\perp }(x) \cY_n^{dc}(x)\,T^c
,\end{align}
%%%
where $Y_n(x)$ and $\cY_n(x)$ are soft Wilson lines in the fundamental and adjoint representations,
%%%
\begin{align} \label{eq:Yin}
Y_n(x) &= P\exp\biggl[\img g\intlim{-\infty}{0}{s} n\sdt A_\soft(x + s\,n) \biggr]
\,,\nn\\
T^c \cY^{cd}_n(x) &= Y_n(x)\, T^d\, Y_n^\dagger(x)
\,.\end{align}
%%%
The symbol $P$ in \eq{Yin} denotes the path ordering of the color generators
along the integration path. The integral limits in \eq{Yin} with the reference
point at $-\infty$ are the natural choice for incoming
particles~\cite{Chay:2004zn}. The final results are always independent of the
choice of reference point, and with the above choice the interpolating fields
for the incoming proton states do not introduce additional Wilson
lines~\cite{Arnesen:2005nk}.

\begin{figure*}[t!]
\includegraphics[scale=0.75]{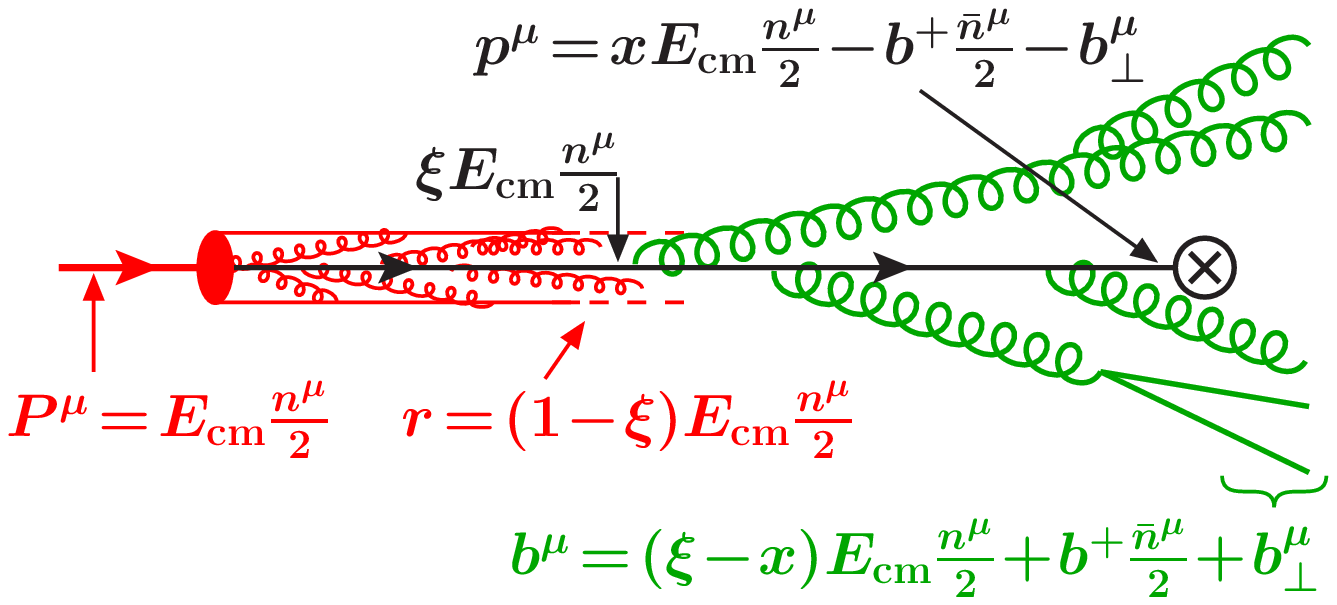}
\caption{Definition of the different collinear momenta related to the incoming beams. The soft radiation
is not shown.}
\label{fig:beam_kinematics}
\end{figure*}

After the field redefinition in \eq{BPS}, the leading-order SCET Lagrangian
separates into the sum of independent $n_i$-collinear and soft Lagrangians,
%%%
\begin{equation} \label{eq:LSCET}
\cL_\mathrm{SCET} = \sum_{n_i} \cL^\zero_{n_i} + \cL_\soft + \dotsb
\,,\end{equation}
%%%
with no interactions between any of the collinear and soft sectors. The
ellipses denote terms that are subleading in the power counting. This decoupling
is what will allow us to factorize the cross section into separate beam and soft
functions. The field redefinition in \eq{BPS} introduces
soft Wilson lines in the operators, which because of \eq{LSCET} can be factored out
of the matrix element and will make up the soft function.

%===============================================================================
\subsection{Kinematics}
\label{subsec:kinematics}
%===============================================================================

Before deriving the factorization theorem, we discuss the relevant kinematics,
as illustrated in \fig{beam_kinematics}. As already mentioned, we introduce a
separate set of collinear fields for each of the beams, with the light-cone
vectors $n_a$ and $n_b$ aligned with the beam directions. To derive the
factorization theorem we work in the center-of-mass frame of the hadronic
collision, so the momenta of the incoming protons are (neglecting the proton
mass)
%%%
\begin{equation} \label{eq:nab_choice}
P_a^\mu = \Ecm\, \frac{n_a^\mu}{2}
\,,\qquad
P_b^\mu = \Ecm\, \frac{n_b^\mu}{2}
\,,\end{equation}
%%%
with $\vec{n}_a = - \vec{n}_b$. In particular, $n_b = \bn_a$ and $n_a \sdt n_b = 2$. We will mostly keep the dependence on the two beam directions explicit, but one should keep in mind that $n_a$ and $n_b$ are related.

The collinear fields in the $n_a$ and $n_b$ directions describe the interactions within each of the beams before and after the collision, and are also responsible for initiating the hard interaction. We define the momenta of the spacelike off-shell partons that go into the hard interaction as
%%%
\begin{align} \label{eq:bain}
p_a^\mu &= x_a \Ecm\, \frac{n_a^\mu}{2} - b_a^+\,\frac{\bn_a^\mu}{2} - b_{a\perp}^\mu
\,,\nn\\
p_b^\mu &= x_b \Ecm\, \frac{n_b^\mu}{2} - b_b^+\,\frac{\bn_b^\mu}{2} - b_{b\perp}^\mu
\,,\end{align}
%%%
where $x_a$ and $x_b$ are the light-cone momentum fractions at which the beam functions will be evaluated.
The power-counting parameters for the collinear sectors are
%%%
\begin{equation} \label{eq:laJab}
\la_a^2
\sim \frac{b_a^+}{x_a\Ecm}
\,,\qquad
\la_b^2
\sim \frac{b_b^+}{x_b\Ecm}
\,,\end{equation}
%%%
where the relevant momenta are those of the off-shell partons in \eq{bain}, because these are the momenta carried by the $n_a$- and $n_b$-collinear fields.

We write the momentum of the incoming partons that are taken out of the proton as
%%%
\begin{equation}
\xi_a \Ecm\, \frac{n_a^\mu}{2} + \ord{\lqcd}
\,,\qquad
\xi_b \Ecm\, \frac{n_b^\mu}{2} + \ord{\lqcd}
\,,\end{equation}
%%%
which defines the light-cone momentum fractions $\xi_{a,b}$ at which the PDFs
are evaluated.  The typical $\perp$-momenta of partons in the proton are
$\ord{\lqcd}$, while the small plus components are $\ord{\lqcd^2/\Ecm}$. These
momenta are much smaller than any soft or residual momenta in \SCETa and are
expanded, which precisely corresponds to the OPE for the beam functions in
\eq{B_fact} when matching them onto \SCETb.

The momentum of the final-state remnant of the proton is thus given by
%%%
\begin{equation} \label{eq:remnant}
r_a^\mu = (1-\xi_a)\Ecm\, \frac{n_a^\mu}{2}
\,,\end{equation}
%%%
while the remnant of the initial-state jet radiated into the final state by the beam function has momentum
%%%
\begin{equation} \label{eq:initialjet}
b_a^\mu = (\xi_a - x_a) \Ecm\, \frac{n_a^\mu}{2} + b_a^+\,\frac{\bn_a^\mu}{2} + b_{a\perp}^\mu
\,,\end{equation}
%%%
and similarly for the $n_b$ direction. The total $n_a$-collinear momentum in the final state is the sum of \eqs{remnant}{initialjet}, or equivalently, the difference between the proton momentum and \eq{bain},
%%%
\begin{equation}
b_a^\mu + r_a^\mu = P_a^\mu - p_a^\mu = (1 - x_a) \Ecm\, \frac{n_a^\mu}{2} + b_a^+\,\frac{\bn_a^\mu}{2} + b_{a\perp}^\mu
\,.\end{equation}
%%%

In addition to the collinear momenta, we define $k_s^\mu$ as the total four-momentum of the soft radiation in the final state. Hence, the total hadronic momentum in the final state is given by
%%%
\begin{equation}
p_X^\mu = (P_a^\mu - p_a^\mu) + (P_b^\mu - p_b^\mu) + k_s^\mu
\,,\end{equation}
%%%
and we can write total momentum conservation $P_a^\mu + P_b^\mu = p_X^\mu + q^\mu$ as
%%%
\begin{equation} \label{eq:momcons}
p_a^\mu + p_b^\mu = q^\mu + k_s^\mu
\,,\end{equation}
%%%
where $q^\mu$ is the total leptonic momentum.

The collinear and soft momenta, $b_a^\mu$, $b_b^\mu$, $k_s^\mu$ are not
experimentally measurable quantities. Instead, the experiments can only measure
hadronic quantities, such as the hemisphere momenta $B_a^+ =n_a\cdot B_a$ and
$B_b^+ = n_b\cdot B_b$ introduced in \subsec{DYfact}. Splitting the total soft
momentum into its contributions from each hemisphere, $k_s^\mu = k_a^\mu +
k_b^\mu$ as shown in \fig{DYkin}, we then have
%%%
\begin{equation}
B_a^\mu = b_a^\mu + r_a^\mu + k_a^\mu
\,,\qquad
B_b^\mu = b_b^\mu + r_b^\mu + k_b^\mu
\,,\end{equation}
%%%
and defining $k_a^+ = n_a\sdt k_a$, $k_b^+ = n_b\sdt k_b$, we get
%%%
\begin{equation}
B_a^+ = n_a\sdt B_a = b_a^+ + k_a^+
\,,\qquad
B_b^+ = n_b\sdt B_b = b_b^+ + k_b^+
\,.\end{equation}
%%%
In particular, the remnant momenta $r_{a,b}^\mu$ do not contribute to
$B_{a,b}^+$. A physical argument for this was discussed in \subsec{DYfact}.

Next, we decompose the total leptonic momentum as
%%%
\begin{equation} \label{eq:qpmperp}
q^\mu = q^-\,\frac{n_a^\mu}{2} + q^+\,\frac{n_b^\mu}{2} + q_\perp^\mu
\,,\end{equation}
%%%
where $q_\perp^\mu$ contains the two components of $q^\mu$ transverse to the
beam direction. Taking the $z$-axis along the $\vec{n}_a$ beam direction, we
have
%%%
\begin{equation}
q^\pm = q^0 \mp q_z
\,,\qquad
q_\perp^\mu = (0, \vec{q}_T, 0)
\,,\end{equation}
%%%
where $\vec{q}_T = (q_x, q_y)$ is a two-vector in the transverse $x$-$y$-plane. The total leptonic invariant mass and rapidity are
%%%
\begin{equation} \label{eq:q2Y_def}
q^2 = q^+ q^- + q_\perp^2 = q^+ q^- - \vec{q}_T^2
\,,\qquad
Y = \frac{1}{2} \ln\frac{q^-}{q^+}
\,,\end{equation}
%%%
with
%%%
\begin{align}
q^\mp &= e^{\pm Y}\,\sqrt{q^2 + \vec{q}_T^2}
\,,\nn\\
\df^4 q &= \frac{1}{2}\, \df q^+\df q^-\, \df^2 \vec{q}_T = \frac{1}{2}\, \df q^2\,\df Y\,\df^2 \vec{q}_T
\,.\end{align}
%%%

As we will see in the next subsection, the derivation of the factorization
theorem requires us to be insensitive to the transverse components $\vec{q}_T$
such that we can freely integrate over them. Therefore, we have to expand the
kinematics in the limit $\vec{q}_T = 0$. This expansion is justified because
from \eq{momcons} we have
%%%
\begin{equation} \label{eq:qperpsmall}
q_\perp^\mu = - p_{X\perp}^\mu = -b_{a\perp}^\mu -b_{b\perp}^\mu - k_{\soft\perp}^\mu \sim \lambda Q
\,.\end{equation}
%%%
A parametrically large $q_\perp^\mu \sim Q$ would require a separate jet at large $p_T \sim Q$ to balance the transverse momentum, which is not allowed in our setup. After integrating over the lepton phase space this expansion incurs power corrections of order $q_\perp^2 \sim \lambda^2Q^2$. The kinematics of the hard matrix element in the factorization theorem is then given by the tree-level partonic kinematics, with the partonic momentum conservation
%%%
\begin{equation} \label{eq:partonmomcons}
x_a \Ecm\,\frac{n_a^\mu}{2} + x_b \Ecm\,\frac{n_b^\mu}{2} = q = q^- \frac{n_a^\mu}{2} + q^+ \frac{n_b^\mu}{2}
\,,\end{equation}
%%%
which implies
%%%
\begin{align} \label{eq:xab}
x_a \Ecm &= q^- = \sqrt{q^2}\, e^Y
\,,\nn\\
x_b \Ecm &= q^+ = \sqrt{q^2}\, e^{-Y}
\,,\nn\\
q^2 &= q^+ q^- = x_a x_b \Ecm^2
\,,\nn\\
Y &= \frac{1}{2} \ln\frac{q^-}{q^+} = \frac{1}{2} \ln\frac{x_a}{x_b}
\,.\end{align}
%%%
Equations~\eqref{eq:qperpsmall} and \eqref{eq:partonmomcons} imply that parametrically the leptons are
back-to-back in the transverse plane. Since $q^+$ and $q^-$ can differ
substantially, the leptons do not need to be back-to-back in three dimensions.

%===============================================================================
\subsection{Derivation of Isolated Factorization Theorem}
\label{subsec:facttheorem}
%===============================================================================

We now proceed to derive the isolated factorization theorem for generic processes $pp\to
XL$, where the hadronic final state $X$ has a restriction on the hemisphere
momenta $B_{a,b}^+$. The derivation is carried out using SCET
without Glauber degrees of freedom. The proof that Glauber effects
are not required is given at the  end of this subsection.

%~~~~~~~~~~~~~~~~~~~~~~~~~~~~~~~~~~~~~~~~~~~~~~~~~~~~~~~~~~~~~~~~~~~~~~~~~~~~~~~
\subsubsection{Cross Section in QCD}
%~~~~~~~~~~~~~~~~~~~~~~~~~~~~~~~~~~~~~~~~~~~~~~~~~~~~~~~~~~~~~~~~~~~~~~~~~~~~~~~

We will generically refer to properties of $L$ as ``leptonic'', even though
$L$ can contain any non-strongly interacting particles.
We only consider processes where the hard
interaction couples the strong and electroweak sectors through one two-particle
QCD current.  (This includes for example Drell-Yan or Higgs production through
gluon fusion with the Higgs decaying non-hadronically, but does not include
electroweak Higgs production via vector-boson fusion.) Then, at leading order in
the electroweak interactions, we can factorize the full-theory matrix element
into its leptonic and hadronic parts
%%%
\begin{align} \label{eq:M}
\mathcal{M}(pp\to X L) = \sum_{J} L_J\,\Mae{X}{J}{pp}
\,.\end{align}
%%%
The sum runs over all relevant color-singlet two-particle QCD currents $J$, and
$L_J$ contains the corresponding electroweak matrix element, including the
electroweak propagator coupling to $J$. For example, for Drell-Yan with $L =
\ell^+\ell^-$, the relevant currents are
%%%
\begin{equation} \label{eq:JDY_QCD}
J_{Vf}^\mu = \bar{q}_f \gamma^\mu q_f
\,,\qquad
J_{Af}^\mu = \bar{q}_f \gamma^\mu \gamma_5 q_f
\,,\end{equation}
%%%
so in this case the sum over $J$ in \eq{M} includes the sums over the two Dirac
structures, the vector index $\mu$, and the quark flavor $f = \{u, d, \ldots\}$.
The corresponding $L_{Vf}^\mu$ and $L_{Af}^\mu$ are given below in \eq{LmuDY}.

The cross section for some hadronic observable $\Obs$ in the center-of-mass frame of the collision, averaged over proton spins, is
%%%
\begin{align} \label{eq:dsigma_dO}
\frac{\df\sigma}{\df q^2\df Y\df\Obs}
&= \frac{1}{2\Ecm^2} \int\!\frac{\df^2 \vec{q}_T}{2(2\pi)^4}
\int\! \df \Phi_L\, (2\pi)^4 \delta^4(q - p_L)\,
\nn\\ & \quad\times
\frac{1}{4}\sum_\mathrm{spins}\sum_{X}\Abs{\mathcal{M}(pp\to XL)}^2 \,
\delta[\Obs - f_\Obs(X)]
\nn\\ & \quad\times
(2\pi)^4 \delta^4(P_a + P_b - q - p_X)
\,.\end{align}
%%%
Here, $P_{a,b}$ are the incoming proton momenta, $p_X$ and $p_L$ are the total
hadronic and leptonic momenta, $\df \Phi_L$ denotes the leptonic phase space,
and the phase-space integrations for the hadronic final states are included in
the sum over $X$. The last $\delta$ function is overall momentum conservation.
The function $f_\Obs(X)$ inside the second $\delta$ function returns the value
of the hadronic observable $\Obs$ for a given hadronic state $X$, so the
$\delta$ function picks out all final states that contribute to a certain value
of $\Obs$. The $\delta^4(q - p_L)$ under the leptonic phase-space integral
defines the measured $q$ as the total leptonic momentum. Expanding this $\delta$
function for $\vec{q}_T = 0$, the leptonic part does not depend on $\vec{q}_T$
at leading order, and using \eq{M}, we can rewrite \eq{dsigma_dO} as
%%%
\begin{align} \label{eq:dsigmadO_LW}
\!\!\frac{\df\sigma}{\df q^2\df Y \df\Obs}
&= \frac{1}{2\Ecm^2} \sum_{J,J'} L_{JJ'}(q^2, Y)\, W_{JJ'}(q^2, Y, \Obs)
\,.\end{align}
%%%
The leptonic tensor is defined as
%%%
\begin{equation} \label{eq:Lmunu_def}
L_{JJ'}(q^2, Y)
= \!\int\! \df \Phi_L\, L_J^\dagger L_{J'}\, (2\pi)^4 \delta^4\Bigl(q^-\frac{n_a}{2} + q^+\frac{n_b}{2} - p_L \Bigr)
,\end{equation}
%%%
where $q^\pm = \sqrt{q^2} e^{\mp Y}$. The hadronic tensor contains the square of the hadronic matrix element
%%%
\begin{align} \label{eq:Wmunu_def}
&W_{JJ'}(q^2, Y, \Obs)
\nn\\ &\qquad
= \int\!\frac{\df^2\vec{q}_T}{2(2\pi)^4}
\sum_X \Mae{pp}{J^\dagger(0)}{X}\Mae{X}{J'(0)}{pp}
\\\nn & \qquad\quad\times
(2\pi)^4 \delta^4(P_a + P_b - q - p_X)\,\delta[O - f_O(X)]
\,,\end{align}
%%%
where as in \sec{beamfunction} we keep the average over proton spins implicit in
the matrix element. Since $W_{JJ'}$ is integrated over $\vec{q}_T$, it can only
depend on $q^2$ and $Y$, as well as the hadronic observable $\Obs$.

We are interested in the hadronic observables $B_{a}^+ = n_{a}\cdot B_{a}$
and $B_{b}^+ = n_{b}\cdot B_{b}$.  The hemisphere hadronic momenta
$B_{a,b}^\mu(X)$ can be obtained from the states $\ket{X}$ using the hemisphere
momentum operators $\hp_{a,b}^\mu$
%%%
\begin{equation} \label{eq:hatpab}
\hp_a^\mu \ket{X} = B_a^\mu(X) \ket{X}
\,,\qquad
\hp_b^\mu \ket{X} = B_b^\mu(X) \ket{X}
\,.\end{equation}
%%%
A field-theoretic definition of $\hp_{a,b}^\mu$ in terms of the energy-momentum tensor of the field theory was given in Ref.~\cite{Bauer:2008jx}. The hadronic tensor for $O \equiv \{B_a^+, B_b^+\}$ is
%%%
\begin{widetext}
\begin{align} \label{eq:Wmunu_QCD}
W_{JJ'}(q^2, Y, B_a^+, B_b^+)
&= \int\!\frac{\df^2\vec{q}_T}{2(2\pi)^4} \int\!\df^4x\,e^{-\img q\cdot x}
\sum_X \Mae{pp}{J^\dagger(x)}{X}\Mae{X}{J'(0)}{pp}
\,\delta[B_a^+ - n_a\sdt B_a(X)]\,\delta[B_b^+ - n_b\sdt B_b(X)]
\nn\\
&= \int\!\frac{\df x^+\df x^-}{(4\pi)^2}\,e^{-\img (q^+ x^- + q^- x^+)/2}\,
\MAe{pp}{J^\dagger\Bigl(x^-\frac{n_a}{2} + x^+\frac{n_b}{2}\Bigr)
\delta(B_a^+ - n_a\sdt\hp_a)\,\delta(B_b^+ - n_b\sdt\hp_b) J'(0)}{pp}
\,.\end{align}
%%%
In the first line we used momentum conservation to shift the position of $J^\dagger$, and in the second line we performed the integral over $\vec{q}_T$, which sets $\vec{x}_T$ to zero. We also used \eq{hatpab} to eliminate the explicit dependence on $X$, allowing us to carry out the sum over all states $X$. The restriction on the states $X$ is now implicit through the operator $\delta$ functions inside the matrix element.

%~~~~~~~~~~~~~~~~~~~~~~~~~~~~~~~~~~~~~~~~~~~~~~~~~~~~~~~~~~~~~~~~~~~~~~~~~~~~~~~
\subsubsection{Matching QCD onto SCET}
%~~~~~~~~~~~~~~~~~~~~~~~~~~~~~~~~~~~~~~~~~~~~~~~~~~~~~~~~~~~~~~~~~~~~~~~~~~~~~~~

In the next step, we match the QCD currents $J$ onto SCET currents by integrating out fluctuations at the hard scale $Q$. At leading order in the power counting, the matching takes the form
%%%
\begin{equation} \label{eq:J_matching}
J(x) = \sum_{n_1, n_2} \int\!\df \w_1\,\df\w_2\, e^{-\img (\lb_1 + \lb_2)\cdot x}
\biggl[\sum_q C_{J q\bar{q}}^{\alpha\beta}(\lb_1, \lb_2)\, O_{q\bar{q}}^{\alpha\beta}(\lb_1, \lb_2;x)
+ C_{J gg}^{\mu\nu}(\lb_1, \lb_2)\, O_{gg\,\mu\nu}(\lb_1, \lb_2;x) \Bigr]
\,,\end{equation}
%%%
where $\alpha$, $\beta$ are spinor indices, $\mu$, $\nu$ are vector indices, and the sum over $q$ runs over all quark flavors $\{u, d, \ldots\}$. The Wilson coefficients and operators depend on the large label momenta
%%%
\begin{equation} \label{eq:labels}
\lb_1^\mu = \w_1\,\frac{n_1^\mu}{2}
\,,\qquad
\lb_2^\mu = \w_2\,\frac{n_2^\mu}{2}
\,.\end{equation}
%%%
They will eventually be set to either $q^- n_a^\mu/2$ or $q^+ n_b^\mu/2$ by
momentum conservation, but at this point are unspecified, and the sums and
integrals over $n_1$, $n_2$ and $\w_1$, $\w_2$ in \eq{J_matching} run over all
sets of distinct collinear directions and large label momenta.  On the
right-hand side of \eq{J_matching}, the full $x$ dependence of the current is
separated into the $x$ dependence appearing in the overall phase factor with
large label momenta and the residual $x$ dependence of the SCET operators.

The SCET operators $O_{q\bar{q}}^{\alpha\beta}(x)$ and $O_{gg}^{\mu\nu}(x)$ are constructed out of the collinear fields in \eq{chiB}. At leading order in the power counting they contain one field for each collinear direction. Since the QCD currents are color singlets, the leading operators that can contribute are
%%%
\begin{equation} \label{eq:Oi_SCET}
O_{q\bar{q}}^{\alpha\beta}(\lb_1,\lb_2; x)
= \bar\chi_{n_1,-\w_1}^{\alpha j}(x)\,\chi_{n_2,\w_2}^{\beta j}(x)
\,,\qquad
O_{gg}^{\mu\nu}(\lb_1,\lb_2; x)
= \sqrt{\w_1\,\w_2}\,\cB_{n_1,-\w_1\perp }^{\mu c}(x)\, \cB_{n_2,-\w_2\perp}^{\nu c}(x)
\,,\end{equation}
%%%
where $j$ and $c$ are color indices in the fundamental and adjoint representations. We included appropriate minus signs on the labels, such that we always have $\w_{1,2} > 0$ for incoming particles. Here, $\chi \equiv \chi_q$ is a quark field of flavor $q$, which for simplicity we keep implicit in our notation. Note that the entire spin and flavor structure of the current $J$ is hidden in the label $J$ on the matching coefficients in \eq{J_matching}. The gluon operator is symmetric under interchanging both $\mu\lra\nu$ and $\lb_1\lra \lb_2$, so its matching coefficient must have the same symmetry,
%%%
\begin{equation} \label{eq:Cgg_symmetry}
C_{J gg}^{\nu\mu}(\lb_2, \lb_1) = C_{J gg}^{\mu\nu}(\lb_1, \lb_2)
\,.\end{equation}
%%%
We define the conjugate quark operator and matching coefficient with the usual factors of $\gamma^0$, i.e.,
%%%
\begin{equation}
O^{\dagger \beta\alpha}_{q\bar q}(\lb_1, \lb_2)
= \bar\chi_{n_2,\w_2}^{\beta j}(x)\,\chi_{n_1,-\w_1}^{\alpha j}(x)
\,,\qquad
\bC^{\beta\alpha}_{J q\bar q}(\lb_1, \lb_2)
= [\gamma^0 C^\dagger_{Jq\bar q}(\lb_1, \lb_2) \gamma^0]^{\beta\alpha}
\,.\end{equation}
%%%

The matching coefficients are obtained by computing the renormalized matrix
elements $\mae{0}{...}{q\bar q}$ and $\mae{0}{...}{gg}$ on both sides of
\eq{J_matching} and comparing the results. In pure dimensional regularization
for UV and IR divergences all loop graphs in SCET are scaleless and vanish,
which means the UV and IR divergences in the bare matrix elements
precisely cancel each other. The renormalized matrix elements of the right-hand
side of \eq{J_matching} are then given by their tree-level expressions plus pure
$1/\eps$ IR divergences, which cancel against those of the full-theory
matrix elements $\mae{0}{J}{q\bar q}$ and $\mae{0}{J}{gg}$ of the left-hand
side. Hence, the matching coefficients in $\overline{\mathrm{MS}}$ are given in
terms of the IR-finite parts of the renormalized full-theory matrix
elements computed in pure dimensional regularization.

%~~~~~~~~~~~~~~~~~~~~~~~~~~~~~~~~~~~~~~~~~~~~~~~~~~~~~~~~~~~~~~~~~~~~~~~~~~~~~~~
\subsubsection{Soft-Collinear Factorization}
%~~~~~~~~~~~~~~~~~~~~~~~~~~~~~~~~~~~~~~~~~~~~~~~~~~~~~~~~~~~~~~~~~~~~~~~~~~~~~~~

The field redefinitions in \eq{BPS} introduce soft Wilson lines into the operators in \eq{Oi_SCET},
%%%
\begin{align} \label{eq:operators_BPSed}
O_{q\bar{q}}^{\alpha\beta}(x)
&= \bar\chi_{n_1,-\w_1}^{\zero \alpha j}(x)\,
T\bigl[Y^\dagger_{n_1}(x)\,Y_{n_2}(x) \bigr]^{jk} \chi_{n_2,\w_2}^{\zero \beta k}(x)
\,,\nn\\
O_{gg}^{\mu\nu}(x)
&= \sqrt{\w_1\,\w_2}\,\cB_{n_1,-\w_1\perp}^{\zero \mu c}(x) T\bigl[\cY_{n_1}^\dagger(x) \cY_{n_2}(x) \bigr]^{cd} \cB_{n_2, -\w_2\perp}^{\zero \nu d}(x)
\,.\end{align}
%%%
The time ordering is required to ensure the proper ordering of the soft gluon fields inside the Wilson lines. It only affects the ordering of the field operators, while the ordering of the color generators is still determined by the (anti)path ordering of the Wilson lines. In the remainder, we use these redefined fields and drop the $(0)$ superscript for convenience.

Since the momentum operator is linear in the Lagrangian, \eq{LSCET} allows us to write the hemisphere momentum operators as the sum of independent operators acting in the separate collinear and soft sectors,
%%%
\begin{equation}
\hp_a = \hp_{a,n_a} + \hp_{a,n_b} + \hp_{a,\soft}
\,,\qquad
\hp_b = \hp_{b,n_a} + \hp_{b,n_b} + \hp_{b,\soft}
\,.\end{equation}
%%%
The $n_a$ ($n_b$) collinear sector cannot contribute momentum in the $n_b$ ($n_a$) hemisphere. Thus, $\hp_{a, n_b} = \hp_{b, n_a} = 0$, while $\hp_{a, n_a} = \hp_{n_a}$ and $\hp_{b, n_b} = \hp_{n_b}$ reduce to the total momentum operators for each of the collinear sectors. For the soft sector, the distinction between the two hemisphere operators is important. We can now write
%%%
\begin{align} \label{eq:hpab_fact}
\delta(B_a^+ - n_a\sdt\hp_a)
&= \int\!\df b_a^+\,\df k_a^+ \,\delta(B_a^+ - b_a^+ - k_a^+)\,\delta(b_a^+ - n_a\sdt\hp_{n_a})\,\delta(k_a^+ - n_a\sdt\hp_{a,\soft})
\,,\nn\\
\delta(B_b^+ - n_b\sdt\hp_b)
&= \int\!\df b_b^+\,\df k_b^+ \,\delta(B_b^+ - b_b^+ - k_b^+)\,\delta(b_b^+ - n_b\sdt\hp_{n_b})\,\delta(k_b^+ - n_b\sdt\hp_{b,\soft})
\,.\end{align}
%%%
Using \eq{J_matching} in the hadronic tensor in \eq{Wmunu_QCD}, the forward matrix element of the product of currents turns into the forward matrix element of the product of the operators in \eq{operators_BPSed}. Since the Lagrangian in \eq{LSCET} contains no interactions between the collinear and soft sectors after the field redefinition, we can use \eq{hpab_fact} to factorize the resulting matrix element into a product of independent $n_a$-collinear, $n_b$-collinear, and soft matrix elements.

We first look at the contribution from $O_{q\bar{q}}$. The $x$ integral of the forward matrix element of $O_{q\bar{q}}$ becomes
%%%
\begin{align} \label{eq:me_fact}
&\int\!\frac{\df x^+\df x^-}{(4\pi)^2}\,e^{-\img (q^+ x^- + q^- x^+)/2}\,e^{\img (\lb_1 + \lb_2)\cdot x}\,
\MAe{p_{n_a} p_{n_b}}{O_{q\bar{q}}^{\dagger \beta\alpha}(x)\,
\delta(B_a^+ - n_a\sdt \hp_a)\,\delta(B_b^+ - n_b\sdt \hp_b)\, O_{q\bar{q}}^{\alpha'\beta'}(0)}{p_{n_a} p_{n_b}}
\nn\\ & \quad
= \int\!\frac{\df x^+\df x^-}{(4\pi)^2}\,e^{-\img (q^+ x^- + q^- x^+)/2}
\int\!\df b_a^+\,\df b_b^+\,\df k_a^+\,\df k_b^+\, \delta(B_a^+ - b_a^+ - k_a^+)\,\delta(B_b^+ - b_b^+ - k_b^+)
\nn\\ & \qquad\times
\int\!\df \w_a\, \df\w_b\, e^{\img (\w_a x^+ + \w_b x^-)/2} \biggl\{
\delta_{n_2 n_a}\, \delta(\w_2 - \w_a)\,\delta_{n_2' n_a}\, \delta(\w_2' - \w_a)\,
\delta_{n_1 n_b}\, \delta(\w_1 - \w_b)\,\delta_{n_1' n_b}\, \delta(\w_1' - \w_b)\,
\nn\\ & \qquad\quad\times
\theta(\w_a)\MAe{p_{n_a}}{\bar\chi_{n_a}^{\beta k}(x)\,\delta(b_a^+ - n_a\sdt \hp_{n_a})\,
\delta(\w_a - \bnP_{n_a})\, \chi_{n_a}^{\beta' k'}(0)}{p_{n_a}}
\nn\\ & \qquad\quad\times
\theta(\w_b) \MAe{p_{n_b}}{\chi_{n_b}^{\alpha j}(x)\,\delta(b_b^+ - n_b\sdt \hp_{n_b})\,
\delta(\w_b - \bnP_{n_b})\,\bar\chi_{n_b}^{\alpha' j'}(0)}{p_{n_b}}\,
\nn\\ & \qquad\quad\times
\MAe{0}{\bT\bigl[Y^\dagger_{n_a}(x)\, Y_{n_b}(x) \bigr]^{kj} \delta(k_a^+ - n_a\sdt\hp_{a,s})\, \delta(k_b^+ - n_b\sdt\hp_{b,s})\,T\bigl[Y^\dagger_{n_b}(0)\,Y_{n_a}(0) \bigr]^{j'k'} }{0} + (a\lra b) \biggr\}
\,.\end{align}
%%%
Here, $\ket{p_{n_a}}$ and $\ket{p_{n_b}}$ are the proton states with momenta
$P_{a,b}^\mu = \Ecm n_{a,b}^\mu/2$ as in \eq{nab_choice}. The two terms in
brackets in \eq{me_fact} arise from the different ways of matching up the fields
with the external proton states. The restriction to have positive labels $\w$
requires the fields in $O_{q\bar{q}}$ to be matched with the incoming proton
states and the fields in $O_{q\bar{q}}^\dagger$ with the outgoing proton states.
In principle, there are two more ways to match the fields and external states,
yielding matrix elements with the structure $\mae{p}{\chi\chi}{p}$ and
$\mae{p}{\bar{\chi}\bar{\chi}}{p}$, which vanish due to quark flavor number
conservation in QCD. For the same reason, in the full product $(\sum_q
O^\dagger_{q\bar q})(\sum_{q'} O_{q'\bar q'})$ only the flavor-diagonal term
with $q = q'$ survives.

We abbreviate the collinear and soft matrix elements in the last three lines of \eq{me_fact} as $M_{\w_a}(x^-)$, $M_{\w_b}(x^+)$, $M_\soft(x^+, x^-)$. The collinear matrix elements only depend on one light-cone coordinate because the label momenta $\w_{a,b}$ are defined to be continuous. We could have also started with discrete label momenta, $\lw_{a,b}$, and then convert to continuous labels by absorbing the residual $k_{n_a}^-$ dependence as follows:
%%%
\begin{equation}
\sum_{\lw_a}\, e^{\img \lw_a x^+/2}\,M_{\lw_a}(x^+, x^-)
= \sum_{\lw_a} \int\!\df k_a^-\,e^{\img (\lw_a + k_{n_a}^-) x^+/2}\,M_{\lw_a + k_{n_a}^-}(x^-)
= \int\!\df\w_a\,e^{\img \w_a x^+/2}\,M_{\w_a}(x^-)
\,,\end{equation}
%%%
and analogously for $M_{\w_b}(x^+)$. In the second step we used that by
reparametrization invariance the Fourier-transformed matrix element can only
depend on the linear combination $\lw_a + k_{n_a}^- = \w_a$.

As an aside, note that in the well-studied case where the collinear matrix
elements are between vacuum states, giving rise to jet functions, the
distinction between discrete and continuous labels is not as relevant. In that
case, the SCET Feynman rules imply that the collinear matrix elements do not
depend on the residual $k^-$ (and $k_\perp$) components, and therefore the label
momenta can be treated in either way. In our case, momentum conservation with
the external state forces the collinear matrix elements to depend on $k^-$.
Therefore, the only way to eliminate the residual $k^-$ dependence is to absorb
it into continuous $\w$ labels. One can easily see this already at tree level.
Replacing the proton states by quark states with momentum $p = \lp + p_r$, we
get
%%%
\begin{equation}
\int\!\frac{\df x^+\,\df x^-}{(4\pi)^2}\,e^{-\img(k^-x^+ + k^+ x^-)/2}\,\Mae{q(p)}{\bar\chi_n(x^+, x^-)\, \delta_{\lw,\bnP_n}\,\chi_n(0)}{q(p)}
= \bar u u\, \delta_{\lw, \lp^-}\,\delta(k^- - p_r^-)\,\delta(k^+ - p_r^+)
\,,\end{equation}
%%%
and the label and residual minus momenta are combined using $\delta_{\lw,
  \lp^-}\,\delta(k^- - p_r^-) = \delta(\w - p^-)$. Our continuous $\w$ is
physical and corresponds to the momentum fraction of the quark in the proton.

Returning to our discussion, to perform the $x$ integral in \eq{me_fact}, we
take the residual Fourier transforms of the matrix elements,
%%%
\begin{align} \label{eq:tildeMi}
M_{\w_a}(x^-)
&= \int\!\frac{\df k^+}{2\pi}\,e^{\img k^+ x^-/2}\,\tM_{\w_a}(k^+)
\,,\qquad
M_{\w_b}(x^+)
= \int\!\frac{\df k^-}{2\pi}\,e^{\img k^- x^+/2}\,\tM_{\w_b}(k^-)
\,,\nn\\
M_\soft(x^-, x^+)
&= \int\!\frac{\df k_s^+\,\df k_s^-}{(2\pi)^2}\,e^{\img(k_s^+ x^+ + k_s^- x^+)/2}\,\tM_\soft(k_s^+, k_s^-)
\,.\end{align}
%%%
Just as $x^\pm$, the residual momenta $k^\pm$ and $k_s^\pm$ here are all defined
with respect to the common $n = n_a$. The $x$ integral in \eq{me_fact} now
becomes
%%%
\begin{align} \label{eq:xintegral}
&\int\!\frac{\df x^+\df x^-}{(4\pi)^2}\,e^{\img (\w_a - q^-) x^+/2}\, e^{\img (\w_b - q^+)x^-/2}
M_{\w_a}(x^-)\,M_{\w_b}(x^+)\,M_\soft(x^+, x^-)
\nn\\ &\qquad
= \int\!\frac{\df k^+}{2\pi}\,\frac{\df k^-}{2\pi}\, \frac{\df k_s^+\,\df k_s^-}{(2\pi)^2}\,
\tM_{\w_a}(k^+)\,\tM_{\w_b}(k^-)\, \tM_\soft(k_\soft^+, k_\soft^-)\,
\delta(\w_a - q^- + k^- + k_\soft^-)\,\delta(\w_b - q^+ + k^+ + k_\soft^+)
\nn\\ &\qquad
= \delta(\w_a - q^-)\,\delta(\w_b - q^+)\,M_{\w_a}(0)\,M_{\w_b}(0)\,M_\soft(0)
\,.\end{align}
%%%
In the last step we expanded $q^\pm - k^\pm - k_\soft^\pm = q^\pm[1 +
\ord{\la^2}]$. The remaining residual integrations are then simply the Fourier
transforms of the matrix elements at $x = 0$.

Note that without the integration over $\vec{q}_T$ in the hadronic tensor
\eq{Wmunu_QCD}, the currents would depend on $x_\perp$, which would require us
to include perpendicular components $b_{a,b\perp}$ in the label momenta, and the
soft matrix element would depend on $x_\perp$, too. (The residual $k_{\perp}$
dependence in the collinear matrix elements can again be absorbed into
continuous $b_{a,b\perp}$.) The corresponding $x_\perp$ integration in
\eq{xintegral} would yield an additional $\delta$ function
$\delta^2(\vec{b}_{a\perp} + \vec{b}_{b\perp} + \vec{q}_T -
\vec{k}_{\soft\perp})$.  Integrating over $\vec{q}_T$ effectively eliminates
this $\delta$ function, which would otherwise force us to introduce an explicit
dependence on $b_{a,b\perp}$ in the beam functions. If one considers the $q_T$
spectrum of the dileptons for $q_T^2\ll q^2$, our analysis here provides a starting
point but requires further study. One cannot just use $p_T$ in place of $B_{a,b}^+$
with our arguments to impose an analogous
restriction on the final state, because at $\ord{\alpha_s^2}$ one can have two jets at high $\vec{p}_T$ that
still have small total $\vec{p}_T$.

The $n_a$-collinear matrix element now reduces to the quark beam functions
defined in \eq{B_def},
%%%
\begin{align}
M_{\w_a}(0) &=
\theta(\w_a) \MAe{p_{n_a}}{\bar\chi_{n_a}^{\beta k}(0)\,
  \delta(b_a^+ - n_a\sdt \hp_{n_a})\,\delta(\w_a - \bnP_{n_a})\, \chi_{n_a}^{\beta' k'}(0)}{p_{n_a}}
\nn\\ & \quad
= \frac{\nslash_a^{\beta'\beta}}{4}\, \frac{\delta^{k'k}}{N_c}
  \theta(\w_a) \MAe{p_{n_a}}{\bar\chi_{n_a}(0)\,\delta(b_a^+ - n_a\sdt \hp_{n_a})\,
\delta(\w_a - \bnP_{n_a})\, \frac{\bnslash_a}{2}  \chi_{n_a}(0)}{p_{n_a}}
\nn\\ & \quad
= \frac{\nslash_a^{\beta'\beta}}{4}\, \frac{\delta^{k'k}}{N_c}\, \theta(\w_a)
 \int\!\frac{\df y^-}{4\pi}\,e^{\img b_a^+ y^-/2}\,
 \MAe{p_{n_a}}{e^{-\img \hp_{n_a}^+ y^-/2}\,e^{\img \hp_{n_a}^+ y^-/2}\,\bar\chi_{n_a}(0)\,
 e^{-\img \hp_{n_a}^+ y^-/2}\,
 \delta(\w_a - \bnP_{n_a})\,\frac{\bnslash_a}{2}\chi_{n_a}(0)}{p_{n_a}}
\nn\\ & \quad
= \frac{\nslash_a^{\beta'\beta}}{4}\, \frac{\delta^{k'k}}{N_c}\, \w_a B_q(\w_a b_a^+, \w_a/P_a^-)
\,.\end{align}
%%%
We abbreviated $\hp_{n_a}^+ = n_a\sdt \hp_{n_a}$, and
in the last step we used $e^{\img \hp_n^+ y^-/2}\,\bar\chi_n(0)\,e^{-\img\hp_n^+ y^-/2} = \bar\chi_n(y^-n/2)$ and $\hp_n^+\ket{p_n} = 0$. Similarly, for the antiquark beam function we have
%%%
\begin{equation}
M_{\w_b}(0) =
\theta(\w_b)\MAe{p_{n_b}}{\chi_{n_b}^{\alpha j}(x)\, \delta(b_b^+ - n_b\sdt \hp_b)\,
\delta(\w_b - \bnP_{n_b})\,\bar\chi_{n_b}^{\alpha' j'}(0)}{p_{n_b}}
= \frac{\nslash_b^{\alpha\alpha'}}{4}\,\frac{\delta^{jj'}}{N_c}\, \w_b B_{\bar q}(\w_b b_b^+ , \w_b/P_b^-)
\,.\end{equation}
%%%

Since the collinear matrix elements are color diagonal, the soft matrix element reduces to an overall color-singlet trace, which defines the $q\bar q$ incoming hemisphere soft function,
%%%
\begin{equation} \label{eq:Sqq_def}
S^{q\bar{q}}_\hemiin(k_a^+, k_b^+)
= \frac{1}{N_c}\, \tr\,\Mae{0}{ \bT\bigl[Y^\dagger_{n_a}(0)\, Y_{n_b}(0) \bigr]
 \delta(k_a^+ - n_a\sdt\hp_{a,s})\, \delta(k_b^+ - n_b\sdt\hp_{b,s})\,
  T\bigl[Y^\dagger_{n_b}(0)\,Y_{n_a}(0) \bigr]}{0}
\,.\end{equation}
%%%
The trace is over color and the factor of $1/N_c$ is included by convention,
such that at tree level we have $S^{q\bar q,\tree}_\hemiin(k_a^+, k_b^+) =
\delta(k_a^+)\,\delta(k_b^+)$.  The soft matrix element in the second term of
\eq{me_fact} with $a \lra b$ interchanged is equal to the above one due to
charge conjugation invariance of QCD. Under charge conjugation, the Wilson lines
transform as $\mathrm{C}^{-1} Y_n^{ij} \mathrm{C} = T[Y_n^{\dagger ji}]$. The
explicit time ordering is required because the fields in $Y_n$ are time-ordered
by default, and charge conjugation only changes the ordering of the color
generators but not of the field operators. For us this is not relevant, because
the ordering of the fields is determined by the overall (anti-)time ordering in
the matrix element. Thus, for the soft matrix element with $a\lra b$
interchanged, we find
%%%
\begin{align}
&\tr \Mae{0}{\bT\bigl[Y^\dagger_{n_b} Y_{n_a} \bigr]
\delta(k_a^+ - n_a\sdt\hp_{a,s})\,\delta(k_b^+ - n_b\sdt\hp_{b,s})\,
  T\bigl[Y^\dagger_{n_a} Y_{n_b} \bigr]}{0}
\nn\\ &\qquad
\stackrel{\mathrm{C}}{=}
\tr \Mae{0}{\bT\bigl[Y^T_{n_b} Y_{n_a}^{\dagger T} \bigr]
\delta(k_a^+ - n_a\sdt\hp_{a,s})\, \delta(k_b^+ - n_b\sdt\hp_{b,s})\,
  T\bigl[Y^T_{n_a} Y_{n_b}^{\dagger T} \bigr] }{0}
= S^{q\bar{q}}_\hemiin(k_a^+, k_b^+)
\,,\end{align}
%%%
where the transpose refers to the color indices. In the last step we used
$\tr[A^T B^T C^T D^T] = \tr[BADC]$ and the fact that the fields in
$Y^\dagger_{n_b}$ and $Y_{n_a}$ are spacelike separated and thus commute.  Under
parity, we have $\mathrm{P}^{-1} Y_{n_a} \mathrm{P} = Y_{n_b}$ and
$\mathrm{P}^{-1} n_a\sdt\hp_{a,s} \mathrm{P} = n_b\sdt\hp_{b,s}$. Therefore, CP
invariance implies that $S_\hemiin^{q\bar q}$ is symmetric in its arguments,
%%%
\begin{equation}
S^{q\bar{q}}_\hemiin(k_a^+, k_b^+)
\stackrel{\mathrm{CP}}{=} \frac{1}{N_c}\, \tr\,\Mae{0}{ \bT\bigl[Y^\dagger_{n_a} Y_{n_b}\bigr]
 \delta(k_a^+ - n_b\sdt\hp_{b,s})\, \delta(k_b^+ - n_a\sdt\hp_{a,s})\,
  T\bigl[Y^\dagger_{n_b} Y_{n_a}\bigr]}{0}
= S^{q\bar{q}}_\hemiin(k_b^+, k_a^+)
\,.\end{equation}
%%%

Having worked out the different terms in \eq{me_fact}, we are ready to include
the remaining pieces from \eqs{J_matching}{Wmunu_QCD}. The $q\bar q$
contribution to the hadronic tensor becomes
%%%
\begin{align} \label{eq:Wqqfactorized}
&W_{JJ' q\bar q}(q^2, Y, B_a^+, B_b^+)
\nn\\ &\qquad
= \int\!\df\w_a\,\df\w_b\,\delta(\w_a - q^-)\,\delta(\w_b - q^+)\,
\!\!\!\sum_{n_1, n_2, n_1', n_2'} \int\!\df\w_1\,\df\w_2\,\df\w_1'\,\df\w_2'
\,\bC_{Jq\bar q}^{\beta\alpha}(\lb_1,\lb_2)\, C_{J' q\bar q}^{\alpha'\beta'} (\lb_1',\lb_2')
\nn\\ & \qquad\quad\times
\biggl\{
\delta_{n_2 n_a}\, \delta(\w_2 - \w_a)\,\delta_{n_2' n_a}\, \delta(\w_2' - \w_a)\,
\delta_{n_1 n_b}\, \delta(\w_1 - \w_b)\,\delta_{n_1' n_b}\, \delta(\w_1' - \w_b)\,
\\ &\qquad\qquad\times
\frac{\nslash_a^{\beta'\beta}}{4}\, \frac{\nslash_b^{\alpha\alpha'}}{4}\,\frac{1}{N_c}\,
\int\! \df k_a^+\,\df k_b^+\,
q^2 B_q[x_a \Ecm (B_a^+ - k_a^+), x_a]\, B_{\bar q}[x_b\Ecm(B_b^+ - k_b^+), x_b]\,
S^{q \bar q}_\hemiin(k_a^+,k_b^+) + (a \leftrightarrow b) \biggr\}
\nn \\\nn &\qquad
= H_{JJ'q \bar q}(\lb_a, \lb_b) \int\!\df k_a^+ \df k_b^+\,
q^2 B_q[x_a \Ecm(B_a^+ - k_a^+), x_a]\, B_{\bar q}[x_b \Ecm(B_b^+ - k_b^+), x_b]\,
 S^{q \bar q}_\hemiin(k_a^+,k_b^+) + (q \lra \bar q)
\,.\end{align}
%%%
All label sums and integrations from \eq{J_matching} eliminate the label $\delta$'s from \eq{me_fact}. In the second step we defined
%%%
\begin{equation} \label{eq:lbxab}
\lb_a^\mu = x_a\Ecm\,\frac{n_a^\mu}{2}
\,,\qquad
\lb_b^\mu = x_b\Ecm\,\frac{n_b^\mu}{2}
\,,\qquad
x_a \equiv \frac{\w_a}{\Ecm} = \frac{q^-}{\Ecm} = \frac{\sqrt{q^2}\, e^Y}{\Ecm}
\,,\qquad
x_b \equiv \frac{\w_b}{\Ecm} = \frac{q^+}{\Ecm} = \frac{\sqrt{q^2}\, e^{-Y}}{\Ecm}
\,,\end{equation}
%%%
as in \eq{xab}, and introduced the hard functions
%%%
\begin{align} \label{eq:Hqq_def}
H_{JJ'q \bar q}(\lb_a, \lb_b)
&= \frac{1}{N_c}\, \frac{1}{4}\,\tr_\mathrm{spins}\Bigl[
\frac{\nslash_a}{2}\, \bC_{J q\bar q}(\lb_b, \lb_a)\,
\frac{\nslash_b}{2}\, C_{J' q\bar q}(\lb_b, \lb_a) \Bigr]
\,,\qquad
H_{JJ'\bar q q}(\lb_a, \lb_b) = H_{JJ' q\bar q}(\lb_b, \lb_a)
\,.\end{align}
%%%
Equation~\eqref{eq:Wqqfactorized} is the final factorized result for the $O_{q\bar{q}}$ contribution
to the hadronic tensor.

Repeating the same steps for $O_{gg}$, we obtain for the forward matrix element
%%%
\begin{align} \label{eq:me_fact_gg}
&\int\!\frac{\df x^+\df x^-}{(4\pi)^2}\,e^{-\img (q^+ x^- + q^- x^+)/2}\,e^{\img (\lb_1 + \lb_2)\cdot x}\,
\MAe{p_{n_a} p_{n_b}}{O_{gg}^{\dagger \nu\mu}(x)\,
\delta(B_a^+ - n_a\sdt \hp_{a,s})\,\delta(B_b^+ - n_b\sdt \hp_{b,s})\, O_{gg}^{\mu'\nu'}(0)}{p_{n_a} p_{n_b}}
\nn\\ & \qquad
= \int\!\df \w_a\, \df\w_b\,\delta(\w_a - q^-)\,\delta(\w_b - q^+)
\int\!\df b_a^+\,\df b_b^+\,\df k_a^+\,\df k_b^+\, \delta(B_a^+ - b_a^+ - k_a^+)\,\delta(B_b^+ - b_b^+ - k_b^+)
\nn\\ & \qquad\quad\times
\bigl[\delta_{n_1 n_a}\, \delta(\w_1 - \w_a)\, \delta_{n_2 n_b}\, \delta(\w_2 - \w_b) + (a \lra b)\bigr]
\bigl[\delta_{n_1'n_a}\, \delta(\w_1' - \w_a)\, \delta_{n_2' n_b}\, \delta(\w_2' - \w_b) + (a \lra b)\bigr]
\nn\\ & \qquad\quad\times
\w_a\,\theta(\w_a)\,\MAe{p_{n_a}}{\cB_{n_a\perp}^{\mu c}(0)\,\delta(b_a^+ - n_a\sdt \hp_{n_a})\,\delta(\w_a - \bnP_{n_a})\, \cB_{n_a\perp}^{\mu' c'}(0)}{p_{n_a}}\,
\nn\\ & \qquad\quad\times
\w_b\,\theta(\w_b)\,\MAe{p_{n_b}}{\cB_{n_b\perp}^{\nu d}(0)\,\delta(b_b^+ - n_b\sdt \hp_{n_b})\,\delta(\w_b - \bnP_{n_b})\,  \cB_{n_b\perp}^{\nu' d'}(0)}{p_{n_b}}
\nn\\ & \qquad\quad\times
\MAe{0}{\overline{T}\bigl[\cY^\dagger_{n_a}(0)\, \cY_{n_b}(0)\bigr]^{cd}\,
\delta(k_a^+ - n_a\sdt\hp_{a,s})\, \delta(k_b^+ - n_b\sdt\hp_{b,s})\,
T\bigl[\cY^\dagger_{n_b}(0)\,\cY_{n_a}(0) \bigr]^{d'c'} }{0}
\,,\end{align}
%%%
where we already performed the integral over $x$. The four terms in the third line correspond
to the four different ways to match up the gluon fields with the incoming proton states.
The collinear matrix elements reduce to the gluon beam function defined in \eq{B_def},
%%%
\begin{align}
&\w_a\,\theta(\w_a)\,\MAe{p_{n_a}}{\cB_{n_a\perp}^{\mu c}(0)\,\delta(b_a^+ - n_a\sdt \hp_{n_a})\,
\delta(\w_a - \bnP_{n_a})\, \cB_{n_a\perp}^{\mu'c'}(0)}{p_{n_a}}
= \frac{g_\perp^{\mu \mu'} }{2}\, \frac{\delta^{cc'}}{N_c^2-1}\, \w_a B_g(\w_a b_a^+, \w_a/P_a^-)
\,.\end{align}
%%%
Including the color traces from the beam functions, the soft matrix element
defines the gluonic incoming hemisphere soft function,
%%%
\begin{align}
S^{gg}_\hemiin(k_a^+, k_b^+)
= \frac{1}{N_c^2 - 1}\,
\MAe{0}{\tr_\mathrm{color}\bigl\{\overline{T}\bigl[\cY^\dagger_{n_a}(0)\, \cY_{n_b}(0) \bigr]\,
\delta(k_a^+ - n_a\sdt\hp_{a,s})\, \delta(k_b^+ - n_b\sdt\hp_{b,s})\, T\bigl[\cY^\dagger_{n_b}(0)\,\cY_{n_a}(0) \bigr] \bigr\}}{0}
\,,\end{align}
%%%
where the normalization is again convention. Putting everything together, the
gluon contribution to the hadronic tensor becomes
%%%
\begin{align}
W_{JJ'gg}(q^2, Y, B_a^+, B_b^+)
&= H_{JJ'gg}(\lb_a, \lb_b) \!\int\!\df k_a^+ \df k_b^+\,
q^2 B_g[x_a \Ecm(B_a^+ - k_a^+), x_a] B_g[x_b \Ecm (B_b^+ - k_b^+), x_b]\,
S^{gg}_\hemiin(k_a^+,k_b^+)
,\end{align}
%%%
with the hard function
%%%
\begin{equation}
H_{JJ'gg}(\lb_a, \lb_b)
= \frac{1}{N_c^2-1}\,
\frac{1}{2}\,(g_{\perp\,\mu\mu'}\, g_{\perp\,\nu\nu'} + g_{\perp\,\mu\nu'}\, g_{\perp\,\nu\mu'})\,
C_{J gg}^{\dagger\,\nu\mu}(\lb_a, \lb_b)\, C_{J' gg}^{\mu'\nu'}(\lb_a, \lb_b)
\,.\end{equation}
%%%
Here we have used the symmetry of the Wilson coefficients in \eq{Cgg_symmetry}
to simplify the four terms that arise from interchanging $a\lra b$ in
\eq{me_fact_gg}.

To obtain the full result for the hadronic tensor all we have to do now is to
add up the contributions from the different quark flavors and the gluon,
%%%
\begin{equation}
W_{JJ'}(q^2, Y, B_a^+, B_b^+)
= \sum_q W_{JJ' q\bar q}(q^2, Y, B_a^+, B_b^+) + W_{JJ'gg}(q^2, Y, B_a^+, B_b^+)
\,.\end{equation}
%%%
Inserting this back into \eq{dsigmadO_LW}, the final result for the factorized cross section becomes
%%%
\begin{equation} \label{eq:dsigma_final}
\frac{\df\sigma}{\df q^2\df Y \df B_a^+\df B_b^+}
= \sum_{ij} H_{ij}(q^2, Y) \int\!\df k_a^+\, \df k_b^+\,
q^2 B_i[x_a \Ecm(B_a^+ - k_a^+), x_a] B_j[x_b \Ecm(B_b^+ - k_b^+), x_b]
S^{ij}_\hemiin(k_a^+,k_b^+)
\,,\end{equation}
\end{widetext}
%%%
with $x_{a,b}\Ecm = \sqrt{q^2} e^{\pm Y}$ as in \eqs{xab}{lbxab} and the hard function
%%%
\begin{align} \label{eq:Hij}
H_{ij}(q^2, Y)
&= \frac{1}{2\Ecm^2} \sum_{J,J'} L_{JJ'}(q^2, Y)
\nn\\ & \quad\times
H_{JJ' ij}\Bigl(x_a\Ecm \frac{n_a}{2}, x_b\Ecm \frac{n_b}{2}\Bigr)
\,.\end{align}
%%%
The sum in \eq{dsigma_final} runs over parton species $ij= \{gg, u\bar u, \bar u
u, d \bar d, \ldots\}$, where $B_i$ is the beam function for parton $i$ in beam
$a$ and $B_j$ for parton $j$ in beam $b$. Equation~\eqref{eq:dsigma_final} is
the final factorization theorem for the isolated $pp\to XL$ and $p\bar{p} \to
XL$ processes. In \subsec{DY_final} below we will apply it to the case of
Drell-Yan, which will yield \eq{DYbeam}.

The beam functions in \eq{dsigma_final} are universal and take into account
collinear radiation for isolated processes with $x$ away from one.
Since the soft function only depends on the color
representation, but not on the specific quark flavor, there are only two
independent soft functions $S^{q\bar q}_\hemiin$ and $S^{gg}_\hemiin$. In the
sum over $ij$ in \eq{dsigma_final}, there are no mixed terms with $ij$ corresponding to
beam functions of two different quark flavors.
Likewise, there are no mixed terms with quark and gluon beam functions.
For example, a graph like \fig{DYBbeamgq} is part of the $ij= q\bar{q}$ term in the
sum. Thus, cross terms between quark and gluon PDFs only appear via the
contributions of different PDFs to a given beam function, as shown in
\eq{B_fact}.

The only process dependence in \eq{dsigma_final} arises through the hard
functions $H_{ij}(q^2, Y)$, and one can study any desired leptonic observables
by inserting the appropriate projections in the leptonic phase-space
integrations inside $L_{JJ'}(q^2, Y)$. Since the hard function $H_{JJ'ij}$
corresponds to the partonic matrix element $\mae{ij}{J^\dagger}{0}\mae{0}{J'}{ij}$ and
$L_{JJ'}$ is given by the square of the relevant electroweak matrix elements
$L_J^\dagger L_{J'}$, $H_{ij}(q^2, Y)$ can be determined from calculations of the
partonic cross section $ij\to L$. Furthermore, $H_{ij}(q^2,Y)$ is identical to
the hard function in threshold factorization theorems and hence in many cases is
known from existing computations.

%~~~~~~~~~~~~~~~~~~~~~~~~~~~~~~~~~~~~~~~~~~~~~~~~~~~~~~~~~~~~~~~~~~~~~~~~~~~~~~~
\subsubsection{Cancellation of Glauber Gluons}
\label{subsubsec:glauber}
%~~~~~~~~~~~~~~~~~~~~~~~~~~~~~~~~~~~~~~~~~~~~~~~~~~~~~~~~~~~~~~~~~~~~~~~~~~~~~~~

In the above derivation we have implicitly assumed that contributions from
Glauber gluons cancel in the final cross section, so that we do not need Glauber
interactions in the effective theory. To complete the proof of factorization, we
now argue that this is indeed the case.

In principle, Glauber interactions add an additional term $\cL_G$ to the SCET
Lagrangian
%%%
\begin{align} \label{eq:LwithG}
\cL_\mathrm{SCET}
&= \cL_{n_a}(\chi_{n_a}, A_\soft) + \cL_{n_b}(\chi_{n_b}, A_\soft)
 + \cL_\soft(A_\soft)
\nn\\ & \quad
+ \cL_G(A_G, \chi_{n_a}, \chi_{n_b}, A_\soft)
\,.\end{align}
%%%
Glauber interactions in SCET have been considered in Refs.~\cite{Idilbi:2008vm,
  Donoghue:2009cq}, but we will not require an explicit construction of $\cL_G$
here.  Our arguments will be based on the one hand, on the consistency with
processes where it has been proven that Glauber interactions cancel, and on the
other hand on systematic scale separation in the language of effective field
theory. The scale separation is valid independently of whether it leads to a
factorization into simple matrix elements, or whether it leads to a
non-factorizable matrix element with complicated dynamics.

The possible danger of the Glauber modes comes from the fact that they couple
the two collinear sectors $n_a$ and $n_b$ with momentum scaling $Q(\lambda^2, 1,
\lambda)$ and $Q(1, \lambda^2, \lambda)$. With $\cL_G$, there will still be
interactions between soft and collinear modes present in the Lagrangian even
after the field redefinition, so we cannot a priori factorize the full matrix
element into independent soft and collinear matrix elements. Therefore, we have
to revisit each step in our derivation with $\cL_G$ in mind.

\begin{figure}[t!]
\includegraphics[scale=0.75]{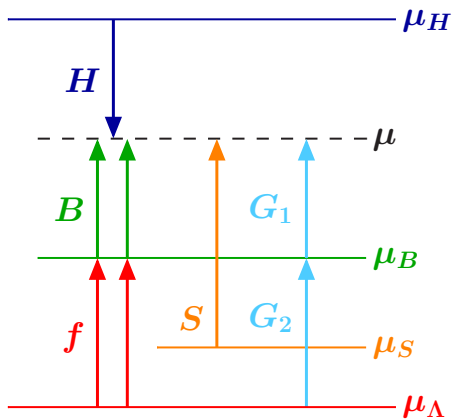}
\caption{RGE running including potential Glauber modes.}
\label{fig:running_drellyan_glauber}
\end{figure}

Our argument will be divided into three steps: (i) above the scale $\mu_B$, (ii)
at the scale $\mu_B$, and (iii) below the scale $\mu_B$. For (i) and (ii) we
have to consider Glauber modes with momentum scaling $Q(\lambda^2, \lambda^2,
\lambda)$, which we call $G_1$ modes. Since they have virtuality $\sim\mu_B^2$
they are integrated out at this scale. Any residual effects of Glauber
interactions below $\mu_B$ could occur from modes with momentum scaling
$Q(\lambda^4, \lambda^4, \lambda^2)$, which we call $G_2$. These modes are
illustrated in \fig{running_drellyan_glauber}.

Above the hard scale $\mu_H \simeq Q$, we have full QCD and no distinction
between different modes is required, so in step (i) we are concerned with
contributions of $G_1$ in the region $\mu_H > \mu > \mu_B$. At the scale
$\mu_H$, we integrate out hard modes with virtualities $Q^2$ or higher by
matching the QCD currents onto SCET currents. For our process, the leading
operators are given in \eq{Oi_SCET}, which contain only one field per collinear
direction. For the theory in \eq{LwithG}, all other possible operators
are power suppressed. The matching onto these currents is valid at an operator
level and can be performed with quark and gluon states. It is independent of the
hadronic matrix element we are going to take later on. The key point is that the
exact same matching calculation and resulting Wilson coefficients $C$ occur for
threshold Drell-Yan and $e^+e^-\to 2$ jets.  For these cases it is
known~\cite{Collins:1981ta, Collins:1989gx} that $G_1$ modes do not
affect the matching of the hard function $H\sim\abs{C}^2$ at $\mu_H$ or the
running of $H$ in the region $\mu_H> \mu > \mu_B$ shown in
Figs.~\ref{fig:running_drellyan}(b) or~\ref{fig:running_drellyan}(c). The hard
function $H$ gives a complete description of the physics down to the scale
$\mu_B$ whether or not the modes in the SCET matrix elements factorize further.
In \fig{running_drellyan_glauber}, this corresponds to taking the scale $\mu = \mu_B$.
Therefore, the $G_1$ modes can give neither large $\ln(\mu_B/\mu_H)$ terms nor
finite contributions above $\mu_B$.

In step (ii), we integrate out modes with virtualities $Q^2\lambda^2$ at the
scale $\mu_B$, which may involve matrix elements with $G_1$ modes exchanged.
This matching affects the $n_a$-collinear, $n_b$-collinear, and $G_1$ modes,
whose momentum scaling below $\mu_B$ changes to $Q(\lambda^4, 1, \lambda^2)$,
$Q(1, \lambda^4, \lambda^2)$, and $Q(\lambda^4, \lambda^4, \lambda^2)$,
respectively. Here we consider the theory right above $\mu_B$ including $G_1$
modes, leaving the discussion of the theory just below $\mu_B$ and $G_2$ modes
to step (iii). Thus, we have to consider the matrix element of the composite operator
%%%
\begin{align}
&\big[\bar\chi_{n_a} \chi_{n_b}\big](x^+, x^-)\,
\delta(B_a^+ - n_a\sdt\hp_a)\,\delta(B_b^+ - n_b\sdt\hp_b)
\nn\\ & \qquad\times
\delta(\w_a - \bnP_{n_a})\, \delta(\w_b - \bnP_{n_b})\,
\big[\bar\chi_{n_b} \chi_{n_a} \big](0)
\,,\end{align}
%%%
where we suppressed all spin and color indices for simplicity, and these
collinear fields still couple to soft fields in their Lagrangians. Since $\mu_B$
is a perturbative scale, we can carry out the matching onto the theory below
$\mu_B$ at the operator level and do not yet have to consider proton states.
Since the Glauber gluons are spacelike, they cannot cross the final-state cut
indicated by the $\delta$ functions and only appear in virtual subdiagrams.
We can therefore make a correspondence with the calculation in step (i) as follows.
For any given final state with collinear and soft particles, the SCET
computation for (ii) is identical to the SCET computation carried out for the
matching in step (i) but using this particular choice of external
states.\footnote{In practice one would never make such a complicated choice, but
  if one does, it must give the same result as picking a minimal state for the
  matching.} Since that SCET computation cannot induce any dependence on $G_1$
in step (i), there can also be no contributions from $G_1$ for the forward
matrix-element computation here. The result of the step (ii) matching is thus
given by a Wilson coefficient times an operator of the form
%%%
\begin{align} \label{eq:G1matching}
&\int\!\df k_a^+\, \df k_b^+\,
C(x^+, x^-, B_a^+ - k_a^+, B_b^+ - k_b^+)
\nn\\* & \qquad \times
\bar\chi_{n_a}'(0)\, \bT\bigl[Y^\dagger_{n_a} Y_{n_b} \bigr] \chi_{n_b}'(0)\,
\delta(k_a^+ - n_a\sdt\hp_a)\,
\nn\\ & \qquad \times
\delta(k_b^+ - n_b\sdt\hp_b)\, \delta(\w_a - \bnP_{n_a})\,
  \delta(\w_b - \bnP_{n_b})
\nn\\ & \qquad \times
\bar\chi_{n_b}'(0)\, T\bigl[Y^\dagger_{n_b} Y_{n_a} \bigr]\chi_{n_a}'(0)
\,,\end{align}
%%%
where the primed collinear fields have scaling $Q(\lambda^4, 1, \lambda^2)$ and
$Q(1, \lambda^4, \lambda^2)$, and the soft fields in the $Y$ Wilson lines have
scaling $Q(\lambda^2, \lambda^2, \lambda^2)$.

For step (iii) below $\mu_B$, we have to consider the $\mae{pp}{\dotsb}{pp}$
matrix element of \eq{G1matching} and possible contributions from $G_2$ Glauber
gluons, which can now also connect to spectator lines in the proton (which are
primed collinear modes). The $G_2$ gluons may spoil the factorization of the two
collinear sectors. To argue that this is not the case, we rely heavily on the
original proof of the cancellation of Glauber gluons for inclusive Drell-Yan in
Ref.~\cite{Collins:1988ig}. By construction, for our observables the $k_{a,b}^+$
variables in \eq{G1matching} are of $\ord{Q\lambda^2}$ and thus only get contributions from the
soft gluons. Hence, we are fully inclusive in the Hilbert space of the primed
collinear fields. Therefore, the $G_2$ modes as well as possible ``ultrasoft''
$Q(\lambda^4, \lambda^4, \lambda^4)$ gluons cancel in the sum over states, just
as in the inclusive case. This discussion for the cancellation of $G_2$ modes is
identical to Ref.~\cite{Aybat:2008ct}, where arguments were presented for the
cancellation of $G_2$ gluons up to the scale induced by the measurement on the
final state, which in our case is $\mu_B$.

Physically, one could imagine that Glauber modes kick the spectators in the
proton remnant such that they can contribute to $B_{a,b}^+$. The above
arguments show that this is not the case, so that our treatment of
the proton and its remnant in the derivation of the factorization
is correct.

Note that the above arguments do not suffice to show that Glauber interactions cancel
when there are additional hard central jets in the final state.

%===============================================================================
\subsection{Final Results for Drell-Yan}
\label{subsec:DY_final}
%===============================================================================

In this subsection, we present the final results for the isolated Drell-Yan
cross section. Our discussion is split into four parts: the leptonic tensor,
the hard function, the soft function, and the final cross section for beam thrust.

%~~~~~~~~~~~~~~~~~~~~~~~~~~~~~~~~~~~~~~~~~~~~~~~~~~~~~~~~~~~~~~~~~~~~~~~~~~~~~~~
\subsubsection{The Leptonic Tensor}
%~~~~~~~~~~~~~~~~~~~~~~~~~~~~~~~~~~~~~~~~~~~~~~~~~~~~~~~~~~~~~~~~~~~~~~~~~~~~~~~

To give an explicit example, we now apply the final factorization result in
\eq{dsigma_final} to the Drell-Yan process with $L = \ell^+\ell^-$. The relevant
QCD currents are the vector and axial-vector currents $J_{hf}^\mu$ with $h =
\{V,A\}$, already given in \eq{JDY_QCD}. The corresponding leptonic
contributions are
%%%
\begin{align} \label{eq:LmuDY}
L^\mu_{Vf}(p_1, p_2) &= \frac{4\pi\alem}{q^2} \Bigl[
- Q_f\, \bar{u}(p_2) \gamma^\mu v(p_1)
\nn\\ & \quad
+ \frac{v_f}{1 - m_Z^2/q^2}\, \bar{u}(p_2)
  \gamma^\mu (v_\ell - a_\ell\gamma_5) v(p_1) \Bigr]
\,,\nn\\
L^\mu_{Af}(p_1, p_2) &= \frac{4\pi\alem}{q^2}\,
\frac{-a_f}{1 - m_Z^2/q^2}
\nn\\ & \quad \times
\bar{u}(p_2) \gamma^\mu (v_\ell - a_\ell\gamma_5) v(p_1)
\,,\end{align}
%%%
where in this subsection $p_1 = p_{\ell^+}$ and $p_2 = p_{\ell^-}$ are the
lepton momenta, $Q_f$ is the quark charge (in units of $\abs{e}$), and
$v_{\ell,f}$, $a_{\ell,f}$ are the standard vector and axial couplings of the
leptons and quarks of flavor $f$ to the $Z$ boson. We will include the
width of the $Z$ later in \eq{H_oneloop}.

The leptonic phase space integral is
%%%
\begin{align} \label{eq:leptonPS}
&\int\!\frac{\df^4 p_1\,\df^4 p_2}{(2\pi)^2}\,\delta(p_1^2) \,\delta(p_2^2)\,
\delta^4\Bigl(q^-\frac{n_a}{2} + q^+\frac{n_b}{2} - p_1 - p_2 \Bigr)
\nn\\ & \qquad
= \frac{1}{32\,\pi^2}\int\!\frac{\df\Delta y\,\df\varphi}{1+\cosh\Delta y}
\,,\end{align}
%%%
where $\varphi$ is the azimuthal angle of the leptons in the transverse plane and $\Delta y$ is the rapidity difference of the two leptons:
%%%
\begin{equation}
y_i = \frac{1}{2}\ln\frac{n_b\sdt p_i}{n_a\sdt p_i}
\,,\quad
\Delta y = y_1 - y_2
\,.\end{equation}
%%%
Since we expanded $\vec{q}_T = 0$, the leptons are back-to-back in the
transverse plane, which implies that at the order we are working
%%%
\begin{align}
\frac{p_1^+}{p_2^-} &= \frac{p_2^+}{p_1^-} = \frac{q^+}{q^-}
\,,&
Y &= \frac{1}{2}\,(y_1 + y_2)
\,,\nn\\
\vec{p}_{1T} &= - \vec{p}_{2T}
\,,&
\vec{p}_{1T}^2 &= \vec{p}_{2T}^2 = \frac{q^2}{2(1+\cosh\Delta y)}
\,.\end{align}
%%%
Thus, the leptonic kinematics is described by the four independent variables
$\{q^2, Y, \Delta y, \varphi\}$, with $\{Y, \Delta y\}$ being equivalent to
$\{y_1, y_2\}$. For simplicity, we assume that we do not distinguish the two
leptons, as one would for example by measuring their rapidities $y_i$ or
transverse momenta $p_{iT}$. We can then integrate over $0\leq\varphi\leq 2\pi$
and $-\infty < \Delta y < \infty$ in \eq{leptonPS}, giving an overall factor of
$4\pi$. The leptonic tensor, \eq{Lmunu_def}, now becomes
%%%
\begin{align} \label{eq:LmunuDY}
&L^{\mu\nu}_{hh'ff'}(q^2, Y)
\nn\\ & \quad
= \frac{1}{32\pi^2} \int\!\frac{\df\Delta y\,\df\varphi}{1+\cosh\Delta y}\,
\sum_\mathrm{spins} L^{\dagger\,\mu}_{hf}(p_1, p_2)\, L^\nu_{h'f'}(p_1, p_2)
\nn\\ &\quad
= \frac{8\pi \alem^2}{3q^2}
   \Bigl(\frac{q^\mu q^\nu}{q^2} - g^{\mu\nu}\Bigr) L_{hh'ff'}(q^2)
\,,\end{align}
%%%
where
%%%
\begin{align} \label{eq:LDY}
L_{VVff'}(q^2)
&= Q_f Q_{f'} - \frac{(Q_f v_{f'} + v_f Q_{f'}) v_\ell}{1-m_Z^2/q^2}
\nn\\ & \quad
+ \frac{v_f v_{f'}(v_\ell^2+a_\ell^2)}{(1-m_Z^2/q^2)^2}
\,,\nn \\
L_{AAff'}(q^2)
&= \frac{a_f a_{f'} (v_\ell^2 + a_\ell^2)}{(1-m_Z^2/q^2)^2}
\,,\nn \\
L_{AVff'}(q^2)
&= \frac{-a_f}{1-m_Z^2/q^2}
\biggl[-Q_{f'} v_\ell + \frac{v_{f'} (v_\ell^2+a_\ell^2)}{1-m_Z^2/q^2}\biggr]
\nn\\
&= L_{VAf'f}(q^2)
\,.\end{align}
%%%

%~~~~~~~~~~~~~~~~~~~~~~~~~~~~~~~~~~~~~~~~~~~~~~~~~~~~~~~~~~~~~~~~~~~~~~~~~~~~~~~
\subsubsection{The Hard Function}
%~~~~~~~~~~~~~~~~~~~~~~~~~~~~~~~~~~~~~~~~~~~~~~~~~~~~~~~~~~~~~~~~~~~~~~~~~~~~~~~

Using parity and charge conjugation invariance of QCD, the matching coefficients
for the vector and axial-vector QCD currents can be written as
%%%
\begin{align} \label{eq:CDY}
C_{V f\,q\bar q}^{\mu\,\alpha\beta}(\lb_a, \lb_b)
&= C_{V fq}(q^2)\, (\gamma^\mu_\perp)^{\alpha\beta}
\,,\nn\\
C_{A f\,q\bar q}^{\mu\,\alpha\beta}(\lb_a, \lb_b)
&= C_{A fq}(q^2)\, (\gamma^\mu_\perp\gamma_5)^{\alpha\beta}
\,,\nn\\
C_{A f\,gg}^{\mu\,\rho\sigma}(\lb_a, \lb_b)
&= C_{Ag}(q^2)\, (\lb_a + \lb_b)^\mu\, \img \eps^{\rho\sigma}{}_{\lambda\kappa} \lb_a^\lambda \lb_b^\kappa
\,.\end{align}
%%%
By Lorentz invariance (or reparametrization invariance of $n_{a,b}$ and
$\bn_{a,b}$~\cite{Manohar:2002fd}), the scalar coefficients can only depend on
$\lb_a\cdot \lb_b = x_a x_b \Ecm^2 = q^2$.  In principle, parity and charge
conjugation would also allow the Dirac structures
$(\lb_a-\lb_b)^\mu\,\delta^{\alpha\beta}$ and
$(\lb_a-\lb_b)^\mu\,(\gamma_5)^{\alpha\beta}$. However, as the vector and
axial-vector currents are chiral even and the matching from QCD conserves
chirality for massless quarks, these cannot be generated. For the gluon
operator, the symmetry of the Wilson coefficient [see \eq{Cgg_symmetry}]
requires it to be proportional to $q^\mu = \lb_a^\mu + \lb_b^\mu$. Current
conservation for the vector current requires $q_\mu C_{Vfq\bar q}^\mu = 0$,
which eliminates this term. Thus, as expected, the only contribution for the
gluon operator is due to the axial anomaly, coming from the diagram in
\fig{DYBglue}. Since we neglect the lepton masses, $q_\mu L^\mu_{Af} = 0$, and
thus $C_{Af gg}$ does not survive the contraction of the leptonic and hadronic
tensors for $L = \ell^+\ell^-$. Hence, the gluon beam functions do not
contribute to Drell-Yan, and the gluon PDF only appears through its contribution
to the quark beam functions. Inserting \eq{CDY} into the general expression for
the hard function in \eq{Hqq_def}, we obtain
%%%
\begin{widetext}
\begin{align} \label{eq:HqqDY}
H_{hh' ff'\,q \bar q}^{\mu \nu}(\lb_a, \lb_b)
&= -\frac{1}{2N_c} \Bigl[g^{\mu\nu} - \frac{1}{2}(n_a^\mu n_b^\nu + n_a^\nu n_b^\mu) \Bigr]
C^*_{hfq}(q^2) C_{h' f'q}(q^2)
&&(\text{for}\quad hh' = \{VV, AA\})
\,,\nn\\
H_{hh'ff'\,q \bar q}^{\mu\nu}(\lb_a, \lb_b)
&= \frac{1}{4N_c} \,\img \epsilon^{\mu\nu}{}_{\lambda\kappa} n_a^\lambda n_b^\kappa\, C^*_{hfq}(q^2) C_{h'f'q}(q^2)
&&(\text{for}\quad hh' = \{VA, AV\})
\,.\end{align}
%%%

At one loop, the vector and axial-vector coefficients are equal and diagonal in
flavor and the SCET matching computation was performed in
Refs.~\cite{Manohar:2003vb, Bauer:2003di}, in agreement with the one-loop form factors
%%%
\begin{equation} \label{eq:CDY_oneloop}
C_{Vfq}(q^2) = C_{A fq}(q^2) = \delta_{f q} C(q^2)
\,,\quad
C(q^2, \mu) = 1 + \frac{\alpha_s(\mu)\,C_F}{4\pi} \biggl[-\ln^2 \Bigl(\frac{-q^2-\img 0}{\mu^2}\Bigr) + 3 \ln \Bigl(\frac{-q^2-\img 0}{\mu^2}\Bigr) - 8 + \frac{\pi^2}{6} \biggr]
\,.\end{equation}
%%%
The vector current coefficient at two loops was obtained in
Refs.~\cite{Idilbi:2006dg, Becher:2006mr} from the known two-loop quark form
factor~\cite{Kramer:1986sg, Matsuura:1987wt, Matsuura:1988sm, Gehrmann:2005pd}.
Starting at three loops, it can have a contribution that is not diagonal in
flavor, i.e., is not proportional to $\delta_{fq}$. The axial-vector coefficient
can also receive additional diagonal and nondiagonal contributions starting at
two loops from the axial anomaly~\cite{Kniehl:1989bb, Kniehl:1989qu,
  Bernreuther:2005rw}. The anomaly contributions cancel in the final result in
the sum over $f$ as long as one sums over massless quark doublets. Therefore,
they will cancel when the hard matching scale is much larger than the top-quark
mass, in which case the top quark can be treated as massless. On the other hand,
they have to be taken into account when the matching scale is below the
top-quark mass, in which case the top quark is integrated out during the
matching step and its mass cannot be neglected.

Combining \eqs{HqqDY}{CDY_oneloop} with \eqs{LmunuDY}{LDY}, the coefficients
$H_{ij}(q^2, Y)$ in \eq{Hij} become
%%%
\begin{equation}
\frac{1}{2\Ecm^2}\,\frac{8\pi\,\alem^2}{3q^2}\,\frac{1}{N_c}\sum_{ff'} \bigl[L_{VV ff'}(q^2) C^*_{Vfq}(q^2) C_{Vf'q}(q^2) + L_{AA ff'}(q^2) C^*_{Afq}(q^2) C_{Af'q}(q^2)
\bigr]
\equiv\sigma_0 H_{q\bar q}(q^2, \mu)
\,,\end{equation}
%%%
where at one loop
%%%
\begin{align} \label{eq:H_oneloop}
\sigma_0 &= \frac{4\pi\alem^2}{3N_c\Ecm^2 q^2}
\,,
& H_{q\bar q}(q^2, \mu) = H_{\bar q q}(q^2, \mu) &=
\biggl[ Q_q^2 + \frac{(v_q^2 + a_q^2) (v_\ell^2+a_\ell^2) - 2 Q_q v_q v_\ell (1-m_Z^2/q^2)}
{(1-m_Z^2/q^2)^2 + m_Z^2 \Gamma_Z^2/q^4} \biggr] \Abs{C(q^2, \mu)}^2
\,,\end{align}
%%%
with $\abs{C(q^2,\mu)}^2$ given by \eq{CDY_oneloop}, and where we also included the
nonzero width of the $Z$. The RGE for the hard function $H_{q\bar q}(q^2, \mu)$
is
%%%
\begin{equation} \label{eq:H_RGE}
\mu\,\frac{\df H_{q\bar q}(q^2, \mu)}{\df\mu} = \gamma_H(q^2, \mu)\, H_{q\bar q}(q^2, \mu)
\,,\qquad
\gamma_H(q^2, \mu) =
2\,\Gamma_\cusp[\alpha_s(\mu)] \ln\frac{q^2}{\mu^2} + \gamma_H[\alpha_s(\mu)]
\,,\end{equation}
%%%
where $\Gamma_\cusp$ is the universal cusp anomalous dimension~\cite{Korchemsky:1987wg}, and the one-loop non-cusp term is $\gamma_H[\alpha_s(\mu)] = - 3\alpha_s(\mu)\, C_F/\pi$~\cite{Manohar:2003vb}. The solution of \eq{H_RGE} has the standard form
%%%
\begin{equation} \label{eq:Hmatchrun}
H_{q\bar q}(q^2, \mu) = H_{q\bar q}(q^2, \mu_0)\, U_H(q^2, \mu_0, \mu)
\,,\qquad
U_H(q^2, \mu_0, \mu) = e^{K_H(\mu_0, \mu)} \Bigl(\frac{q^2}{\mu_0^2}\Bigr)^{\eta_H(\mu_0, \mu)}
\,,\end{equation}
%%%
where $K_H(\mu_0, \mu)$ and $\eta_H(\mu_0, \mu)$ are analogous to \eq{KetaB_def},
%%%
\begin{equation} \label{eq:KetaH}
K_H(\mu_0, \mu)
= \int_{\alpha_s(\mu_0)}^{\alpha_s(\mu)}\!\frac{\df\alpha_s}{\beta(\alpha_s)}\,
\biggl[-4\, \Gamma_\cusp(\alpha_s) \int_{\alpha_s(\mu_0)}^{\alpha_s} \frac{\df \alpha_s'}{\beta(\alpha_s')}
   + \gamma_H(\alpha_s) \biggr]
\,,\qquad
\eta_H(\mu_0, \mu)
= 2 \int_{\alpha_s(\mu_0)}^{\alpha_s(\mu)}\!\frac{\df\alpha_s}{\beta(\alpha_s)}\, \Gamma_\cusp(\alpha_s)
\,.\end{equation}
%%%
Together, \eqs{Hmatchrun}{KetaH} sum the large logarithms occurring in isolated Drell-Yan between the scales $\mu_H$ and $\mu_B$. Electroweak corrections to the hard function $H_{q\bar q}(q^2, \mu)$ can be included using the results of Refs.~\cite{Chiu:2007dg, Chiu:2008vv, Chiu:2009mg}.

%~~~~~~~~~~~~~~~~~~~~~~~~~~~~~~~~~~~~~~~~~~~~~~~~~~~~~~~~~~~~~~~~~~~~~~~~~~~~~~~
\subsubsection{The $q\bar{q}$ Soft Function}
%~~~~~~~~~~~~~~~~~~~~~~~~~~~~~~~~~~~~~~~~~~~~~~~~~~~~~~~~~~~~~~~~~~~~~~~~~~~~~~~

The incoming hemisphere soft function contains incoming Wilson lines stretching
from $-\infty$ to $0$ along $n_a$ and $n_b$. Under time reversal, each incoming
Wilson line transforms into a corresponding outgoing Wilson line stretching from
$0$ to $\infty$ along the opposite direction,
%%%
\begin{equation}
\mathrm{T}^{-1} Y_{n_a} \mathrm{T}
= \overline{P}\exp\biggl[-\img g\intlim{0}{\infty}{s} n_b\sdt A_\soft(s\,n_b) \biggr]
= \widetilde{Y}_{n_b}
\,,\end{equation}
%%%
where $\overline{P}$ denotes anti-path ordering. Since $\mathrm{T}$ itself does
not affect the original ordering of the field operators, time ordering turns
into anti-time ordering and vice versa. In addition $\mathrm{T}\,
n_a\sdt\hp_{a,s}\, \mathrm{T}^{-1} = n_b\sdt\hp_{b,s}$. Therefore, time-reversal
invariance implies
%%%
\begin{align} \label{eq:Sqq_T}
S_\hemiin^{q\bar q}(k_a^+, k_b^+)
&\stackrel{\mathrm{T}}{=}
\frac{1}{N_c}\, \tr\,\Mae{0}{ T\bigl[\widetilde{Y}^\dagger_{n_b} \widetilde{Y}_{n_a} \bigr]
 \delta(k_a^+ - n_b\sdt\hp_{b,s})\, \delta(k_b^+ - n_a\sdt\hp_{a,s})\,
  \bT\bigl[\widetilde{Y}^\dagger_{n_a} \widetilde{Y}_{n_b} \bigr]}{0}^*
\nn\\
&= \frac{1}{N_c}\, \tr\,\Mae{0}{ T\bigl[\widetilde{Y}^\dagger_{n_a} \widetilde{Y}_{n_b} \bigr]
 \delta(k_a^+ - n_a\sdt\hp_{a,s})\, \delta(k_b^+ - n_b\sdt\hp_{b,s})\,
  \bT\bigl[\widetilde{Y}^\dagger_{n_b} \widetilde{Y}_{n_a} \bigr]}{0}
\,.\end{align}
%%%
In the second step, the complex conjugation has no effect since the matrix
element is real, and we used parity to switch $n_{b,a}$ back to $n_{a,b}$. For
comparison, the hemisphere soft function with outgoing Wilson appearing in the
double-differential hemisphere invariant-mass distribution in $e^+e^- \to 2$
jets~\cite{Korchemsky:1999kt, Korchemsky:2000kp, Fleming:2007qr, Schwartz:2007ib,
Fleming:2007xt, Hoang:2008fs} is
%%%
\begin{equation}
S_\hemiout^{q\bar q}(k_a^+, k_b^+)
= \frac{1}{N_c}\, \tr\,\Mae{0}{ \bT\bigl[\widetilde{Y}^\dagger_{n_a} \widetilde{Y}_{n_b} \bigr]
 \delta(k_a^+ - n_a\sdt\hp_{a,s})\, \delta(k_b^+ - n_b\sdt\hp_{b,s})\,
  T\bigl[\widetilde{Y}^\dagger_{n_b} \widetilde{Y}_{n_a} \bigr]}{0}
\,.\end{equation}
%%%
This is almost the same as \eq{Sqq_T}, the only difference being the opposite
time ordering. Thus, $S_\hemiin$ and $S_\hemiout$ are equal at one loop, where
the time ordering is still irrelevant. Beyond one loop, $S_\hemiin$ and
$S_\hemiout$ may in general be different.
However, since the beam and jet functions have the same anomalous dimension,
the combined anomalous dimension of the hard and beam functions in isolated
Drell-Yan agrees with that of the hard and jet functions for the $e^+e^-$
hemisphere invariant-mass distribution. The consistency of the RGE in both cases
then requires that $S_\hemiin$ and $S_\hemiout$ have the same anomalous
dimension to all orders in perturbation theory. In addition, the purely virtual
contributions, obtained by inserting the vacuum state, are the same in both
cases,
%%%
\begin{equation}
S_\hemiin^{q\bar q,\mathrm{virtual}}(k_a^+, k_b^+)
= \frac{1}{N_c}\, \delta(k_a^+)\, \delta(k_b^+)\,
\tr\,\Abs{\Mae{0}{ T\bigl[\widetilde{Y}^\dagger_{n_a} \widetilde{Y}_{n_b} \bigr] }{0}}^2
= S_\hemiout^{q\bar q,\mathrm{virtual}}(k_a^+, k_b^+)
\,.\end{equation}
%%%
Hence, $S_\hemiin$ and $S_\hemiout$ can only differ by finite real-emission
corrections at each order in perturbation theory.

Using the one-loop results for $S_\hemiout^{q\bar q}$ from
Refs.~\cite{Schwartz:2007ib, Fleming:2007xt}, we have
%%%
\begin{align} \label{eq:S_oneloop}
S^{q \bar q}_\hemiin(k_a^+, k_b^+) &= \delta(k_a^+)\,\delta(k_b^+)
+ \delta(k_a^+)\,S^\oneloop(k_b^+) + S^\oneloop(k_a^+)\,\delta(k_b^+)
\,,\nn\\
S^\oneloop(k^+) &= \frac{\alpha_s(\mu)\,C_F}{4\pi} \biggl\{
-\frac{8}{\mu} \biggl[\frac{\theta(k^+ / \mu) \ln(k^+ / \mu)}{k^+ / \mu} \biggr]_+ +
\frac{\pi^2}{6}\, \delta(k^+) \biggr\}
\,.\end{align}
%%%
The plus distribution is defined in \eq{plus_def}. The one-loop soft function
for beam thrust in \eq{SB} then becomes $S_B(k^+, \mu) = \delta(k^+) + 2\,
S^\oneloop(k^+)$.

%~~~~~~~~~~~~~~~~~~~~~~~~~~~~~~~~~~~~~~~~~~~~~~~~~~~~~~~~~~~~~~~~~~~~~~~~~~~~~~~
\subsubsection{Final Cross Section for Beam Thrust}
%~~~~~~~~~~~~~~~~~~~~~~~~~~~~~~~~~~~~~~~~~~~~~~~~~~~~~~~~~~~~~~~~~~~~~~~~~~~~~~~

The differential cross section for beam thrust in \eq{dsigma_tauB} including the RGE running is
%%%
\begin{align} \label{eq:dsigma_tauB_final}
\frac{\df\sigma}{\df q^2\, \df Y\, \df \tau_B}
&= \sigma_0 \sum_{ij} H_{ij}(q^2, \mu_H)\, U_H(q^2, \mu_H, \mu_S)
\int\!\df t_a\,\df t_b\,Q\,S_B\Bigl(Q\,\tau_B - \frac{t_a + t_b}{Q}, \mu_S \Bigr)
\nn\\ &\quad \times
\int\!\df t_a'\, B_i(t_a - t_a', x_a, \mu_B)\, U_B(t_a', \mu_B, \mu_S)
\int\!\df t_b'\, B_j(t_b - t_b', x_b, \mu_B)\, U_B(t_b', \mu_B, \mu_S)
\,.\end{align}
%%%
For simplicity, we evolve the hard and beam functions from their respective hard
and beam scales, $\mu_H$ and $\mu_B$, down to the common scale $\mu = \mu_S$ of the soft
function. In this way, we do not need to consider the running of the soft function
separately. Different choices for $\mu$ are all equivalent, as we discussed in \subsec{RGE}.
At LL, we include the
one-loop cusp anomalous dimension in the evolution kernels $U_H$ and $U_B$, and
at NLL we include the two-loop cusp and one-loop non-cusp anomalous dimensions.
In both cases we use the LO results as initial conditions.

We also consider the fixed-order $\alpha_s$ expansion. To our knowledge
$\df\sigma/\df q^2 \df Y \df \tau_B$ has not been considered in perturbation
theory in full QCD even at one loop. To obtain an expression for $\df\sigma/\df
q^2 \df Y \df \tau_B$ at NLO in $\alpha_s$ and leading order in the power
counting, we drop the evolution factors $U_H$ and $U_B$ and expand all functions
to NLO at a common scale $\mu$. From the above NLO results for the hard and soft
functions and the NLO results for the beam functions from \sec{beamfunction}, we
find
%%%
\begin{align} \label{eq:dsigma_tauB_NLO}
\frac{\df\sigma}{\df q^2\, \df Y\, \df \tau_B}
&= \sigma_0 \sum_{i,j}
\biggl[ Q_i^2 + \frac{(v_i^2 + a_i^2) (v_\ell^2+a_\ell^2) - 2 Q_i v_i v_\ell (1-m_Z^2/q^2)}
{(1-m_Z^2/q^2)^2 + m_Z^2 \Gamma_Z^2/q^4} \biggr]
\nn\\ & \quad \times
\int\! \frac{\df \xi_a}{\xi_a}\, \frac{\df \xi_b}{\xi_b}\,
C_{ij}\Bigl(\frac{x_a}{\xi_a},\frac{x_b}{\xi_b}, q^2, \tau_B, \mu \Bigr)\, f_{i/a}(\xi_a,\mu)\, f_{j/b}(\xi_b,\mu)
\,.\end{align}
%%%
Here, $f_{i/a}(\xi_a, \mu)$ and $f_{j/b}(\xi_b, \mu)$ are the PDFs for parton $i$ in proton $a$ and parton $j$ in (anti-)proton $b$. At tree level, the nonzero coefficients are
%%%
\begin{equation}
C_{q\bar q}^\tree(z_a, z_b, q^2, \tau_B, \mu) = C_{\bar q q}^\tree(z_a, z_b, q^2, \tau_B, \mu)
= \delta(\tau_B)\,\delta(1 - z_a)\,\delta(1 - z_b)
\,.\end{equation}
%%%
At one loop, we obtain
%%%
\begin{align}
C^\oneloop_{q \bar q}(z_a,z_b, q^2, \tau_B, \mu)
&= \frac{\alpha_s(\mu)\, C_F}{2\pi}\, \delta(1 - z_a)\,\theta(z_b) \biggl\{
\biggl[-2 \biggl[\frac{\theta(\tau_B) \ln\tau_B}{\tau_B}\biggr]_+
- \frac{3}{2} \biggl[\frac{\theta(\tau_B)}{\tau_B}\biggr]_+
- \delta(\tau_B) \Bigl(4 - \frac{\pi^2}{2}\Bigr) \biggr] \delta(1 - z_b)
\nn\\ &\quad
+ \biggl[\biggl[\frac{\theta(\tau_B)}{\tau_B}\biggr]_+ + \delta(\tau_B) \ln\frac{q^2}{\mu^2} \biggr] \biggl[\theta(1-z_b)\frac{1 + z_b^2}{1-z_b}\biggr]_+
\nn\\ &\quad
  + \delta(\tau_B) \biggl[
   \biggl[\frac{\theta(1-z_b) \ln(1-z_b)}{1-z_b}\biggr]_+(1+z_b^2)
   + \theta(1-z_b)\biggl(1-z_b - \frac{1 + z_b^2}{1-z_b} \ln z_b\biggr)
   \biggr] \biggr\}  + (z_a\lra z_b)
\,,\nn\\
C^\oneloop_{\bar qq}(z_a,z_b, q^2, \tau_B, \mu) &= C^\oneloop_{q \bar q}(z_a,z_b, q^2, \tau_B, \mu)
\,,\nn\\[1ex]
C^\oneloop_{qg}(z_a,z_b, q^2, \tau_B, \mu)
&= \frac{\alpha_s(\mu)\,T_F}{2\pi}\, \delta(1 - z_a)\,\theta(z_b)\, \theta(1 - z_b) \biggl\{ \biggl[
  \biggl[\frac{\theta(\tau_B)}{\tau_B}\biggr]_+
  + \delta(\tau_B) \ln\frac{q^2}{\mu^2} \biggr]\bigl[z_b^2 + (1-z_b)^2\bigr]
\nn\\ &\quad
  + \delta(\tau_B)\biggl[\ln\frac{1 - z_b}{z_b}\bigl[z_b^2 + (1-z_b)^2\bigr] + 2z_b(1-z_b) \biggr] \biggr\}
\,,\nn\\
C^\oneloop_{\bar q g}(z_a,z_b, q^2, \tau_B, \mu) &= C^\oneloop_{q g}(z_a,z_b, q^2, \tau_B, \mu)
\,,\nn\\
C^\oneloop_{gq}(z_a, z_b, q^2, \tau_B, \mu) &= C^\oneloop_{g \bar q}(z_a,z_b, q^2, \tau_B, \mu)
= C^\oneloop_{q g}(z_b,z_a, q^2, \tau_B, \mu)
\,.\end{align}
\end{widetext}
%%%
The coefficient $C_{gg}$ only starts to contribute at two loops. The single
logarithms of $q^2/\mu^2$ are multiplied by the QCD splitting kernels and are
resummed by the PDFs. Thus, in fixed-order perturbation theory the PDFs should
be evaluated at the hard scale $\mu = Q$, such that there are no large
logarithms when integrating over $0 \leq \tau_B \lesssim 1$. However, if the
integration is restricted to $\tau_B \leq \tau_B^\cut \ll 1$, the plus
distributions in $\tau_B$ produce large logarithms $\ln^2\tau_B^\cut$ and
$\ln\tau_B^\cut$, which make a fixed-order expansion unreliable. These are
precisely the logarithms that are resummed by the
combined RGE of hard, jet, and soft functions in \eq{dsigma_tauB_final}.

%%%%%%%%%%%%%%%%%%%%%%%%%%%%%%%%%%%%%%%%%%%%%%%%%%%%%%%%%%%%%%%%%%%%%%%%%%%%%%%%
\section{Isolated Drell-Yan Cross Section}
\label{sec:summary}
%%%%%%%%%%%%%%%%%%%%%%%%%%%%%%%%%%%%%%%%%%%%%%%%%%%%%%%%%%%%%%%%%%%%%%%%%%%%%%%%

\begin{figure*}[t!]
\includegraphics[width=\columnwidth]{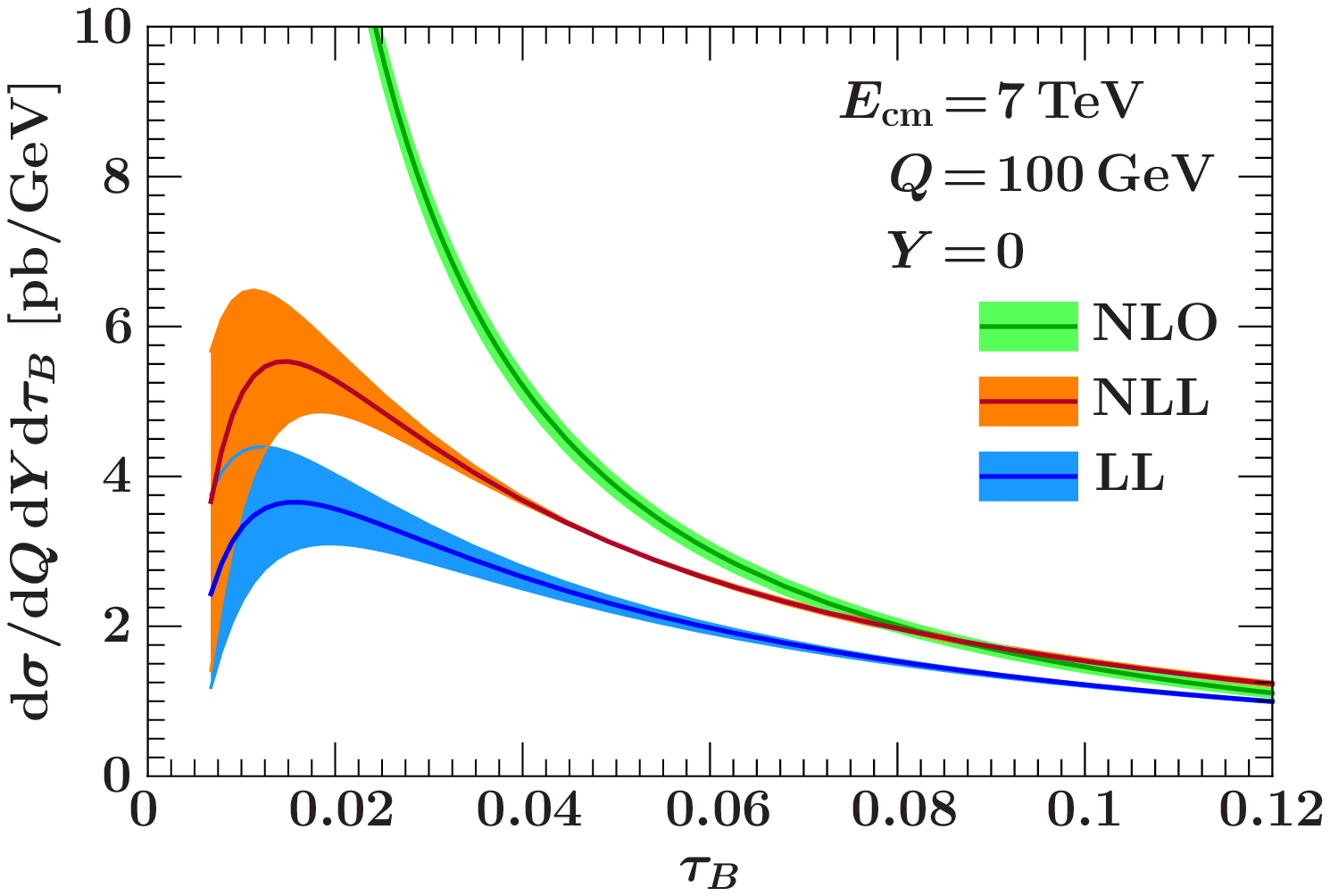}%
\hspace{\columnsep}%
\includegraphics[width=\columnwidth]{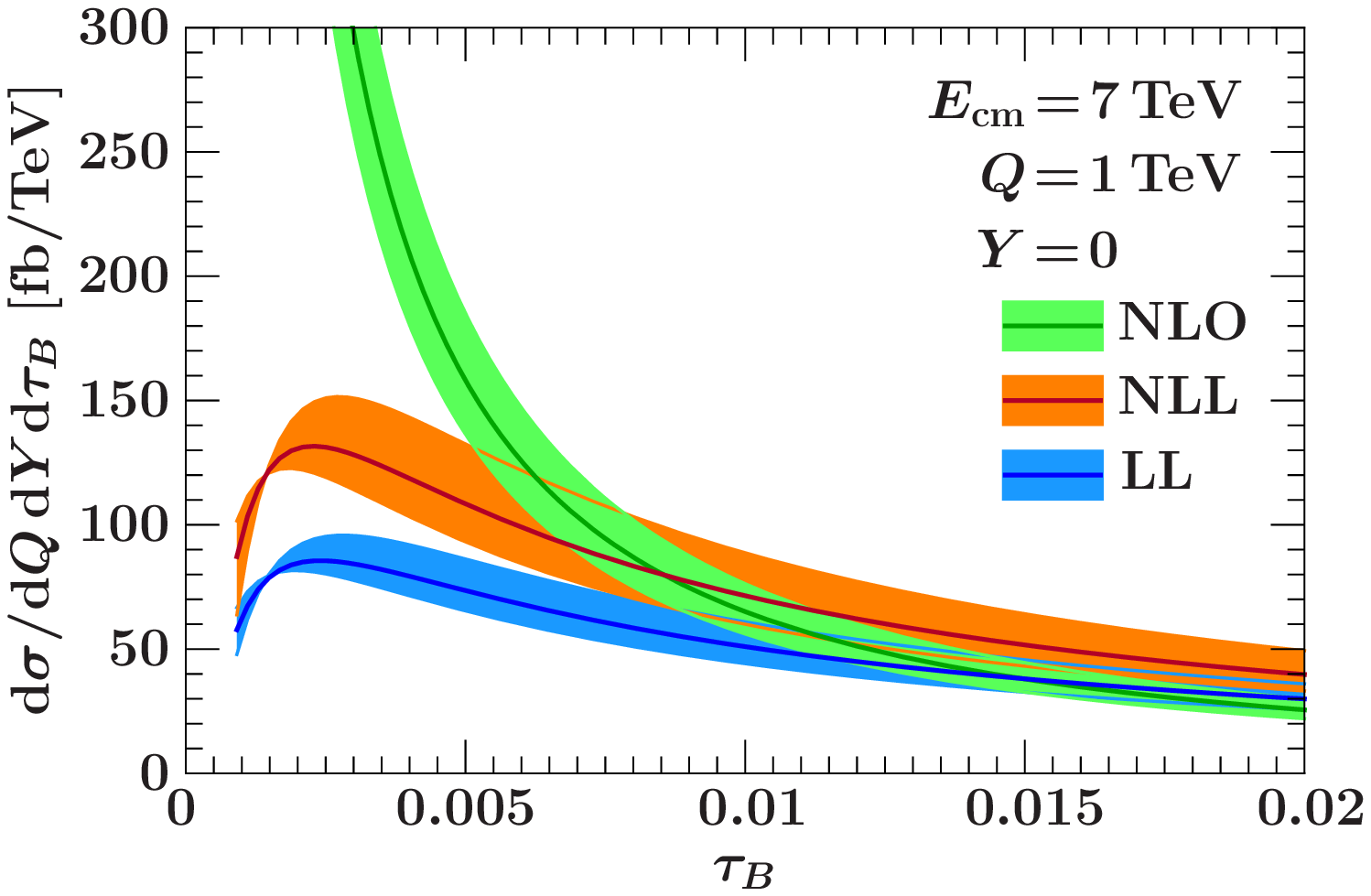}
\includegraphics[width=\columnwidth]{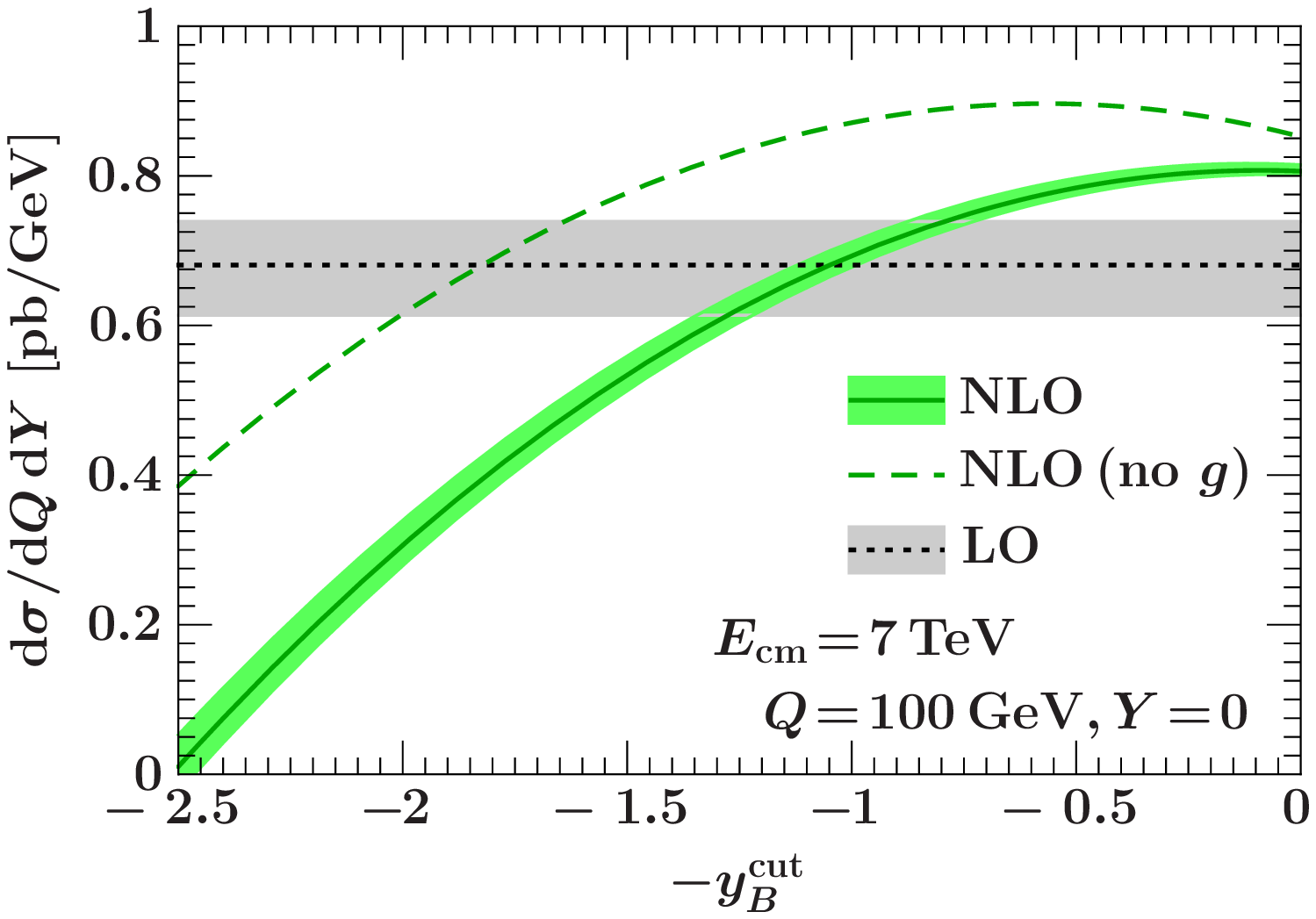}%
\hspace{\columnsep}%
\includegraphics[width=\columnwidth]{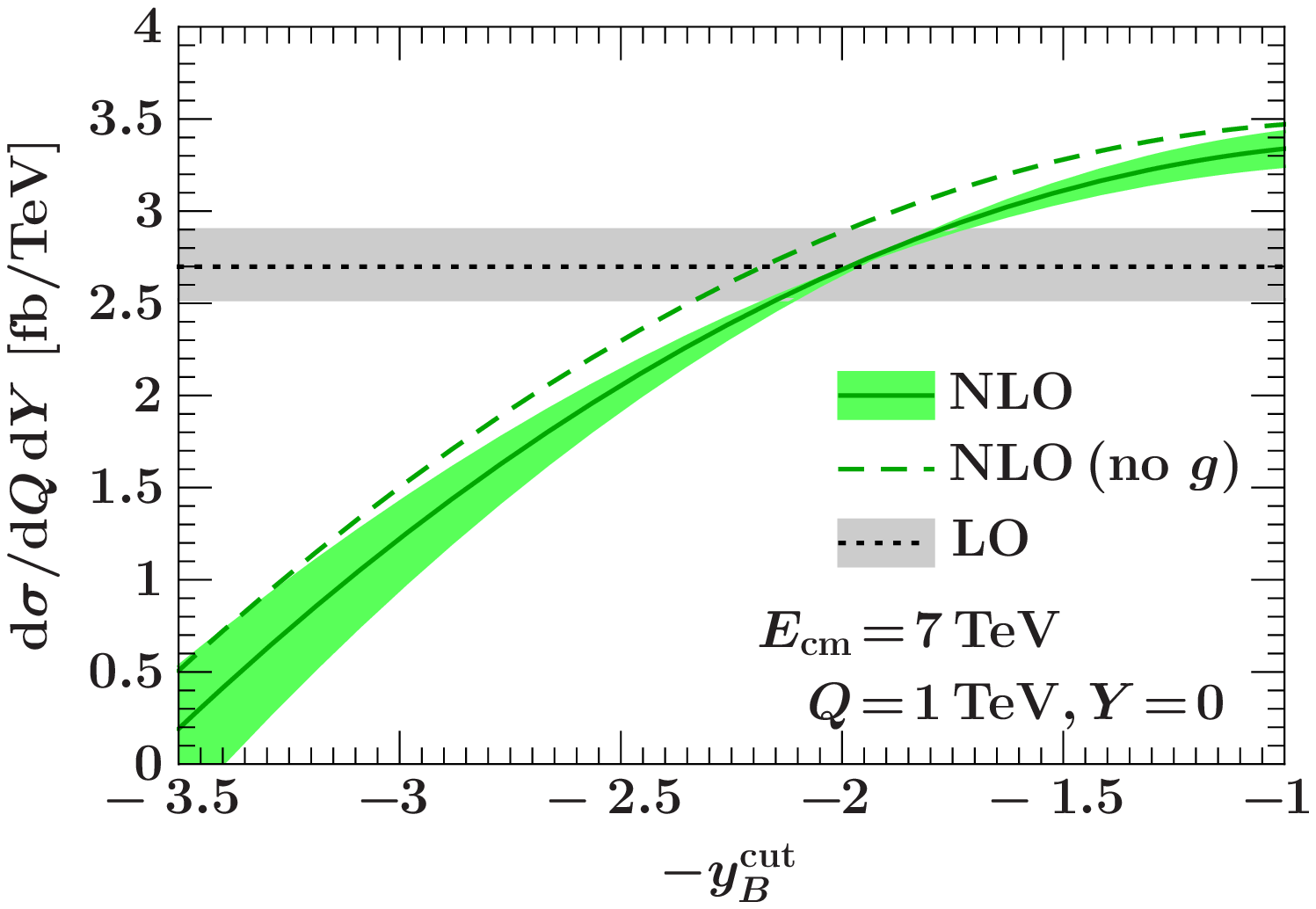}
\includegraphics[width=\columnwidth]{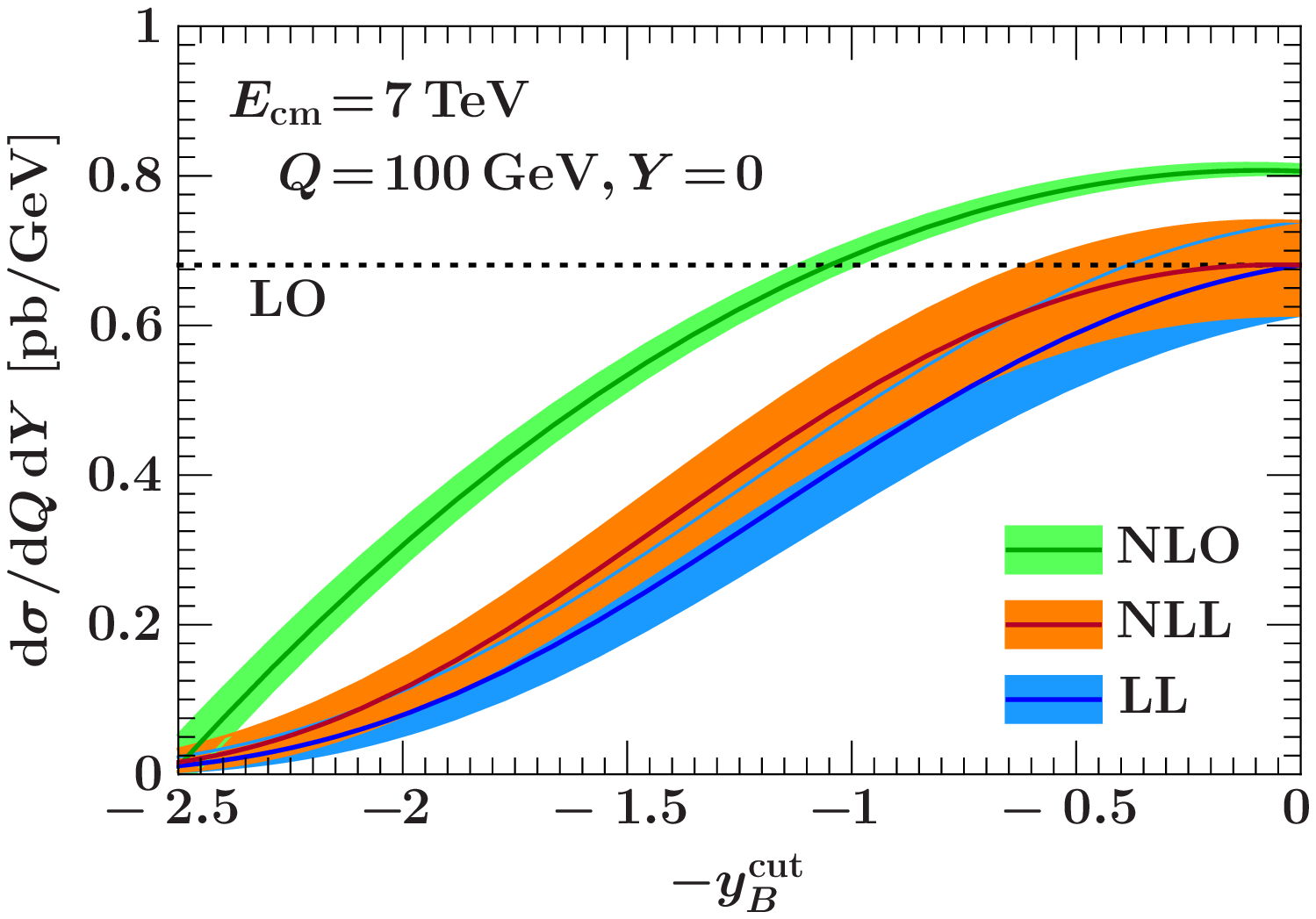}%
\hspace{\columnsep}%
\includegraphics[width=\columnwidth]{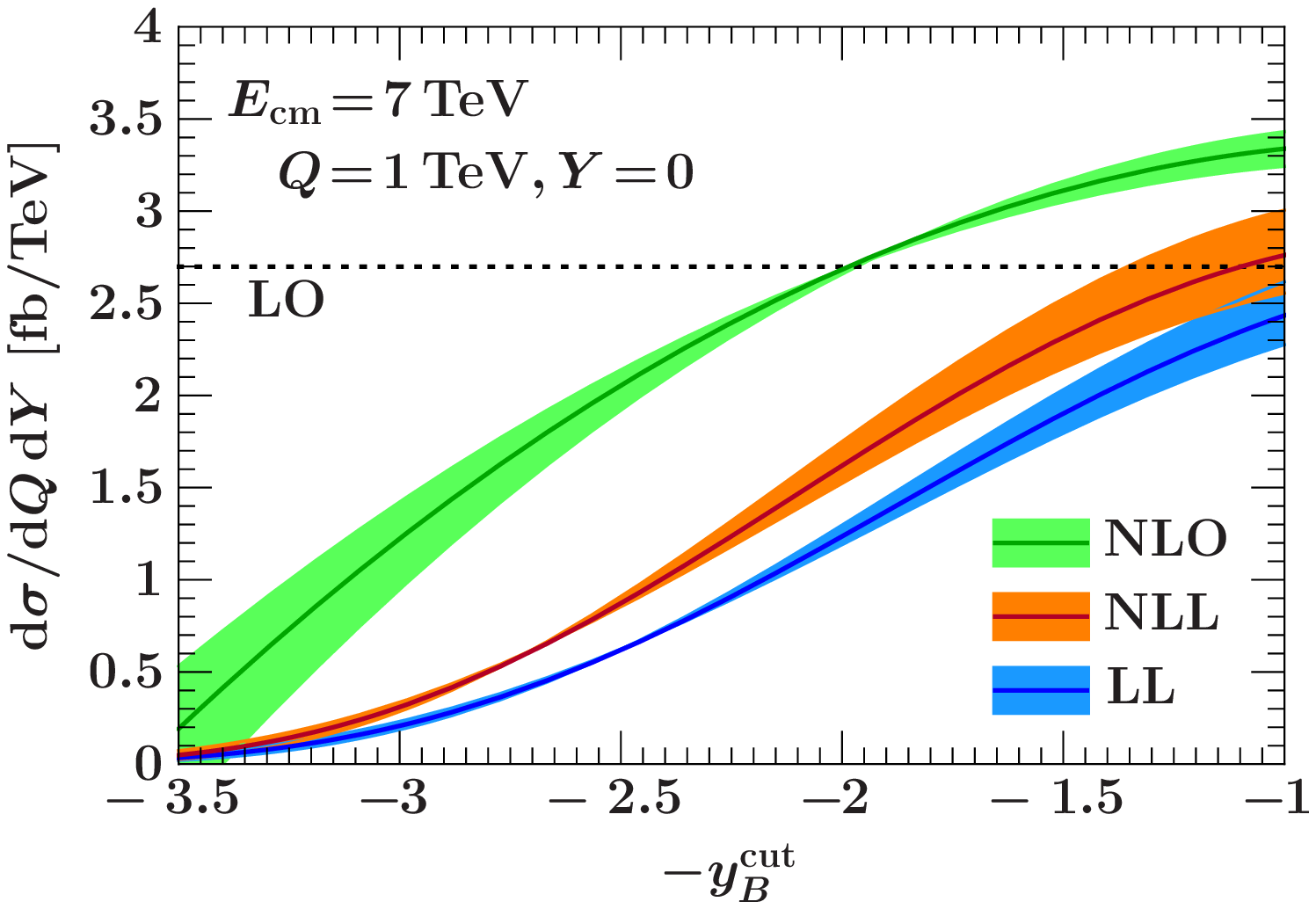}
\caption{Cross sections for beam thrust at the LHC with $\Ecm = 7\TeV$ at $Y = 0$ and $Q = 100\GeV$ (left column) and $Q = 1\TeV$ (right column). Top row: The cross section differential in $\tau_B$ at NLO, LL, and NLL. Middle and bottom rows: The cross section integrated up to $\tau_B \leq \exp(-2y_B^\cut)$ at LO, NLO, LL, and NLL.}
\label{fig:sigmaLHC}
\end{figure*}

\begin{figure*}
\includegraphics[width=\columnwidth]{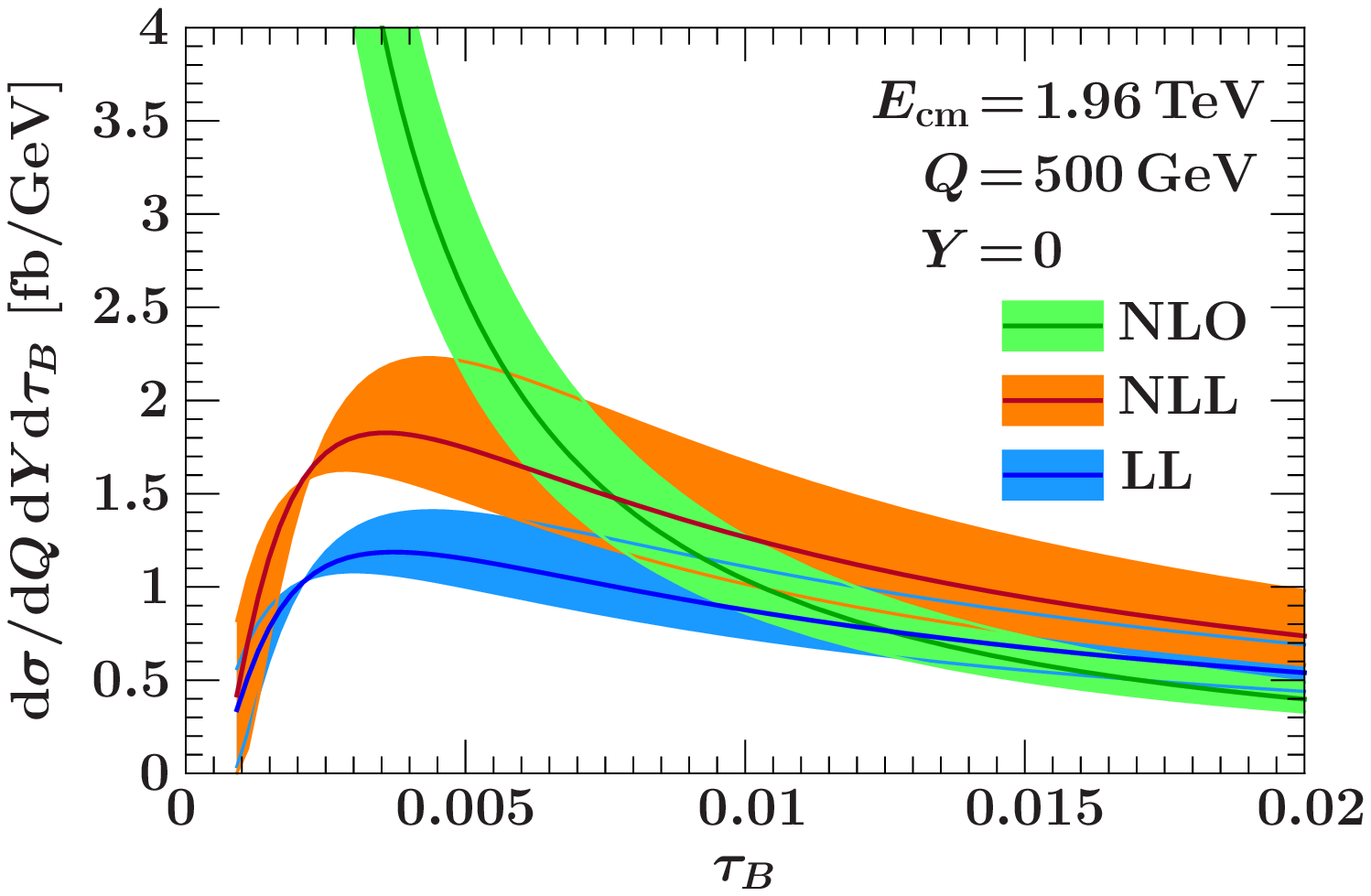}%
\hspace{\columnsep}%
\includegraphics[width=\columnwidth]{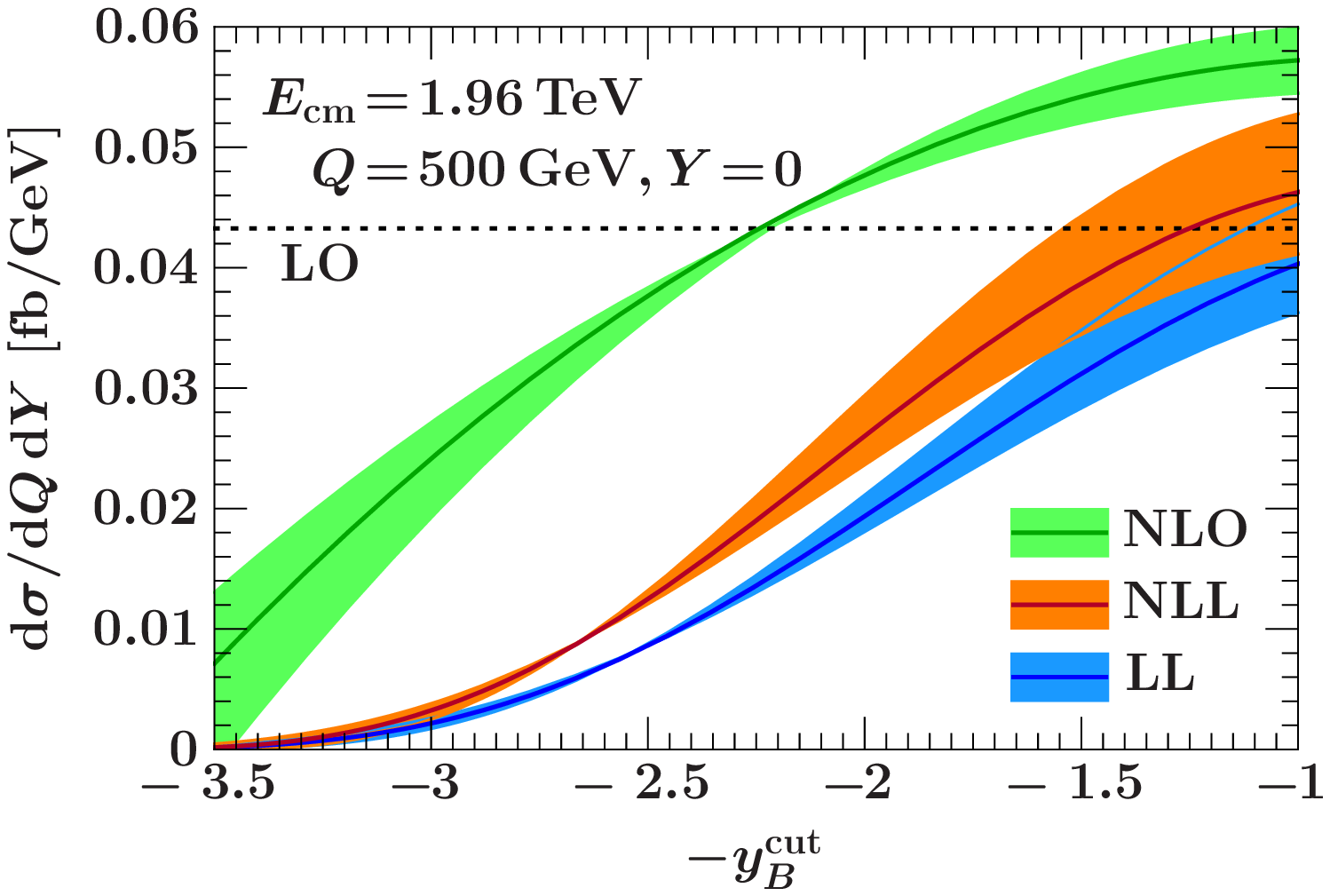}
\caption{The cross section for beam thrust at the Tevatron at $Y = 0$ and $Q = 500\GeV$. Left panel: The cross section differential in $\tau_B$ at NLO, LL, and NLL. Right panel: The cross section integrated up to $\tau_B \leq \exp(-2y_B^\cut)$ at LO, NLO, LL, and NLL.}
\label{fig:sigmaTevatron}
\end{figure*}

In this section we illustrate our results with plots of the isolated Drell-Yan cross section.
Rather than considering the cross section as a function of two
variables, $B_a^+$ and $B_b^+$, we use the beam thrust $\tau_B$ defined in \subsec{generalobs}.
We consider both the differential cross section $\df\sigma/\df Q\df Y\df \tau_B$
as a function of $\tau_B$, see \eq{dsigma_tauB}, as
well as the cross section integrated over $0\le \tau_B \le \exp(-2
y_B^\cut)$ as a function of $y_B^\cut$, see \eqs{tauBycut}{sigma_yBcut}.
We fix $Q = \sqrt{q^2}$ to a few representative
values. We also restrict our discussion to back-to-back leptons, $Y = 0$,
since measuring $\tau_B$ mainly affects the normalization and not the shape
of the rapidity distribution.

For our cross-section predictions, we always use the NLO PDFs from MSTW2008~\cite{Martin:2009iq},
with the corresponding $\alpha_s(m_Z) = 0.1202$
and two-loop five-flavor running for $\alpha_s(\mu)$.

We show results for the cross section both in the fixed-order expansion at LO
and NLO [see \eq{dsigma_tauB_NLO}] and in the resummed expansions at LL and NLL
[see \eq{dsigma_tauB_final}]. For the resummed results we always choose the
hard, beam, and soft scales as $\mu_H = \mu$, $\mu_B = \mu \sqrt{\tau_B}$,
$\mu_S = \mu\, \tau_B$, and for the fixed-order results we use a common fixed
scale $\mu$. The central values correspond to $\mu = Q$, and the bands show the
scale variation obtained by varying $\mu = 2Q$ and $\mu = Q/2$.  Since our
purpose here is to illustrate the main features of the factorized cross section
with beam functions, we will limit ourselves to the LL and NLL results without
also including additional fixed-order corrections to the hard, beam, and soft
functions. A complete analysis combining both NLO corrections and NNLL
resummation, and a more detailed analysis of scale uncertainties, is left for
future work.

In \fig{sigmaLHC}, we show results for $pp$ collisions at the LHC with $\Ecm =
7\TeV$ and two different values $Q = 100\GeV$ and $Q = 1\TeV$. For $Y=0$ this
corresponds to $x_a = x_b = 0.014$ and $x_a = x_b  = 0.14$ respectively, representing
two typical values for the isolated factorization theorem.  In
\fig{sigmaTevatron}, we show results for $p\bar p$ collisions at the Tevatron for
$Q = 500\GeV$ corresponding to $x_a = x_b = 0.26$.

The top row in \fig{sigmaLHC} depicts the LHC cross section as a function of
$\tau_B$ at several different orders. At tree level, there is only a
$\delta$ function and no radiation. The green (light) curve and band show the
NLO-expanded cross section, which grows for decreasing $\tau_B$, showing that
the radiation is peaked in the forward direction.  As $\tau_B\to 0$ it exhibits
the expected singular behavior, and this IR singularity is canceled in the
integrated cross section by a corresponding $\delta$ function at $\tau_B = 0$.
The blue (dark) and orange (medium) curves are the resummed results at LL and
NLL, respectively. As expected, the resummation has a large effect at small
$\tau_B$ and effectively regulates the IR singularity in the NLO result. The curves
are not plotted for $\tau_B\le 0.007$ in the left panel and $\tau_B\le 0.001$ in
the right panel, because at this point the soft scale drops below $1\GeV$. Near
this cutoff, the soft function becomes nonperturbative and so our purely
perturbative results should not be taken too seriously.  Nevertheless, it is
interesting to see that the resummed results show a characteristic turnover that
one would also expect from nonperturbative corrections. These nonperturbative
corrections scale as $\lqcd/(Q\tau_B)$ and hence become less relevant than the
perturbative corrections for $\tau_B\gtrsim 0.02$ and $\tau_B\gtrsim 0.002$ in
the left and right panels, respectively.

In the middle row of \fig{sigmaLHC}, we compare our results expanded to
fixed-order at LO (gray horizontal band) and NLO (green band with solid central
line) as a function of $y_B^\cut$.  Since we plot the integrated cross section,
the tree-level $\delta$ function gives a constant contribution in $y_B^\cut$. As
expected, the NLO corrections to the LO result become very large as we move away
from $y_B^\cut = 0$ because of the large double logarithms in the fixed-order
expressions. Hence, a fixed-order expansion is not reliable here. The green
dashed line illustrates what happens if we take the NLO result given by the solid
line but exclude the contribution of the gluon PDF to the quark beam
function. The gluon contribution has a bigger effect at smaller $Q$, because it
corresponds to smaller $x$.  Overall, it reduces the integrated cross section,
which is the same effect we already observed in the integrated beam function in
\sec{beamfunction}. The plots in the bottom row of \fig{sigmaLHC} show the LL,
NLL, and NLO results, which are the integrated versions of the curves in the top
row. Here, larger $y_B^\cut$ towards the left of the plot corresponds to smaller $\tau_B$.
A large $y_B^\cut$ implies a stronger constraint on the final state and hence a
smaller cross section.  In this region, the logarithmic resummation is necessary
and suppresses the cross section. For $y_B^\cut \to 0$ the LL and NLL approach
the LO result as they must.

Figure~\ref{fig:sigmaTevatron} shows the corresponding plots for the Tevatron.
The left panel and right panel are equivalent to the top row and bottom row of
\fig{sigmaLHC}, respectively showing the differential and integrated cross
sections. The plots show similar features overall, though the overall scale
uncertainties are slightly larger here, because the PDFs are evaluated at a
larger $x$. In the Tevatron case, even though we are at larger $x$ compared to
the right panels in \fig{sigmaLHC}, the cross section is larger due to the
presence of valence antiquarks (note the different units in the two plots).

%%%%%%%%%%%%%%%%%%%%%%%%%%%%%%%%%%%%%%%%%%%%%%%%%%%%%%%%%%%%%%%%%%%%%%%%%%%%%%%%
\section{Conclusions}
\label{sec:conclusions}
%%%%%%%%%%%%%%%%%%%%%%%%%%%%%%%%%%%%%%%%%%%%%%%%%%%%%%%%%%%%%%%%%%%%%%%%%%%%%%%%

Experimental measurements at the LHC or Tevatron are typically characterized by
two conditions. First, the dominant part of the cross section arises from
parton momentum fractions $x$ away from one, and second, to probe the hard scattering
the measurements impose restrictions on the final state to identify and isolate
hard leptons or jets.

Factorization is required to separate the perturbatively calculable pieces from
the nonperturbative parton distribution functions, and is a key ingredient for
the resummation of large logarithms that occur due to phase-space restrictions.
The most well-known factorization theorem for inclusive Drell-Yan applies for
generic momentum fractions, but the hadronic final state is completely summed
over, only subject to overall momentum conservation. This requires an inclusive
experimental measurement, with only mild restrictions on the final state. On
the other hand, threshold factorization theorems for Drell-Yan or dijet
production take into account phase-space restrictions and resum resulting large
logarithms, but they are only valid in the limit $x\to 1$. Thus each of these
cases satisfies only one of the above experimental conditions.

In this paper we have studied factorization for generic $x$ and with explicit
restrictions on the hadronic final state. We considered the simplest situation,
namely Drell-Yan production $pp\to X\ell^+\ell^-$ at generic $x$ with a
restriction on $X$ that vetoes hard central jets, which we call isolated
Drell-Yan. The restriction on the hadronic final state is implemented by
dividing the total hadronic momentum into two hemispheres and requiring the
components $B_a^+$ and $B_b^+$ of the resulting hemisphere momenta to be small.
For this situation, we prove a factorization theorem for the cross section
differential in the hadronic variables $B_a^+$ and $B_b^+$ (as well as the
leptonic phase-space variables). It allows us to systematically resum large
double logarithms of $B_{a,b}^+/Q$ arising from the phase-space restrictions.
The factorization theorem also applies to other isolated processes of the form
$pp\to XL$, like $H\to \gamma\gamma$ or $H\to 4\ell$.

The main conclusion from our analysis is that PDFs alone are insufficient to
properly describe the initial state in the collision. The restriction on the
hadronic final state effectively probes the proton prior to the hard interaction
by constraining the virtuality of the colliding hard parton. At that point, the
parton cannot be confined to the proton anymore and must properly be described
as part of an incoming initial-state jet. This description is given by quark and
gluon beam functions, which replace the PDFs in the factorization theorem. In
addition to the parton's momentum fraction, the beam functions depend on the
virtuality of the colliding parton and live at an intermediate beam scale
$\mu_B$ of the order of this virtuality.  At the scale $\mu=\mu_B$, the beam
functions can be matched onto PDFs evaluated at this $\mu$ times perturbatively
calculable corrections. In this way, our factorization theorem unambiguously
determines the proper scale at which the PDFs must be evaluated. This scale
$\mu_B$ is much smaller than the hard scale of the partonic collision.  As a
result, the evolution of the initial state below $\mu_B$ is governed by the
PDFs, while above $\mu_B$ it is governed by the beam function, whose evolution
depends on the parton's virtuality rather than momentum fraction. Other
differences compared to the PDF evolution are that the beam-function evolution
sums double logarithms and does not involve parton mixing.

In a Monte Carlo setting the corresponding physical effects should be described
by the initial-state parton shower. Our factorization theorem thus provides a
way to explicitly check whether the initial-state parton shower resums the
correct double logarithms, which is left for future work. We believe that
experimental measurements of the isolated Drell-Yan spectrum will provide a
direct method for testing the initial-state shower in Monte Carlo. This spectrum
combined with the results reported here, is therefore useful for tuning the
Monte Carlo with early LHC data.

Even though our derivation is only rigorous for the specific case of isolated
Drell-Yan, we argued that the necessity for beam functions is more general,
essentially applying to any process where the hadronic final state is restricted
in a similar way. For example, the consistency of the renormalization group
evolution implies that the description for any threshold process can be extended
to a respective isolated case by supplementing it with corresponding variables
$B_{a,b}^+$, replacing the PDFs by beam functions, and replacing the threshold
soft function by an appropriate isolated soft function.

We briefly discussed the extension of isolated Drell-Yan to isolated dijet
production. We also pointed out that if the hadronic final state is constrained
by different global variables, one can expect to find different beam functions
to encode these constraints, which will then sum double logarithms in these
variables.

Ultimately, we hope that this type of factorization theorem with beam functions
will bridge the gap between experimentally realistic cuts for LHC measurements
and systematically improvable theoretical results that go beyond fixed-order
calculations.

%%%%%%%%%%%%%%%%%%%%%%%%%%%%%%%%%%%%%%%%%%%%%%%%%%%%%%%%%%%%%%%%%%%%%%%%%%%%%%%%
\begin{acknowledgments}
  We thank Matthew Schwartz for urging us to keep it simple.
  We also thank Carola Berger, Zoltan Ligeti, Andre Hoang, Aneesh Manohar, Ira Rothstein,
  Kerstin Tackmann, and Jesse Thaler for comments on the manuscript.
  I.S. and F.T. thank the MPI Munich and F.T. thanks the
  CERN theory group for hospitality during part of this work.  This work was
  supported in part by the Office of Nuclear Physics of the U.S.\ Department of
  Energy under the Contract DE-FG02-94ER40818, and by a Friedrich Wilhelm Bessel
  award from the Alexander von Humboldt foundation.
\end{acknowledgments}
%%%%%%%%%%%%%%%%%%%%%%%%%%%%%%%%%%%%%%%%%%%%%%%%%%%%%%%%%%%%%%%%%%%%%%%%%%%%%%%%

%%%%%%%%%%%%%%%%%%%%%%%%%%%%%%%%%%%%%%%%%%%%%%%%%%%%%%%%%%%%%%%%%%%%%%%%%%%%%%%%

\bibliographystyle{../physrev4}
\bibliography{../pp}

\providecommand{\href}[2]{#2}\begin{thebibliography}{100}

\bibitem{Collins:1989gx}
J.~C. Collins, D.~E. Soper, and G.~Sterman,
\newblock Adv. Ser. Direct. High Energy Phys. {\bf 5}, 1 (1988),
  [\href{http://arXiv.org/abs/hep-ph/0409313}{hep-ph/0409313}].
%%CITATION = HEP-PH/0409313;%%

\bibitem{:1999fr}
ATLAS Collaboration,
\newblock (1999),
\newblock CERN-LHCC-99-15.
%%CITATION = ATLAS-TDR-15;%%

\bibitem{:1999fq}
ATLAS Collaboration,
\newblock (1999),
\newblock CERN-LHCC-99-14.
%%CITATION = ATLAS-TDR-14;%%

\bibitem{Bayatian:2006zz}
CMS Collaboration, G.~L. Bayatian {\em et~al.},
\newblock CERN-LHCC-2006-001.

\bibitem{Ball:2007zza}
CMS Collaboration, G.~L. Bayatian {\em et~al.},
\newblock J. Phys. G {\bf 34}, 995 (2007).
%%CITATION = JPHGB,G34,995;%%

\bibitem{Sterman:1986aj}
G.~Sterman,
\newblock Nucl. Phys. B {\bf 281}, 310 (1987).
%%CITATION = NUPHA,B281,310;%%

\bibitem{Catani:1989ne}
S.~Catani and L.~Trentadue,
\newblock Nucl. Phys. B {\bf 327}, 323 (1989).
%%CITATION = NUPHA,B327,323;%%

\bibitem{Kidonakis:1998bk}
N.~Kidonakis, G.~Oderda, and G.~Sterman,
\newblock Nucl. Phys. B {\bf 525}, 299 (1998),
  [\href{http://arXiv.org/abs/hep-ph/9801268}{hep-ph/9801268}].
%%CITATION = HEP-PH/9801268;%%

\bibitem{Bonciani:1998vc}
R.~Bonciani, S.~Catani, M.~L. Mangano, and P.~Nason,
\newblock Nucl. Phys. {\bf B529}, 424 (1998),
  [\href{http://arXiv.org/abs/hep-ph/9801375}{hep-ph/9801375}].
%%CITATION = HEP-PH/9801375;%%

\bibitem{Laenen:1998qw}
E.~Laenen, G.~Oderda, and G.~Sterman,
\newblock Phys. Lett. {\bf B438}, 173 (1998),
  [\href{http://arXiv.org/abs/hep-ph/9806467}{hep-ph/9806467}].
%%CITATION = HEP-PH/9806467;%%

\bibitem{Catani:2003zt}
S.~Catani, D.~de~Florian, M.~Grazzini, and P.~Nason,
\newblock JHEP {\bf 07}, 028 (2003),
  [\href{http://arXiv.org/abs/hep-ph/0306211}{hep-ph/0306211}].
%%CITATION = HEP-PH/0306211;%%

\bibitem{Idilbi:2006dg}
A.~Idilbi, X.~dong Ji, and F.~Yuan,
\newblock Nucl. Phys. B {\bf 753}, 42 (2006),
  [\href{http://arXiv.org/abs/hep-ph/0605068}{hep-ph/0605068}].
%%CITATION = HEP-PH/0605068;%%

\bibitem{Becher:2007ty}
T.~Becher, M.~Neubert, and G.~Xu,
\newblock JHEP {\bf 07}, 030 (2008),
  [\href{http://arXiv.org/abs/arXiv:0710.0680}{arXiv:0710.0680}].
%%CITATION = 0710.0680;%%

\bibitem{Chiu:2008vv}
J.-y. Chiu, R.~Kelley, and A.~V. Manohar,
\newblock Phys. Rev. D {\bf 78}, 073006 (2008),
  [\href{http://arXiv.org/abs/arXiv:0806.1240}{arXiv:0806.1240}].
%%CITATION = 0806.1240;%%

\bibitem{Campbell:2006wx}
J.~M. Campbell, J.~W. Huston, and W.~J. Stirling,
\newblock Rept. Prog. Phys. {\bf 70}, 89 (2007),
  [\href{http://arXiv.org/abs/hep-ph/0611148}{hep-ph/0611148}].
%%CITATION = HEP-PH/0611148;%%

\bibitem{Trott:2006bk}
M.~Trott,
\newblock Phys. Rev. D {\bf 75}, 054011 (2007),
  [\href{http://arXiv.org/abs/hep-ph/0608300}{hep-ph/0608300}].
%%CITATION = HEP-PH/0608300;%%

\bibitem{Bauer:2008jx}
C.~W. Bauer, A.~Hornig, and F.~J. Tackmann,
\newblock Phys. Rev. D {\bf 79}, 114013 (2009),
  [\href{http://arXiv.org/abs/arXiv:0808.2191}{arXiv:0808.2191}].
%%CITATION = 0808.2191;%%

\bibitem{Banfi:2004nk}
A.~Banfi, G.~P. Salam, and G.~Zanderighi,
\newblock JHEP {\bf 08}, 062 (2004),
  [\href{http://arXiv.org/abs/hep-ph/0407287}{hep-ph/0407287}].
%%CITATION = HEP-PH/0407287;%%

\bibitem{Banfi:2005mt}
A.~Banfi {\em et~al.},
\newblock \href{http://arXiv.org/abs/hep-ph/0508096}{hep-ph/0508096}.
%%CITATION = HEP-PH/0508096;%%

\bibitem{Dasgupta:2001sh}
M.~Dasgupta and G.~P. Salam,
\newblock Phys. Lett. B {\bf 512}, 323 (2001),
  [\href{http://arXiv.org/abs/hep-ph/0104277}{hep-ph/0104277}].
%%CITATION = HEP-PH/0104277;%%

\bibitem{Berger:2001ns}
C.~F. Berger, T.~Kucs, and G.~Sterman,
\newblock Phys. Rev. D {\bf 65}, 094031 (2002),
  [\href{http://arXiv.org/abs/hep-ph/0110004}{hep-ph/0110004}].
%%CITATION = HEP-PH/0110004;%%

\bibitem{Appleby:2003sj}
R.~B. Appleby and M.~H. Seymour,
\newblock JHEP {\bf 09}, 056 (2003),
  [\href{http://arXiv.org/abs/hep-ph/0308086}{hep-ph/0308086}].
%%CITATION = HEP-PH/0308086;%%

\bibitem{Forshaw:2006fk}
J.~R. Forshaw, A.~Kyrieleis, and M.~H. Seymour,
\newblock JHEP {\bf 08}, 059 (2006),
  [\href{http://arXiv.org/abs/hep-ph/0604094}{hep-ph/0604094}].
%%CITATION = HEP-PH/0604094;%%

\bibitem{Bauer:2000ew}
C.~W. Bauer, S.~Fleming, and M.~E. Luke,
\newblock Phys. Rev. D {\bf 63}, 014006 (2000),
  [\href{http://arXiv.org/abs/hep-ph/0005275}{hep-ph/0005275}].
%%CITATION = HEP-PH 0005275;%%

\bibitem{Bauer:2000yr}
C.~W. Bauer, S.~Fleming, D.~Pirjol, and I.~W. Stewart,
\newblock Phys. Rev. D {\bf 63}, 114020 (2001),
  [\href{http://arXiv.org/abs/hep-ph/0011336}{hep-ph/0011336}].
%%CITATION = HEP-PH 0011336;%%

\bibitem{Bauer:2001ct}
C.~W. Bauer and I.~W. Stewart,
\newblock Phys. Lett. B {\bf 516}, 134 (2001),
  [\href{http://arXiv.org/abs/hep-ph/0107001}{hep-ph/0107001}].
%%CITATION = HEP-PH 0107001;%%

\bibitem{Bauer:2001yt}
C.~W. Bauer, D.~Pirjol, and I.~W. Stewart,
\newblock Phys. Rev. D {\bf 65}, 054022 (2002),
  [\href{http://arXiv.org/abs/hep-ph/0109045}{hep-ph/0109045}].
%%CITATION = HEP-PH 0109045;%%

\bibitem{Collins:1988ig}
J.~C. Collins, D.~E. Soper, and G.~Sterman,
\newblock Nucl. Phys. B {\bf 308}, 833 (1988).
%%CITATION = NUPHA,B308,833;%%

\bibitem{Aybat:2008ct}
S.~M. Aybat and G.~Sterman,
\newblock Phys. Lett. B {\bf 671}, 46 (2009),
  [\href{http://arXiv.org/abs/arXiv:0811.0246}{arXiv:0811.0246}].
%%CITATION = 0811.0246;%%

\bibitem{Fleming:2006cd}
S.~Fleming, A.~K. Leibovich, and T.~Mehen,
\newblock Phys. Rev. D {\bf 74}, 114004 (2006),
  [\href{http://arXiv.org/abs/hep-ph/0607121}{hep-ph/0607121}].
%%CITATION = HEP-PH/0607121;%%

\bibitem{Gribov:1972ri}
V.~N. Gribov and L.~N. Lipatov,
\newblock Sov. J. Nucl. Phys. {\bf 15}, 438 (1972).
%%CITATION = SJNCA,15,438;%%

\bibitem{Georgi:1951sr}
H.~Georgi and H.~D. Politzer,
\newblock Phys. Rev. D {\bf 9}, 416 (1974).
%%CITATION = PHRVA,D9,416;%%

\bibitem{Gross:1974cs}
D.~J. Gross and F.~Wilczek,
\newblock Phys. Rev. D {\bf 9}, 980 (1974).
%%CITATION = PHRVA,D9,980;%%

\bibitem{Altarelli:1977zs}
G.~Altarelli and G.~Parisi,
\newblock Nucl. Phys. B {\bf 126}, 298 (1977).
%%CITATION = NUPHA,B126,298;%%

\bibitem{Dokshitzer:1977sg}
Y.~L. Dokshitzer,
\newblock Sov. Phys. JETP {\bf 46}, 641 (1977).
%%CITATION = SPHJA,46,641;%%

\bibitem{Sjostrand:2006za}
T.~Sj{\"o}strand, S.~Mrenna, and P.~Skands,
\newblock JHEP {\bf 05}, 026 (2006),
  [\href{http://arXiv.org/abs/hep-ph/0603175}{hep-ph/0603175}].
%%CITATION = HEP-PH/0603175;%%

\bibitem{Sjostrand:2007gs}
T.~Sj{\"o}strand, S.~Mrenna, and P.~Skands,
\newblock Comput. Phys. Commun. {\bf 178}, 852 (2008),
  [\href{http://arXiv.org/abs/arXiv:0710.3820}{arXiv:0710.3820}].
%%CITATION = 0710.3820;%%

\bibitem{Corcella:2000bw}
G.~Corcella {\em et~al.},
\newblock JHEP {\bf 01}, 010 (2001),
  [\href{http://arXiv.org/abs/hep-ph/0011363}{hep-ph/0011363}].
%%CITATION = HEP-PH/0011363;%%

\bibitem{Bahr:2008pv}
M.~Bahr {\em et~al.},
\newblock Eur. Phys. J. C {\bf 58}, 639 (2008),
  [\href{http://arXiv.org/abs/arXiv:0803.0883}{arXiv:0803.0883}].
%%CITATION = 0803.0883;%%

\bibitem{Sjostrand:2004pf}
T.~Sj{\"o}strand and P.~Z. Skands,
\newblock JHEP {\bf 03}, 053 (2004),
  [\href{http://arXiv.org/abs/hep-ph/0402078}{hep-ph/0402078}].
%%CITATION = HEP-PH/0402078;%%

\bibitem{Sjostrand:2004ef}
T.~Sj{\"o}strand and P.~Z. Skands,
\newblock Eur. Phys. J. C {\bf 39}, 129 (2005),
  [\href{http://arXiv.org/abs/hep-ph/0408302}{hep-ph/0408302}].
%%CITATION = HEP-PH/0408302;%%

\bibitem{Butterworth:1996zw}
J.~M. Butterworth, J.~R. Forshaw, and M.~H. Seymour,
\newblock Z. Phys. C {\bf 72}, 637 (1996),
  [\href{http://arXiv.org/abs/hep-ph/9601371}{hep-ph/9601371}].
%%CITATION = HEP-PH/9601371;%%

\bibitem{Bahr:2008dy}
M.~Bahr, S.~Gieseke, and M.~H. Seymour,
\newblock JHEP {\bf 07}, 076 (2008),
  [\href{http://arXiv.org/abs/arXiv:0803.3633}{arXiv:0803.3633}].
%%CITATION = 0803.3633;%%

\bibitem{Affolder:2001xt}
CDF Collaboration, A.~A. Affolder {\em et~al.},
\newblock Phys. Rev. D {\bf 65}, 092002 (2002).
%%CITATION = PHRVA,D65,092002;%%

\bibitem{Acosta:2004wqa}
CDF Collaboration, D.~E. Acosta {\em et~al.},
\newblock Phys. Rev. D {\bf 70}, 072002 (2004),
  [\href{http://arXiv.org/abs/hep-ex/0404004}{hep-ex/0404004}].
%%CITATION = HEP-EX/0404004;%%

\bibitem{Kar:2008zza}
D.~Kar,
\newblock FERMILAB-THESIS-2008-54  (2008).

\bibitem{Sjostrand:1985xi}
T.~Sj{\"o}strand,
\newblock Phys. Lett. B {\bf 157}, 321 (1985).
%%CITATION = PHLTA,B157,321;%%

\bibitem{Bengtsson:1986gz}
M.~Bengtsson, T.~Sjostrand, and M.~{van Zijl},
\newblock Z. Phys. C {\bf 32}, 67 (1986).
%%CITATION = ZEPYA,C32,67;%%

\bibitem{Gribov:1984tu}
L.~V. Gribov, E.~M. Levin, and M.~G. Ryskin,
\newblock Phys. Rept. {\bf 100}, 1 (1983).
%%CITATION = PRPLC,100,1;%%

\bibitem{Stewart:2010qs}
I.~W. Stewart, F.~J. Tackmann, and W.~J. Waalewijn,
\newblock \href{http://arXiv.org/abs/arXiv:1002.2213}{arXiv:1002.2213}.
%%CITATION = 1002.2213;%%

\bibitem{Bodwin:1984hc}
G.~T. Bodwin,
\newblock Phys. Rev. D {\bf 31}, 2616 (1985).
%%CITATION = PHRVA,D31,2616;%%

\bibitem{Collins:1985ue}
J.~C. Collins, D.~E. Soper, and G.~Sterman,
\newblock Nucl. Phys. B {\bf 261}, 104 (1985).
%%CITATION = NUPHA,B261,104;%%

\bibitem{Altarelli:1979ub}
G.~Altarelli, R.~K. Ellis, and G.~Martinelli,
\newblock Nucl. Phys. B {\bf 157}, 461 (1979).
%%CITATION = NUPHA,B157,461;%%

\bibitem{Hamberg:1990np}
R.~Hamberg, W.~L. van Neerven, and T.~Matsuura,
\newblock Nucl. Phys. B {\bf 359}, 343 (1991).
%%CITATION = NUPHA,B359,343;%%

\bibitem{Harlander:2002wh}
R.~V. Harlander and W.~B. Kilgore,
\newblock Phys. Rev. Lett. {\bf 88}, 201801 (2002),
  [\href{http://arXiv.org/abs/hep-ph/0201206}{hep-ph/0201206}].
%%CITATION = HEP-PH/0201206;%%

\bibitem{Anastasiou:2003yy}
C.~Anastasiou, L.~J. Dixon, K.~Melnikov, and F.~Petriello,
\newblock Phys. Rev. Lett. {\bf 91}, 182002 (2003),
  [\href{http://arXiv.org/abs/hep-ph/0306192}{hep-ph/0306192}].
%%CITATION = HEP-PH/0306192;%%

\bibitem{Anastasiou:2003ds}
C.~Anastasiou, L.~J. Dixon, K.~Melnikov, and F.~Petriello,
\newblock Phys. Rev. D {\bf 69}, 094008 (2004),
  [\href{http://arXiv.org/abs/hep-ph/0312266}{hep-ph/0312266}].
%%CITATION = HEP-PH/0312266;%%

\bibitem{Magnea:1990qg}
L.~Magnea,
\newblock Nucl. Phys. B {\bf 349}, 703 (1991).
%%CITATION = NUPHA,B349,703;%%

\bibitem{Korchemsky:1992xv}
G.~P. Korchemsky and G.~Marchesini,
\newblock Nucl. Phys. B {\bf 406}, 225 (1993).
%%CITATION = HEP-PH 9210281;%%

\bibitem{Catani:1996yz}
S.~Catani, M.~L. Mangano, P.~Nason, and L.~Trentadue,
\newblock Nucl. Phys. B {\bf 478}, 273 (1996),
  [\href{http://arXiv.org/abs/hep-ph/9604351}{hep-ph/9604351}].
%%CITATION = HEP-PH/9604351;%%

\bibitem{Belitsky:1998tc}
A.~V. Belitsky,
\newblock Phys. Lett. B {\bf 442}, 307 (1998),
  [\href{http://arXiv.org/abs/hep-ph/9808389}{hep-ph/9808389}].
%%CITATION = HEP-PH/9808389;%%

\bibitem{Moch:2005ky}
S.~Moch and A.~Vogt,
\newblock Phys. Lett. B {\bf 631}, 48 (2005),
  [\href{http://arXiv.org/abs/hep-ph/0508265}{hep-ph/0508265}].
%%CITATION = HEP-PH/0508265;%%

\bibitem{Korchemsky:1996iq}
G.~P. Korchemsky,
\newblock \href{http://arXiv.org/abs/hep-ph/9610207}{hep-ph/9610207}.
%%CITATION = HEP-PH/9610207;%%

\bibitem{Berger:2003iw}
C.~F. Berger, T.~Kucs, and G.~Sterman,
\newblock Phys. Rev. D {\bf 68}, 014012 (2003),
  [\href{http://arXiv.org/abs/hep-ph/0303051}{hep-ph/0303051}].
%%CITATION = HEP-PH/0303051;%%

\bibitem{Sterman:1978bj}
G.~Sterman,
\newblock Phys. Rev. D {\bf 17}, 2789 (1978).
%%CITATION = PHRVA,D17,2789;%%

\bibitem{Hoang:2007vb}
A.~H. Hoang and I.~W. Stewart,
\newblock Phys. Lett. B {\bf 660}, 483 (2008),
  [\href{http://arXiv.org/abs/arXiv:0709.3519}{arXiv:0709.3519}].
%%CITATION = 0709.3519;%%

\bibitem{Ligeti:2008ac}
Z.~Ligeti, I.~W. Stewart, and F.~J. Tackmann,
\newblock Phys. Rev. D {\bf 78}, 114014 (2008),
  [\href{http://arXiv.org/abs/arXiv:0807.1926}{arXiv:0807.1926}].
%%CITATION = 0807.1926;%%

\bibitem{Catani:1992ua}
S.~Catani, L.~Trentadue, G.~Turnock, and B.~R. Webber,
\newblock Nucl. Phys. B {\bf 407}, 3 (1993).
%%CITATION = NUPHA,B407,3;%%

\bibitem{Fleming:2007qr}
S.~Fleming, A.~H. Hoang, S.~Mantry, and I.~W. Stewart,
\newblock Phys. Rev. D {\bf 77}, 074010 (2008),
  [\href{http://arXiv.org/abs/hep-ph/0703207}{hep-ph/0703207}].
%%CITATION = HEP-PH/0703207;%%

\bibitem{Fleming:2007xt}
S.~Fleming, A.~H. Hoang, S.~Mantry, and I.~W. Stewart,
\newblock Phys. Rev. D {\bf 77}, 114003 (2008),
  [\href{http://arXiv.org/abs/arXiv:0711.2079}{arXiv:0711.2079}].
%%CITATION = 0711.2079;%%

\bibitem{Manohar:2006nz}
A.~V. Manohar and I.~W. Stewart,
\newblock Phys. Rev. D {\bf 76}, 074002 (2007),
  [\href{http://arXiv.org/abs/hep-ph/0605001}{hep-ph/0605001}].
%%CITATION = HEP-PH/0605001;%%

\bibitem{Lee:2006nr}
C.~Lee and G.~Sterman,
\newblock Phys. Rev. D {\bf 75}, 014022 (2007),
  [\href{http://arXiv.org/abs/hep-ph/0611061}{hep-ph/0611061}].
%%CITATION = HEP-PH/0611061;%%

\bibitem{Bauer:2002nz}
C.~W. Bauer, S.~Fleming, D.~Pirjol, I.~Z. Rothstein, and I.~W. Stewart,
\newblock Phys. Rev. D {\bf 66}, 014017 (2002),
  [\href{http://arXiv.org/abs/hep-ph/0202088}{hep-ph/0202088}].
%%CITATION = HEP-PH 0202088;%%

\bibitem{Korchemsky:1987wg}
G.~P. Korchemsky and A.~V. Radyushkin,
\newblock Nucl. Phys. B {\bf 283}, 342 (1987).
%%CITATION = NUPHA,B283,342;%%

\bibitem{Balzereit:1998yf}
C.~Balzereit, T.~Mannel, and W.~Kilian,
\newblock Phys. Rev. D {\bf 58}, 114029 (1998),
  [\href{http://arXiv.org/abs/hep-ph/9805297}{hep-ph/9805297}].
%%CITATION = HEP-PH/9805297;%%

\bibitem{Neubert:2004dd}
M.~Neubert,
\newblock Eur. Phys. J. C {\bf 40}, 165 (2005),
  [\href{http://arXiv.org/abs/hep-ph/0408179}{hep-ph/0408179}].
%%CITATION = HEP-PH/0408179;%%

\bibitem{Martin:2009iq}
A.~D. Martin, W.~J. Stirling, R.~S. Thorne, and G.~Watt,
\newblock Eur. Phys. J. C {\bf 63}, 189 (2009),
  [\href{http://arXiv.org/abs/arXiv:0901.0002}{arXiv:0901.0002}].
%%CITATION = 0901.0002;%%

\bibitem{Bauer:2002aj}
C.~W. Bauer, D.~Pirjol, and I.~W. Stewart,
\newblock Phys. Rev. D {\bf 67}, 071502(R) (2003),
  [\href{http://arXiv.org/abs/hep-ph/0211069}{hep-ph/0211069}].
%%CITATION = HEP-PH 0211069;%%

\bibitem{Bauer:2002ie}
C.~W. Bauer, A.~V. Manohar, and M.~B. Wise,
\newblock Phys. Rev. Lett. {\bf 91}, 122001 (2003),
  [\href{http://arXiv.org/abs/hep-ph/0212255}{hep-ph/0212255}].
%%CITATION = HEP-PH/0212255;%%

\bibitem{Bauer:2003di}
C.~W. Bauer, C.~Lee, A.~V. Manohar, and M.~B. Wise,
\newblock Phys. Rev. D {\bf 70}, 034014 (2004),
  [\href{http://arXiv.org/abs/hep-ph/0309278}{hep-ph/0309278}].
%%CITATION = HEP-PH/0309278;%%

\bibitem{Chay:2004zn}
J.~Chay, C.~Kim, Y.~G. Kim, and J.-P. Lee,
\newblock Phys. Rev. D {\bf 71}, 056001 (2005),
  [\href{http://arXiv.org/abs/hep-ph/0412110}{hep-ph/0412110}].
%%CITATION = HEP-PH/0412110;%%

\bibitem{Arnesen:2005nk}
C.~M. Arnesen, J.~Kundu, and I.~W. Stewart,
\newblock Phys. Rev. D {\bf 72}, 114002 (2005),
  [\href{http://arXiv.org/abs/hep-ph/0508214}{hep-ph/0508214}].
%%CITATION = HEP-PH/0508214;%%

\bibitem{Idilbi:2008vm}
A.~Idilbi and A.~Majumder,
\newblock Phys. Rev. D {\bf 80}, 054022 (2009),
  [\href{http://arXiv.org/abs/arXiv:0808.1087}{arXiv:0808.1087}].
%%CITATION = 0808.1087;%%

\bibitem{Donoghue:2009cq}
J.~F. Donoghue and D.~Wyler,
\newblock \href{http://arXiv.org/abs/arXiv:0908.4559}{arXiv:0908.4559}.
%%CITATION = 0908.4559;%%

\bibitem{Collins:1981ta}
J.~C. Collins and G.~Sterman,
\newblock Nucl. Phys. B {\bf 185}, 172 (1981).
%%CITATION = NUPHA,B185,172;%%

\bibitem{Manohar:2002fd}
A.~V. Manohar, T.~Mehen, D.~Pirjol, and I.~W. Stewart,
\newblock Phys. Lett. B {\bf 539}, 59 (2002),
  [\href{http://arXiv.org/abs/hep-ph/0204229}{hep-ph/0204229}].
%%CITATION = HEP-PH 0204229;%%

\bibitem{Manohar:2003vb}
A.~V. Manohar,
\newblock Phys. Rev. D {\bf 68}, 114019 (2003),
  [\href{http://arXiv.org/abs/hep-ph/0309176}{hep-ph/0309176}].
%%CITATION = HEP-PH/0309176;%%

\bibitem{Becher:2006mr}
T.~Becher, M.~Neubert, and B.~D. Pecjak,
\newblock JHEP {\bf 01}, 076 (2007),
  [\href{http://arXiv.org/abs/hep-ph/0607228}{hep-ph/0607228}].
%%CITATION = HEP-PH/0607228;%%

\bibitem{Kramer:1986sg}
G.~Kramer and B.~Lampe,
\newblock Z. Phys. C {\bf 34}, 497 (1987),
\newblock [Erratum-ibid.\ C {\bf 42}, 504 (1989)].
%%CITATION = ZEPYA,C34,497;%%

\bibitem{Matsuura:1987wt}
T.~Matsuura and W.~L. {van Neerven},
\newblock Z. Phys. C {\bf 38}, 623 (1988).
%%CITATION = ZEPYA,C38,623;%%

\bibitem{Matsuura:1988sm}
T.~Matsuura, S.~C. {van der Marck}, and W.~L. {van Neerven},
\newblock Nucl. Phys. B {\bf 319}, 570 (1989).
%%CITATION = NUPHA,B319,570;%%

\bibitem{Gehrmann:2005pd}
T.~Gehrmann, T.~Huber, and D.~Maitre,
\newblock Phys. Lett. B {\bf 622}, 295 (2005),
  [\href{http://arXiv.org/abs/hep-ph/0507061}{hep-ph/0507061}].
%%CITATION = HEP-PH/0507061;%%

\bibitem{Kniehl:1989bb}
B.~A. Kniehl and J.~H. Kuhn,
\newblock Phys. Lett. B {\bf 224}, 229 (1989).
%%CITATION = PHLTA,B224,229;%%

\bibitem{Kniehl:1989qu}
B.~A. Kniehl and J.~H. Kuhn,
\newblock Nucl. Phys. B {\bf 329}, 547 (1990).
%%CITATION = NUPHA,B329,547;%%

\bibitem{Bernreuther:2005rw}
W.~Bernreuther {\em et~al.},
\newblock Nucl. Phys. B {\bf 723}, 91 (2005),
  [\href{http://arXiv.org/abs/hep-ph/0504190}{hep-ph/0504190}].
%%CITATION = HEP-PH/0504190;%%

\bibitem{Chiu:2007dg}
J.-y. Chiu, F.~Golf, R.~Kelley, and A.~V. Manohar,
\newblock Phys. Rev. D {\bf 77}, 053004 (2008),
  [\href{http://arXiv.org/abs/arXiv:0712.0396}{arXiv:0712.0396}].
%%CITATION = 0712.0396;%%

\bibitem{Chiu:2009mg}
J.-y. Chiu, A.~Fuhrer, R.~Kelley, and A.~V. Manohar,
\newblock Phys. Rev. D {\bf 80}, 094013 (2009),
  [\href{http://arXiv.org/abs/arXiv:0909.0012}{arXiv:0909.0012}].
%%CITATION = 0909.0012;%%

\bibitem{Korchemsky:1999kt}
G.~P. Korchemsky and G.~Sterman,
\newblock Nucl. Phys. B {\bf 555}, 335 (1999),
  [\href{http://arXiv.org/abs/hep-ph/9902341}{hep-ph/9902341}].
%%CITATION = HEP-PH/9902341;%%

\bibitem{Korchemsky:2000kp}
G.~P. Korchemsky and S.~Tafat,
\newblock JHEP {\bf 10}, 010 (2000),
  [\href{http://arXiv.org/abs/hep-ph/0007005}{hep-ph/0007005}].
%%CITATION = HEP-PH/0007005;%%

\bibitem{Schwartz:2007ib}
M.~D. Schwartz,
\newblock Phys. Rev. D {\bf 77}, 014026 (2008),
  [\href{http://arXiv.org/abs/arXiv:0709.2709}{arXiv:0709.2709}].
%%CITATION = 0709.2709;%%

\bibitem{Hoang:2008fs}
A.~H. Hoang and S.~Kluth,
\newblock \href{http://arXiv.org/abs/arXiv:0806.3852}{arXiv:0806.3852}.
%%CITATION = 0806.3852;%%

\end{thebibliography}

\end{document}